\documentclass[aps,prx,twocolumn,superscriptaddress,showpacs,floatfix]{revtex4-2}
\usepackage{comment}
\usepackage{physics}
\usepackage{amsmath,amsthm,amssymb,amsfonts, color, comment, graphicx}%, 
\usepackage{xcolor}
\usepackage{float}
\usepackage{graphicx}
\usepackage{dcolumn}% Align table columns on decimal point
\usepackage{bm}% bold math
\UseRawInputEncoding
\usepackage{hyperref}% add hypertext capabilities

\usepackage[normalem]{ulem}

\newcommand{\bea}{\begin{eqnarray}}
\newcommand{\eea}{\end{eqnarray}}

\newcommand{\br}{\mathbf{r}}

\newcommand{\be}{\begin{equation}}
\newcommand{\ee}{\end{equation}}
\newcommand{\bk}{{{\bf{k}}}}

\newcommand{\bq}{{\bf{q}}}

\newcommand{\beal}{\begin{align}}
\newcommand{\eeal}{\end{align}}

\newcommand{\pdg}{{\phantom\dagger}}

\newcommand{\btjstrw}{\mathrel{{\rotatebox[origin=c]{90}
{$\bowtie$}}\kern-0.18em\raisebox{-.95ex}{$\bullet$}
\kern-0.5em\raisebox{.97ex}{$\bullet$}
\kern-1.12em\raisebox{.97ex}{$\bullet$}
\kern-0.52em\raisebox{-.95ex}{$\bullet$}}}

\newcommand{\btjnbrR}{{\mathrel{\rotatebox[origin=c]{90}
{$\bowtie$}}\kern-0.22em\raisebox{.9ex}{$\bullet$}
\kern-1.em\raisebox{-.8ex}{$\bullet$}}}
\newcommand{\btjnbrL}{{\mathrel{\rotatebox[origin=c]{90}
{$\bowtie$}}\kern-0.22em\raisebox{-.8ex}{$\bullet$}
\kern-1.em\raisebox{+.9ex}{$\bullet$}}}

\def\a{\alpha}
\def\b{\beta}

\def\d{\delta}
\def\e{\epsilon}

\def\k{\kappa}

\def\s{\sigma}

\def\w{\omega}

\newcommand{\llangle}[1][]{\savebox{\@brx}{\(\m@th{#1\langle}\)}%
  \mathopen{\copy\@brx\kern-0.5\wd\@brx\usebox{\@brx}}}
\newcommand{\rrangle}[1][]{\savebox{\@brx}{\(\m@th{#1\rangle}\)}%
  \mathclose{\copy\@brx\kern-0.5\wd\@brx\usebox{\@brx}}}

\usepackage{graphicx}
\usepackage[caption=false]{subfig}

\begin{document}

\preprint{APS/123-QED}

\title{Orbital selective order and $\mathbb{Z}_3$ Potts nematicity from a 
	non-Fermi liquid}% Force line breaks with \\

\author{YuZheng Xie}
\author{Andrew Hardy}
\author{Arun Paramekanti}
\affiliation{Department of Physics, University of Toronto, 60 St. George Street, Toronto, ON, M5S 1A7 Canada}

\date{\today}% It is always \today, today,
%  but any date may be explicitly specified

\begin{abstract}
Motivated by systems where a high temperature non-Fermi liquid  gives way to low temperature $\mathbb{Z}_3$ Potts  nematic order, we studied a three-orbital Sachdev-Ye-Kitaev (SYK) model in the large-$N$ limit. 
In the single-site limit, this model exhibits a spontaneous orbital-selective transition which preserves average particle-hole symmetry, with two orbitals becoming insulators while the third orbital remains a non-Fermi liquid down to zero temperature. 
We extend this study to lattice models of three-orbital SYK dots, exploring uniform symmetry broken states on the triangular and cubic lattices. At high temperature, these lattice models exhibit an isotropic non-Fermi liquid metal phase.
On the three-dimensional (3D) cubic lattice, the low temperature uniform $\mathbb{Z}_3$ nematic state corresponds to an orbital selective layered state which preserves particle-hole symmetry at small hopping and spontaneously breaks the particle-hole symmetry at large hopping.
Over a wide range of temperature, the transport in this layered state shows metallic in-plane resistivity but insulating out-of-plane resistivity.
On the 2D triangular lattice, the low temperature state with uniform orbital order is also a correlated $\mathbb{Z}_3$ nematic with orbital-selective transport but it remains metallic in both principal directions.
We discuss a Landau theory with $\mathbb{Z}_3$ clock terms which captures salient features of the phase diagram and nematic order in all these models. 
We also present results on the approximate wavevector dependent orbital susceptibility of the isotropic non-Fermi liquid states.
\end{abstract}

\maketitle

%\tableofcontents
%\section{Introduction}

The breaking of the crystalline rotational symmetry for electrons in solids, 
termed `electronic nematic order' \cite{Nematic_Review_Fradkin_ARCMP2010}
in analogy with liquid crystals, has been reported in diverse quantum materials ranging from semiconductor 2D electron gases in the quantum Hall regime \cite{Nematic_Review_Fradkin_ARCMP2010} to  unconventional superconductors such as  heavy-fermion systems \cite{Okazaki2011, Seo2020},
CuO-,  and Fe- based superconductors \cite{Kivelson2003, Hinkov2008, Chu2012,bohmer2022, fernandes2014}, and disordered Sr$_3$Ru$_2$O$_7$ \cite{Nematic327_Borzi_Science2007,SDW327_Lester_NatMat2015,nematic_kee_PRB2005,nematic_327_raghu_PRB2009}.
The microscopic origin for nematicity could be 
underlying orbital order or Fermi surface Pomeranchuk instabilities \cite{nematic_kee_PRB2004,nematic_metzner_PRB2006} or it could be the
vestigial remnant 
of unidirectional spin-density wave or charge-density wave orders \cite{Vestigial_Cuprates_Nie_PRB2017,VestigialNematic_Review_Fernandes_ARCMP2019}.
Many of these correlated materials also exhibit a high temperature non-Fermi liquid (nFL) regime, which may or may not be directly tied to nematic fluctuations \cite{biermannNonFermiLiquidBehavior2005, bergMonteCarloStudies2019, Licciardello2019}. 
It is, therefore, interesting to explore the phenomenology of model systems which exhibit nFL behavior at high temperature but form broken symmetry nematic states upon cooling below a critical temperature. \par

While the above quantum materials exhibit Ising-nematic ($\mathbb{Z}_2$) order, which continues to be thoroughly investigated \cite{Sachdev_2011}, the crystalline symmetry of other materials allows for a 
more complex $\mathbb{Z}_3$ symmetry breaking order \cite{chakraborty2023}.
Such $\mathbb{Z}_3$ nematics may be described by effective Potts models or clock models, and have recently been invoked in a number of materials including 2D electron gases at (111) surfaces or interfaces  \cite{Boudjada2018,KTO_Nematic_PRB2021,KTO_Nematic_AnnPhys_2021}, 
topological insulator Bi$_2$Se$_3$ \cite{Sun2019,cho2020}, van der Waals magnets
FePS$_3$ and Fe$_{1/3}$NbS$_2$ \cite{little2020, kimOrbitalselectiveMottPhase2022, nie2023},and other frustrated
magnets \cite{Z3_Mulder_PRB2010, Z3_Orth_PRB2023, Z3_Vaknin_CaMn2P2_PRB2023}. Recently, several intriguing metallic nematic orders have been observed in twisted Moir\'e crystals of both graphene and TMDs \cite{Nematic_Moire_Abhay_Nature2019,jiangChargeOrderBroken2019,cao_nematicity_2021, rubio-verdu_moire_2022, Nematic_Moire_NadjPerge_NatPhys2019,jinStripePhasesWSe22021,xu_moireZ3_prb2020}. The microscopic origin of this  $\mathbb{Z}_3$ nematicity eludes the community currently. The effective model proposed here may provide some insight into such an enigmatic phase.
The interplay of $\mathbb{Z}_3$ symmetry breaking and nFL physics remains  underexplored.
For instance, new types of fermion induced quantum critical points \cite{Li2017, Zhen2020, xu_moireZ3_prb2020} have been recently shown to occur for
Dirac fermions coupled to $\mathbb{Z}_3$ orders, with quantum fluctuations tuning from classical first-order transitions to continuous transitions. Such Potts nematic orders coupled to fermions are therefore of great interest.
In addition, the increased number of degrees of freedom allows for the possibility of spontaneous orbital selective transitions, 
similar to an orbital selective Mott transition (OSMT) \cite{koga2004,mediciOrbitalSelectiveMott2005,mediciOrbitalSelectiveMottTransition2009, Werner2007,Werner2009, Yu2018}. 
These questions require an approach that provides information about the dynamical nature of the excitations as well as the mean-field 
order. \par 

The Sachdev-Ye-Kitaev (SYK) model with random interactions between $N$ electronic `flavors' is a paradigmatic example of a nFL in the $N = \infty$ limit, featuring a linear-in-T scattering rate and a residual entropy at low temperature \cite{sachdev1993gapless, Maldacena2016, Chowdhury2021b}. 
Combinations of SYK dots can lead to a variety of phase transitions which alter the order and dynamics of these coupled systems \cite{Banerjee2017, Haldar2018, sahoo2020, Chowdhury_PRR2020, lantagne2021}. 
When such SYK dots are arranged on a lattice with uniform or random hopping, they exhibit a crossover from a high temperature nFL to a low temperature FL \cite{Song2017a, Chowdhury2018}.
Motivated by exploring the physics of multi-orbital nFLs, we previously explored a two-orbital variant of the SYK model on the square lattice which might be thought to mimic $d_{xz}$ and $d_{yz}$ orbitals \cite{hardy2023}.
In addition to the nFL-FL crossover upon cooling, this model exhibits a low temperature Ising nematic phase. 
Weak uniaxial strain of $B_{1g}$ symmetry splits the orbital degeneracy, and  was shown to lead to a strong enhancement of elastoresistivity in the vicinity of the nematic transition, as experimentally observed in the iron pnictides and selenides 
\cite{kuo2013, kuo_strain_science2016, palmstrom2022}.

In this paper, we extend our earlier work and discuss the physics of three-orbital SYK models as a route to exploring $\mathbb{Z}_3$ nematic orders and its interplay with
nFL physics.
Our main results are the following. (i) For a single three-orbital SYK dot model with particle-hole symmetry (PHS), we find that the 
high temperature phase is a fully symmetric nFL. Below a critical temperature $T_{\rm orb}$, the orbital symmetry is broken. This
leads to an orbital selective nFL phase at low  $T < T_{\rm orb}$, 
where one orbital remains a gapless nFL at half-filling, while the two other orbitals become gapped as their densities symmetrically deviate away from half-filling. 
We generically label such phases as $[+,0,-]$ to
represent the density deviation from half-filling of the three orbitals which preserves average PHS. 
At larger inter-orbital interaction, we also
find an insulating phase which spontaneously breaks PHS and leads to a spontaneous density deviation from half-filling. We generically label such phases as $[+,-,-]$ to denote that the density of two
orbitals stays equal but they become distinct from that of the third. We distinguish `Insulating' and `Metallic' phases
of this type as $[+,-,-]_{\rm I}$ and $[+,-,-]_{\rm M}$. The $[+,-,-]_{\rm I}$ phase exhibits strong PHS breaking, 
%a gapped spectral function, 
and a strong deviation of the average density from half-filling. The $[+,-,-]_{\rm M}$ remains a gapless
phase with weak PHS breaking and a weak deviation of the average density from half-filling.
(ii) Viewing the three orbitals as `$t_{2g}$' type orbitals,
we studied a cubic lattice model of such three-orbital SYK dots, where each orbital has two strong hopping
directions (easy-plane hopping $t_{\parallel}$) and one direction with weaker hopping (hard-axis hopping $t_\perp \ll t_\parallel$). 
We explore possible translationally 
invariant symmetry broken phases of this model, finding that the high $T$ symmetry nFL phase gives way to 
an orbital selective transition: depending on the model interactions, the 
orbital order is into a $[+,0,-]$ phase or a $[+,-,-]_{\rm M}$ phase. 
These nematic phases correspond to the spontaneous layering of the three-dimensional (3D) cubic lattice into 2D layers, leading to strongly anisotropic transport. Over a wide range of $T$, the
$T$ dependence of the resistivity is metallic within the layers and strongly insulating across the layers, a dichotomy
reminiscent of the phenomenology seen in cuprate high $T$ superconductors and in dynamical mean field theory studies of Hubbard-type 
models \cite{caxis_Kumar_ModPhysLett1997,caxis_IoffeMillis_PRB1998,caxis_Tremblay_PRB2013}. 
Eventually, for $T \ll T_{\rm orb}$, the hard-axis resistivity exhibits a metallic downturn, and
we enter a highly anisotropic 3D metal phase with resistive anisotropy 
scaling as $\rho_\perp/\rho_\parallel \sim (t_\parallel/t_\perp)^2$.
(iii) We next study an effective 2D triangular lattice model of such three-orbital SYK dots. This lattice model may be 
viewed as an effective model of a
(111) layer of the 3D cubic lattice which has been used to describe 2D electron gases at
oxide surfaces and interfaces \cite{Boudjada2018, KTO_Nematic_PRB2021}. 
In this case, the lattice dispersion breaks PHS, and we find that for $T < T_{\rm orb}$
this model generically transitions to an $[a,b,c]$ phase, where none of the orbital densities are
equal and all triangular lattice
symmetries are broken.
This phase displays metallic transport in both directions in the plane, but with a resistive anisotropy tied to the orbital
order.  The transitions of these models follow similar phenomenology of general nematic transitions \cite{Kee2003, Khavkine2004}. They exhibit first-order transitions at sharp density of states, with tricritical points separating a continuous transition at broader density of states. The broadening occurs due to either hopping $t$ or temperature $T$.
(iv) We construct a Landau theory, for the $\mathbb{Z}_3$
clock or Potts order, which captures salient features of our numerical phase diagrams.

An outline of our paper is as follows.
We begin in Sec.~\ref{Section I: three orbital dot}  with a discussion of the isolated three-orbital SYK dot, followed 
by a discussion in Sec.~\ref{Section III: Lattice Extensions} of three-orbital SYK dots
arranged on a lattice, outlining the free energy, spectral function, and transport calculations. 
Sec.~\ref{Section III: Cubic lattice model} and Sec.~\ref{Section IV: Triangular lattice model} discuss results on the phase diagram, orbital selective transitions,
spectral functions, and transport on the 3D cubic lattice and 2D triangular lattice respectively.
In Sec.~\ref{Section V: Landau Theory of Z3 nematic order}, we present a Landau theory, which captures salient features of the 
phase diagram and nematic order in all the models we
 studied. Sec.~\ref{Section VI: Susceptibility} discusses the wavevector-dependent susceptibility for non-uniform orbital ordering and its
implications. We end in Sec.~\ref{Section VII: Outlook} with a summary of our main results.

\section{Three Orbital SYK-Dot} \label{Section I: three orbital dot}
\subsection{Model and large-$N$ solution}
We begin with a three-orbital SYK model with each orbital filled with $N$ fermion modes, described by random all-to-all intraorbital 
couplings. The three orbitals are further coupled by SYK-like interorbital interactions. The Hamiltonian is of the form
\begin{eqnarray}
	\label{SYKDot}
	\!\!\!\!\! H_{\rm SYK} &=& H_{\text{SYK}}^{\rm intra} + H_{\text{SYK}}^{\rm inter} - \mu \sum_{s,i} c_{s,i}^\dag c_{s,i}^\pdg , \notag \\
	\!\!\!\!\! H_{\text{SYK}}^{\rm intra} &=& \!\!\!\! \sum_{s,(ijkl)} \!\!\!\!\! J^{(s)}_{ij;kl} c_{s,i}^\dag c_{s,j}^\dag c^\pdg_{s,k} c^\pdg_{s,l} , \notag \\
	\!\!\!\!\! H_{\text{SYK}}^{\rm inter} &=& \!\!\!\!\!\!\!\!\! \sum_{s < s',(ijkl)} \!\!\!\!\!\!\!\! [V^{(s s')}_{ij;kl} c_{s,i}^\dag c_{s,j}^\dag c^\pdg_{s',k} c^\pdg_{s',l} \!+\! {\rm h.c.}] .
\end{eqnarray}
Here, $s \!=\! 1,2,3$ labels the orbitals, and the fermionic modes are denoted by $i,j,k,l \!=\! 1,\! \cdots, \! N$. The system has a single conserved $U(1)$ charge, corresponding to the total density $n =  \sum_{s,i} \langle c_{s,i}^\dag c_{s,i} \rangle/N$, which can be tuned by a chemical potential $\mu$. 
 We focus on  pair-hopping terms \cite{Banerjee2017, Haldar2018}. Other symmetry-allowed density-density terms do not produce spontaneous orbital order, rather forming various condensates under the appropriate disorder averaging procedure \cite{sahoo2020, Chowdhury_PRR2020,lantagne2021, HardyBose2023} which may
be interpreted in terms of `wormholes' \cite{sahoo2020, Chowdhury_PRR2020,lantagne2021}. 
$J_{ij;kl}^{(s)}, V_{ij;kl}^{(s,s')}$ are uncorrelated random complex numbers having Gaussian distributions with zero mean, and variance set by
\begin{eqnarray}
	\overline{ J_{ij;kl}^{(s) *} J^{(s')}_{ij;kl}} &=& \frac{J^2}{(2N)^3} \delta_{s,s'} , \notag \\
	\overline{ V_{ij;kl}^{(s,s') *}V^{(s,s')}_{ij;kl}} &=& \frac{V^2}{(2N)^3} (1 - \delta_{s, s'}) .
\end{eqnarray}
Each coupling $J_{ij;kl}^{(s)}, V_{ij;kl}^{(s,s')}$ follows the correct anti-symmetry properties.  The $J$ and $V$ in the model describe the magnitude of the intraorbital and interorbital interactions, respectively. 
\par

The appeal of this construction is that the model is solvable at the large $N$ limit after disorder averaging over the distributions. We disorder average over $\{ J_{ij;kl}^{(s)}, V_{ij;kl}^{(s,s')} \}$ via the replica trick and consider replica-diagonal solutions. The behavior of such a large $N$ point is also much better understood and controlled than non-disordered formulations \cite{leeRecentDevelopmentsNonFermi2018, Chowdhury2021b}.
This procedure produces self-consistent Dyson equations given as 
\begin{subequations}\label{DS for 0D}
\begin{align}
	\Sigma_s(\tau) = &-\big(J\big)^2G_s^2(\tau)G_s(-\tau) \notag \\
	&\quad - \sum_{s': s'\neq s} \big(V\big)^2 G_{s'}^2(\tau)G_{s}(-\tau), \label{DS1 for 0D} \\
	G_s(i\omega_n) =& \big[ i\omega_n + \mu  - \Sigma_s(i\omega_n) \big]^{-1} . \label{DS2 for 0D}
\end{align}
\end{subequations}
These equations are solved numerically through an iterative process.  We investigate the model in the Grand Canonical ensemble at $\mu = 0$, where the model in Eq.\eqref{SYKDot} has a particle-hole symmetry (PHS) in the $N \to \infty$ limit.

\subsection{Phase Diagram} \label{Section I B: phase diagram}
\begin{figure}
	\centering
	\includegraphics[width=1\linewidth]{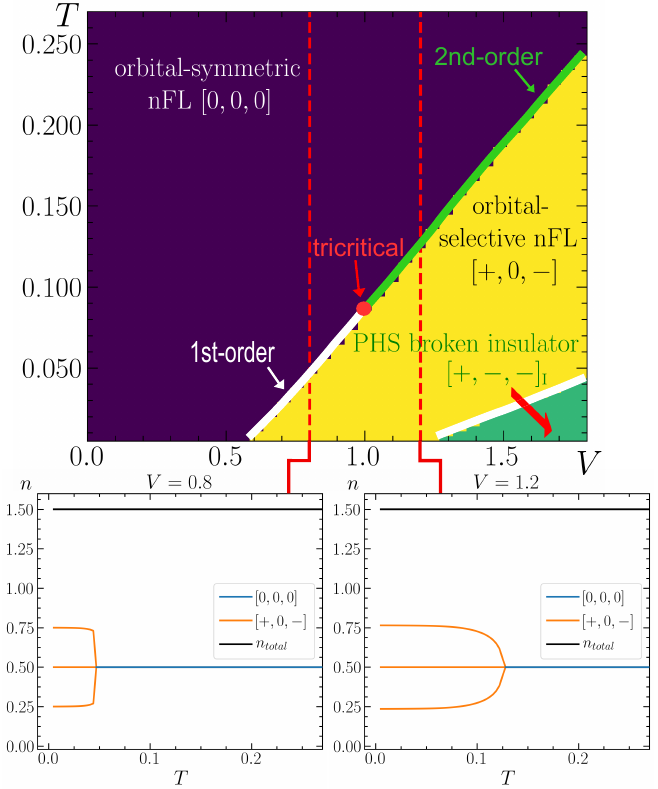} \hfill
	\caption{The upper panel shows the $T$-$V$ phase diagram for the three-orbital SYK model in the Grand Canonical Ensemble. The lower shows the orbital-resolved density $n_s(T)$ and the total density $n(T) = \sum_{s} n_s$ as a function of temperature for $V = 0.8$ and $V = 1.2$. Colors are used to distinguish different phases: $[0,0,0]$ (blue), and $[+,0,-]$ (yellow). In the $[0,0,0]$ phase, the three orbitals are degenerate, and the three orbitals split while entering the $[+,0,-]$ phase.}
	\label{0D phase diagram}
\end{figure}

We work in units where $J=1$; in these units, $V \equiv V/J$ and the temperature $T \equiv T/J$.
Fig.~\ref{0D phase diagram} shows the $T$-$V$ phase diagram of the three-orbital SYK model, which
exhibits three thermodynamically stable phases.

(i) The high $T$ orbital-symmetric nFL phase $[0,0,0]$ is a gapless phase with three degenerate half-filled orbitals,
i.e, $G_s(\tau) \equiv G(\tau)$ independent of $s$.  
It features a large residual entropy $S_{\rm SYK3} (T\to 0) =3 \times S_{\text{SYK1}}(T\to 0) $, thrice 
the residual entropy of the single orbital SYK dot.

(ii) At low $T$ and intermediate $V$, we find the orbital-selective nFL $[+,0,-]$. In this phase, one orbital remains a gapless nFL
at half-filling, while the other two are gapped with symmetrical deviations  from half-filling. The densities are $0.5 \pm \delta n$, preserving PHS on average. 
The $[+,0,-]$ phase has the same residual entropy as the single-dot SYK model. This is elaborated in Appendix \ref{Appendix: Further discussion on the three-orbital SYK dot}

(iii) Finally, at low $T$ and large $V$, we find 
the $[+,-,-]_{\rm I}$ phase, which has two degenerate orbitals, with density  less than $1/2$
while the third orbital has a distinct density.
This phase exhibits strong spontaneous PHS breaking, where the total filling deviate from half-filling with 
$\sum_s \langle n_s \rangle = 1.0$ at all $T$. 
Here, $n_s = G_s(\tau = 0^-)$ represents the orbital-resolved density. 
This phase is characterized by a spectral gap,
a vanishing entropy $S(T) = -\left( \partial \Omega/\partial T \right)_{\mathcal{V},\mu}$, and a vanishing compressibility $\kappa(T) = n^{-2}\left(\partial^2 \Omega/\partial \mu^2\right)_{\mathcal{V},T}$, indicating a
correlated insulator.

Fig.~\ref{0D phase diagram} also presents two illustrative cuts through the phase diagram at $V=0.8$ and $V=1.2$, 
showing the temperature dependence of the orbital
densities $n_s(T)$. Both these cuts depict two phases: the symmetric $[0,0,0]$ phase and the orbital selective
$[+,0,-]$ phase. However, the transition
is first-order for $V = 0.8$ and continuous for $V = 1.2$, as indicated by the (dis)continuity of the orbital densities
at the transition. Further details of the phase diagram, such as the
orbital polarizations and other thermodynamic quantities such as $S(T)$, $C_V(T)$, and $\k(T)$  are provided in Appendix \ref{Appendix: Further discussion on the three-orbital SYK dot}. 
\par

We also find a fourth
competing gapless phase $[+,-,-]_{\rm M}$ over a wide range of $V$ and $T$. However, this phase is never a true 
minimum of $\Omega$ for any $V,T$. Nevertheless, this phase plays an important role in the lattice models 
that will be discussed in Sec.~\ref{Section III: Cubic lattice model} and \ref{Section IV: Triangular lattice model}. 
The details of the procedure for obtaining the phase diagram are provided in Appendix \ref{Appendix: Thermodynamic Numerical Procedure}. \par

\subsection{Orbital Resolved Spectral Functions}\label{Section I C: Spectral Functions}
One intriguing effect of the $\mathbb{Z}_3$ ordering, not present in the two-orbital SYK dot, is the possibility of spontaneous orbital selective behavior. Similar to the phenomenology of Hund's metals, the orbitals here are `differentiated' in that their spectral functions and resulting physical properties differ  \cite{Georges2013}.
Unlike the OSMT, this behavior is not determined by an imposed
crystal field splitting or lattice induced `heavy' or `light' orbitals \cite{mediciOrbitalSelectiveMott2005, mediciOrbitalSelectiveMottTransition2009,kuglerOrbitalSelectiveMottPhase2022}. Rather, the spontaneous orbital ordering induces this differentiation.  This transition is also distinct from a Mott transition in that the gap emerges from orbital order rather than from
an orbital liquid.
The nature of the orbital ordering in turn changes the orbital behavior. To study these effects, we compute the real frequency orbital-resolved local spectral functions $A_s(\w)$.  \par

The $N \rightarrow \infty$ limit provides a major computational advantage. 
To obtain $A_s(\w)$, we follow the method in \cite{Banerjee2017} and analytically continue the self-consistent Dyson equations (Eq.\eqref{DS1 for 0D} and \eqref{DS2 for 0D}) to real frequency. 
This allows us to numerically solve the self-consistent equations directly in real-frequency without needing to resort to ill-conditioned numerical analytic continuation \cite{Levy2017}. \par

\begin{figure}
	\centering
	\subfloat[$V = 0.8$]{\includegraphics[width=0.485\linewidth]{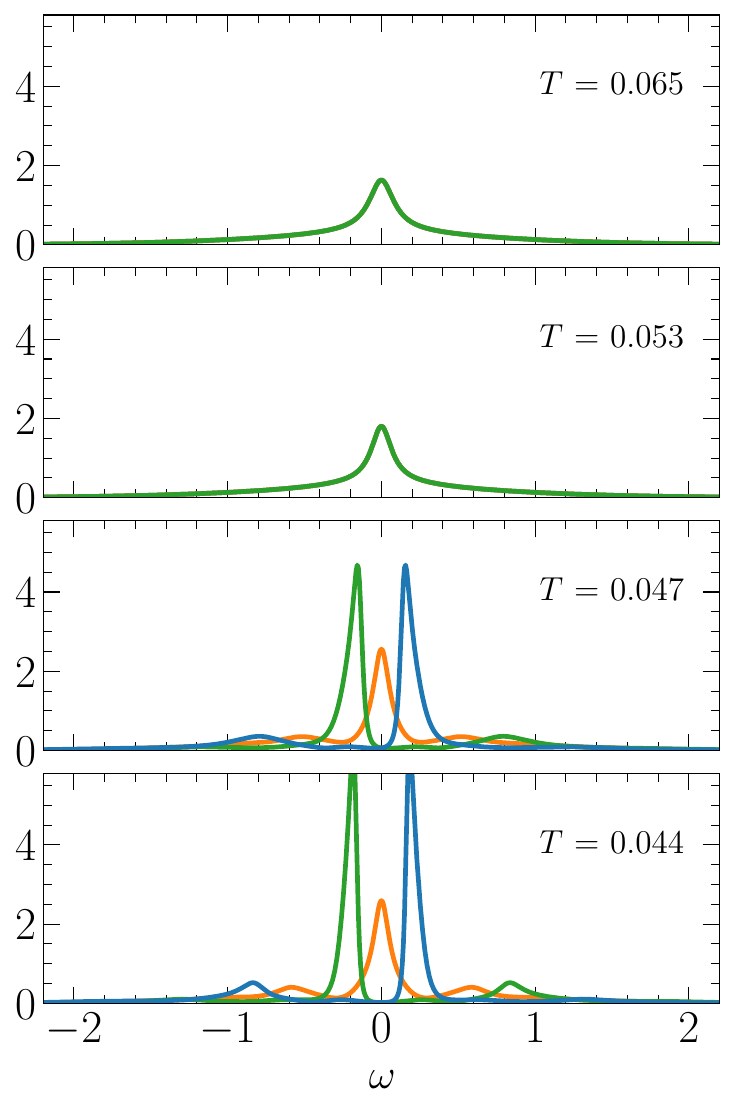}} \hfill
	\subfloat[$V = 1.2$]{ \includegraphics[width=0.485\linewidth]{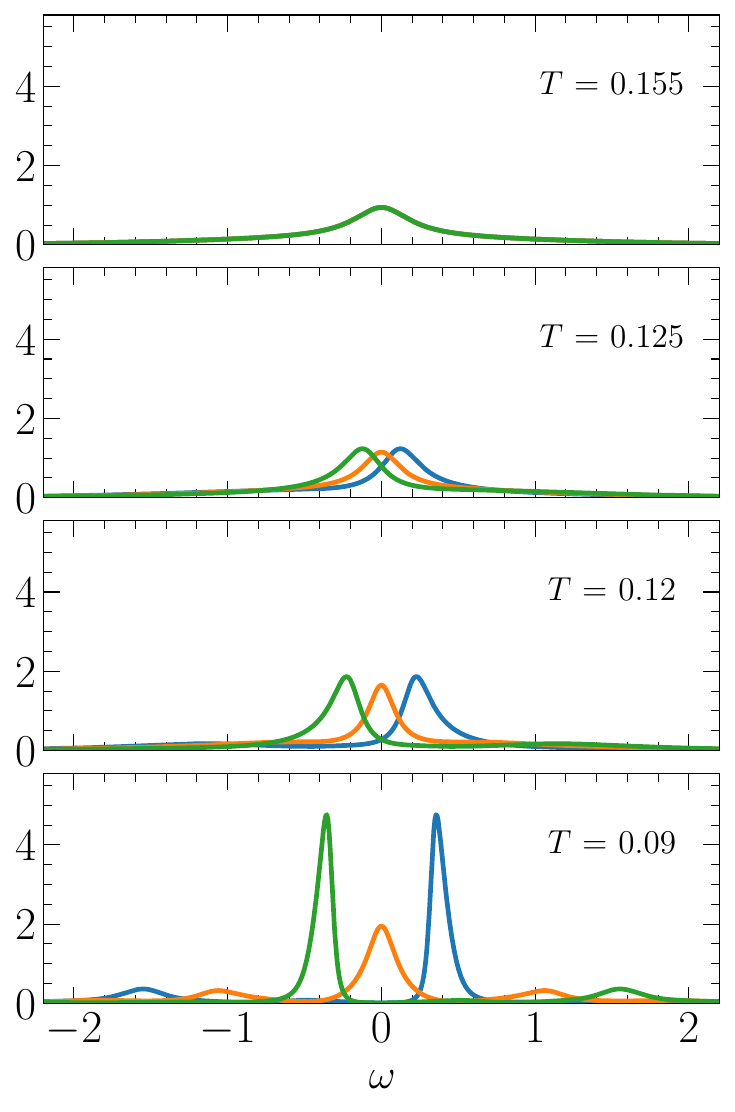} }\hfill
	\caption{The spectral function $A_s(\w)$ for the three-orbital SYK model. Colors distinguish $A_s(\w)$ for each of the three orbitals. The symmetric $[+,0,-]$ phase is clear at lower $T$. }
	\label{Examples of spectral functions for 0D model}
\end{figure}

Fig.~\ref{Examples of spectral functions for 0D model} shows the orbital-resolved spectral functions for both $V = 0.8$ 
and $V = 1.2$ at four temperatures.
The spectral functions are obtained using the retarded Green's function $A_s(\omega) \equiv  -\frac{1}{\pi}\text{Im}G_s(\omega)$. 
At high $T$, both $V=0.8$ and $V=1.2$ have an orbital-symmetric spectral function peaked at $\omega = 0$. 
Upon cooling, they indicate a $[0,0,0]$ to $[+,0,-]$ transition where spectral functions of the three orbitals split. 
This orbital order shifts $A_1(\omega) = A_3(-\omega)$ away from $\omega = 0$ in a symmetric way,
while the third orbital spectral function $A_2(\omega)$ 
remains symmetric and peaked at $\omega = 0$. At the lowest $T$ shown, the two 
orbitals are fully gapped, while the third orbital remain gapless, which indicates a spontaneous orbital-selective 
transition. \par

Within the $[0,0,0]$ phase, the spectral function of each orbital has PHS, so $A_s(\omega) = A_s(-\omega)$, and an 
orbital-exchange symmetry for any pair of orbitals, e.g. $A_1(\omega) = A_3(\omega)$.
In the orbitally ordered $[+,0,-]$ phase, this symmetry reduces to a combined symmetry under particle-hole transformation
and specific orbital exchange $A_1(\omega) = A_3(-\omega)$.
However, the total spectral function $A(\omega) \equiv \sum_s A_s(\omega)$ remains particle-hole symmetric.

\section{Lattice Extensions}\label{Section III: Lattice Extensions}
\subsection{Model Hamiltonians}

We turn to  higher-dimensional lattice-extensions of the three-orbital model discussed in Sec.~\ref{Section I: three orbital dot}. Each lattice site is decorated by three orbitals with both on-site intraorbital and interorbital SYK interactions. 
Electrons of the same orbital can hop to the nearest sites with some definite set of site-independent amplitudes $\{t_{s, \vec{\delta}} \}$. 
The corresponding Hamiltonian is given by
\begin{gather}
\label{SYK Model Hamiltonian}
	\begin{aligned}
		H &= H_{\text{SYK}}^{\rm intra} + H_{\text{SYK}}^{\rm inter} + H_{\rm kin} - \mu \sum_{\mathbf{r},s} \sum_{i} c_{\mathbf{r},s,i}^\dag c^\pdg_{\mathbf{r},s,i}, \\
		H_{\text{SYK}}^{\rm intra} &= \sum_{\mathbf{r},s,(ijkl)} J^{(s)}_{ij;kl}(\mathbf{r}) c_{\mathbf{r},s,i}^\dag c_{\mathbf{r},s,j}^\dag c^\pdg_{\mathbf{r},s,k} c^\pdg_{\mathbf{r},s,l} , \\
		H_{\text{SYK}}^{\rm inter} &= \sum_{s < s'} \sum_{\mathbf{r},(ijkl)} V_{ij;kl}^{(s,s')}(\mathbf{r}) c_{\mathbf{r},s,i}^\dag c_{\mathbf{r},s,j}^\dag c^\pdg_{\mathbf{r},s',k} c^\pdg_{\mathbf{r},s',l} +\text{h.c} ,  \\
		H_{\rm kin} &= - \sum_{\mathbf{r},\vec{\delta}} \sum_{s,i} t_{s, \vec{\delta}} c_{\mathbf{r},s,i}^\dag 
  c_{\mathbf{r}+\vec{\delta},s,i} +\text{h.c} \\
		&= \sum_{\mathbf{k},s,i} \varepsilon^\pdg_s(\mathbf{k}) c_{\mathbf{k},s,i}^\dag c^\pdg_{\mathbf{k},s,i}.
	\end{aligned}
\end{gather}
Here $\mathbf{r}$ labels the position of lattice sites, $\mathbf{k}$ denotes lattice momentum, and $\varepsilon_s(\mathbf{k})$ are the orbital-dependent dispersions obtained from the set of hopping amplitudes $\{t_{s, \vec{\delta}} \}$. 
The remainder of the notation and formulation are identical to Eq.\eqref{SYKDot}.
The higher dimensional extension adds the constraint in the disorder averaging that 
\begin{eqnarray}
    \overline{ J_{ij;kl}^{(s)}(\mathbf{r})^{^{*}} J^{(s')}_{ij;kl}(\mathbf{r}') } &=& \delta_{ss'}\delta_{\mathbf{r}\mathbf{r}'}~J^2/(2N)^3, \notag \\
    \overline{ V_{ij;kl}^{(s,s')}(\mathbf{r})^{^{*}} V_{ij;kl}^{(s,s')}(\mathbf{r}')} &=& \delta_{\mathbf{r}\mathbf{r}'} 
    (1 - \delta_{s,s'})~V^2/(2N)^3 .
\end{eqnarray}
The system has a single conserved $U(1)$ charge, corresponding to the total fermion number 
$\sum_{\br,s,i} c_{\br,s,i}^\dag c^\pdg_{\br,s,i}$, which can be tuned by a chemical potential $\mu$.
We investigate the model on a cubic lattice in Sec.~\ref{Section III: Cubic lattice model} and on a triangular lattice in Sec.~\ref{Section IV: Triangular lattice model}. 
In each of the lattices, we consider anisotropic nearest-neighbour hopping with each orbital preferring particular directions, characterized by a hopping amplitude $t$ and an anisotropy parameter $\delta t/t > 0$. 
The details of the hopping will be discussed separately for each lattice. 
\subsection{Large-N Theory }

We consider uniform solutions upon disorder averaging where the local imaginary-time Green's function $G_{\mathbf{r}, \mathbf{r}, s}(\tau, \tau') \equiv -\frac{1}{N}\sum_i \big\langle T_\tau \big( c_{\mathbf{r},s,i}(\tau) c_{\mathbf{r},s,i}^\dag(\tau') \big) \big \rangle$ and the self-energy $\Sigma_{\mathbf{r},s,i}$ are site-independent, i.e, $G_{\mathbf{r}, \mathbf{r}, s}(\tau, \tau') = G_{s}(\tau, \tau'), \, \Sigma_{\mathbf{r},s}(\tau, \tau') = \Sigma_{s}(\tau, \tau')$.
Due to time-translational invariance , $G_s(\tau, \tau') = G_s(\tau - \tau')$ and $\Sigma_s(\tau, \tau') = \Sigma_s(\tau - \tau')$. 
The saddle point solution leads to the following self-consistent Dyson equations (for details, see Appendix
\ref{Appendix A: Effective action}) ,
\begin{subequations}
\begin{align}
	\Sigma_s(\tau) = &- J^2 G_s^2(\tau)G_s(-\tau) \notag \\
	&\quad - \sum_{s': s'\neq s} V^2 G_{s'}^2(\tau)G_{s}(-\tau), \label{DS1} \\
	G_s(i\omega_n) =& \int d\varepsilon \, g_s(\varepsilon) \big[ i\omega_n + \mu - \varepsilon - \Sigma_s(i\omega_n) \big]^{-1} , \label{DS2} 
\end{align}
\end{subequations}
where $g_s(\varepsilon) \equiv \nu \int \frac{d^d \mathbf{k}}{(2\pi)^d} \delta(\varepsilon - \varepsilon_s(\mathbf{k}))$ is the orbital-dependent lattice density of states, with $\nu$ being the volume of the corresponding primitive lattice. 
These equations are solved numerically through an iterative procedure. The disorder averaged free energy density $\Omega = -T\overline{\ln Z}$ (per fermionic mode $N$
per lattice site $N_L$) evaluated at the saddle point level is given by 
\begin{eqnarray}\label{grand potential functional}
		\Omega &=&  \sum_{s} \bigg[ -\frac{1}{\beta} \sum_{i\omega_n} \int_{\varepsilon} g_s(\varepsilon) \ln \big[ i\omega_n 
        + \mu - \varepsilon - \Sigma_s(i\omega_n) \big] \nonumber \\
		&+& \int_{0}^{\beta}\!\!\! d\tau \Sigma_s(\tau) G_s(\beta - \tau)
		- \frac{J^{2}}{4} \int_{0}^{\beta}\!\!\! d\tau G_s^2(\beta - \tau) G_s^2(\tau) \bigg] \nonumber \\
		&-& \sum_{s,s': s<s'} \frac{V^2}{2} \int_{0}^{\beta} \!\!\! d\tau G_{s}^2 (\beta - \tau) G_{s'}^2(\tau).
\end{eqnarray}

\subsection{Real Frequency Calculations and Transport}
As in Sec.~\ref{Section I C: Spectral Functions}, the $N \rightarrow \infty$ allows for the calculations of $A_s(\w)$ as well as the momentum-resolved $A_s(\bk,\w)$ (see Appendix \ref{Appendix: Real Frequency Self-Consistent Solutions}).
The $N \rightarrow \infty$ limit also provides the simplification that the vertex corrections to the conductivity vanish with a local $\Sigma(\w)$ \cite{Georges1996, Chowdhury2018}. 
The resulting expression for the conductivity is 
\begin{gather}
\begin{aligned}
\sigma_{\alpha \beta}(\omega, T) =& \frac{1}{\omega} \sum_{s} \int_{\mathbf{k},\omega'} v_{s,\alpha}(\mathbf{k}) v_{s,\beta}(\mathbf{k}) A_s(\mathbf{k},\omega')  \\
 &A_s(\mathbf{k},\omega+\omega')\big[ n_F(\omega') - n_F(\omega + \omega') \big] ,
 \end{aligned}
\end{gather}
where $\alpha,\beta$ denote directions, $\mathbf{v}_{s} = \grad_{\mathbf{k}} \varepsilon_{s}(\mathbf{k})$ is the velocity for orbital $s$ and $n_F(\cdots)$ is the Fermi-Dirac function. 
This is used to calculate the total and orbital-resolved conductivity in Sec.~\ref{Section III D: Transport} and Sec.~\ref{Section IV D: Transport}, which we use to extract the d.c.\ resistivity tensor $\rho$ in each sections. 

\section{Cubic Lattice Model} \label{Section III: Cubic lattice model}
\begin{figure*}
	\centering
	\subfloat[phase diagram, $V = 0.8$]{\label{3-orbital phase diagram on cubic lattice with V = 0.8}\includegraphics[width=0.35\textwidth, height = 0.35\textwidth]{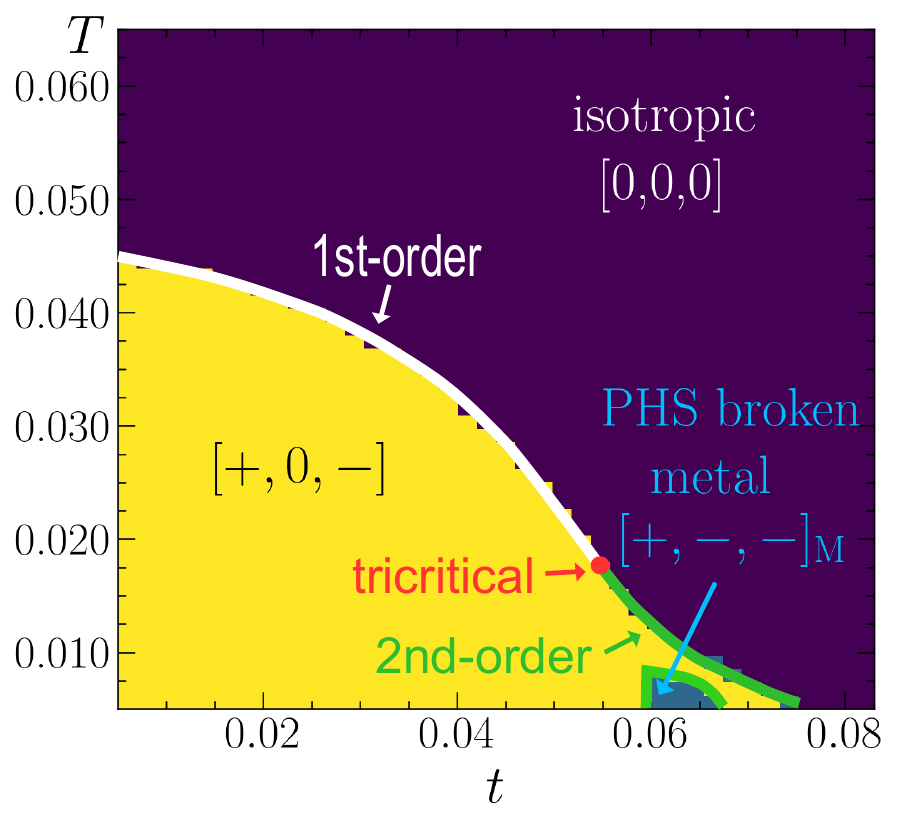}}
        \subfloat[phase diagram, $V = 1.2$]{\label{3-orbital phase diagram on cubic lattice with V = 1.2}\includegraphics[width=0.35\textwidth, height = 0.35\textwidth]{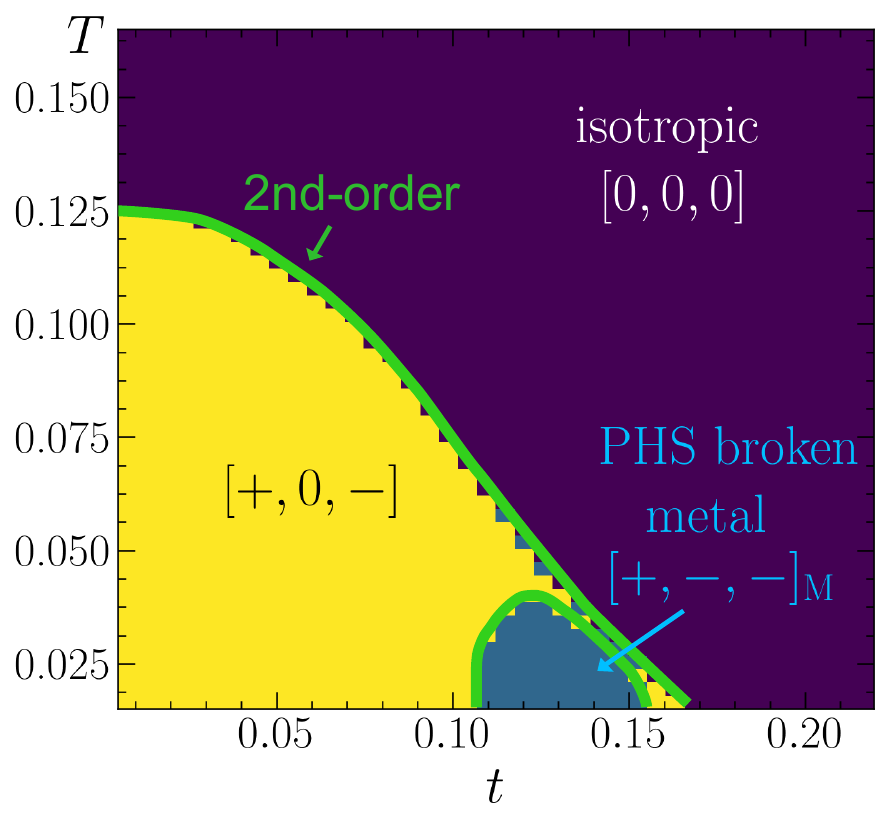}}
	\caption{Phase diagram of the three-orbital cubic lattice SYK model for fixed anisotropy $\delta t /t = 0.8$ and $\mu=0$, 
 with
 (a) $V = 0.8$ and (b) $V = 1.2$.
 We find an orbital-symmetric nFL at high $T$, an orbital-ordered metal $[+,0,-]$ at low $T$ upto moderate $t$ 
 (which matches the three-orbital dot as $t \to 0$), and a metallic $[+,-,-]_{\rm M}$ phase at higher $t$. 
 While the $[+,-,-]_{\rm M}$ is a robust phase at large $t$ and 
low $T$, the thin sliver of the $[+,0,-]$ phase between the $[+,-,-]_{\rm M}$ and 
$[0,0,0]$ phase might be an artifact of our numerics since the free energy difference between the $[+,-,-]_{\rm M}$ 
and $[+,0,-]$ phases is very small in this regime; there could be a direct transition here from $[+,-,-]_{\rm M}$
to $[0,0,0]$ upon heating or increasing $t$.}
	\label{3-orbital phase diagram data on cubic lattice}
\end{figure*}

\subsection{Lattice Model and Symmetries}
On the cubic lattice, we assume that each orbital has a distinct easy-plane ($yz,zx,xy$) with an 
in-plane hopping $t_\parallel = -(t + \delta t/2)$, and a hard-axis ($x,y,z$ respectively) that is orthogonal to the easy-plane 
with an out-of-plane hopping $t_\perp = (t - \delta t)$. The sign choice corresponds to that appropriate
for the $d_{xy}, d_{yz}$ and $d_{xz}$ orbitals. 
This lattice SYK model is symmetric under a $C_3$ rotation about the body diagonal of the cube and 
$C_4$ symmetry about the cubic axes
followed by a suitable orbital permutation.
We investigate the model in the Grand Canonical ensemble at $\mu = 0$, where this Hamiltonian is 
particle-hole symmetric. \\
\par

\subsection{Phase Diagram} \label{Section III B: Phase diagram}
In order to highlight the effects of orbital anisotropy on the phase diagram and physical properties, 
we focus here on the regime of large anisotropy, and present results for 
$\delta t /t = 0.8$, which corresponds to $t_\perp/t_\parallel=-1/7$.
Using the same convention as for the three-orbital dot, we fix $J=1$. We vary the hopping amplitude $t \equiv t/J$ and temperature 
$T \equiv T/J$,
and discuss numerical results for two values of $V$: $V = 0.8 < J$ and $V = 1.2 > J$.  
Fig.~\ref{3-orbital phase diagram on cubic lattice with V = 0.8} and Fig.~\ref{3-orbital phase diagram on cubic lattice with V = 1.2}
show the $T$-$t$ phase diagrams obtained by minimizing the free energy density $\Omega(T)$ in Eq.\eqref{grand potential functional}.
Similar to the $0$D model in Sec.~\ref{Section I: three orbital dot}, we uncover three phases in
the phase diagram for both $V = 0.8$ and $V= 1.2$.

 (i) At high $T$, we find an orbital-symmetric state $[0,0,0]$, which is a nFL with linear-in-$T$ resistivity.
 
 (ii) For low $T$ and small to moderate hopping $t$,
 we find a broken symmetry orbital-ordered state $[+,0,-]$. This orbital order is reflected in transport nematicity
 as discussed later. This state crosses over from a NFL to a FL at a $V$-independent scale $T \lessapprox t^2/J$ \cite{Chowdhury2021b}, reminiscent of the single-orbital model (see Appendix \ref{Appendix: NFL}. 
 
 (iii) At larger $t$, we find a $[+,-,-]_{\rm M}$ phase with weak spontaneous PHS breaking where the total filling
 weakly deviates from half-filling, with two orbitals having equal densities, which is lower from that of the third orbital.
 This nematic metal is distinct from the insulating $[+,-,-]_{\rm I}$ phase found in the SYK dot.

\begin{figure}
	\centering
	\subfloat[$P_{21}$]{\includegraphics[width=0.235\textwidth, height = 0.22\textwidth]{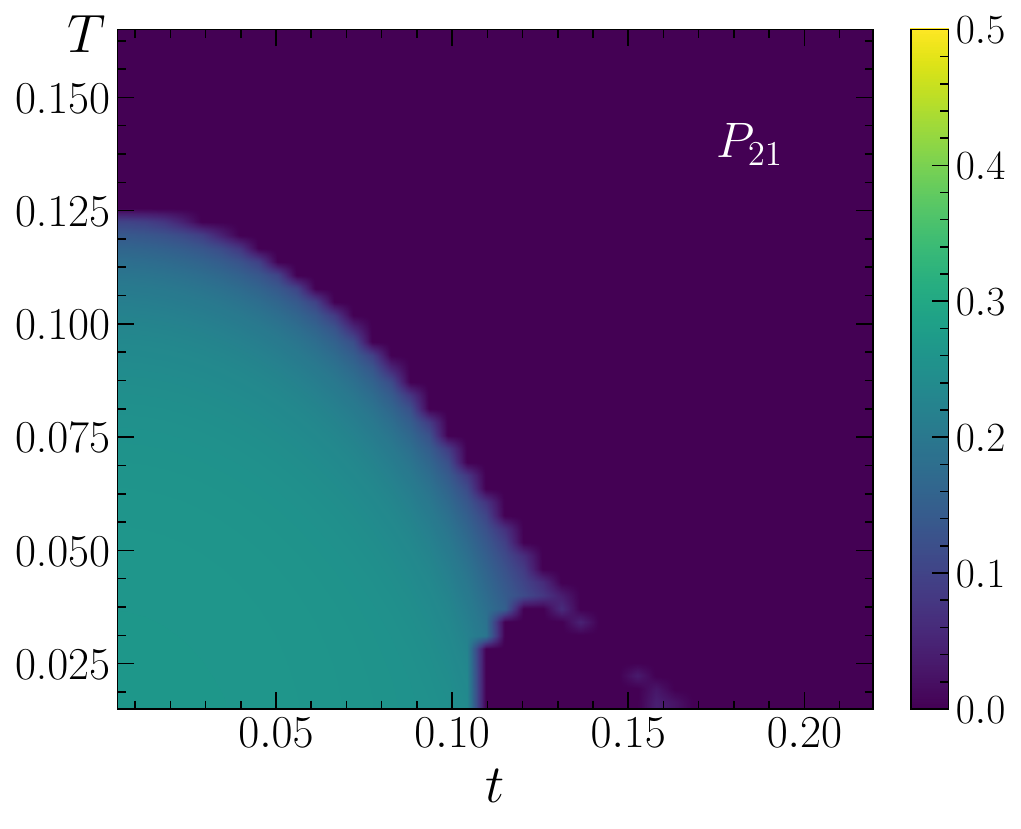}\label{Cubic lattice P21}} \hfill
	\subfloat[$P_{32}$]{ \includegraphics[width=0.235\textwidth, height = 0.22\textwidth]{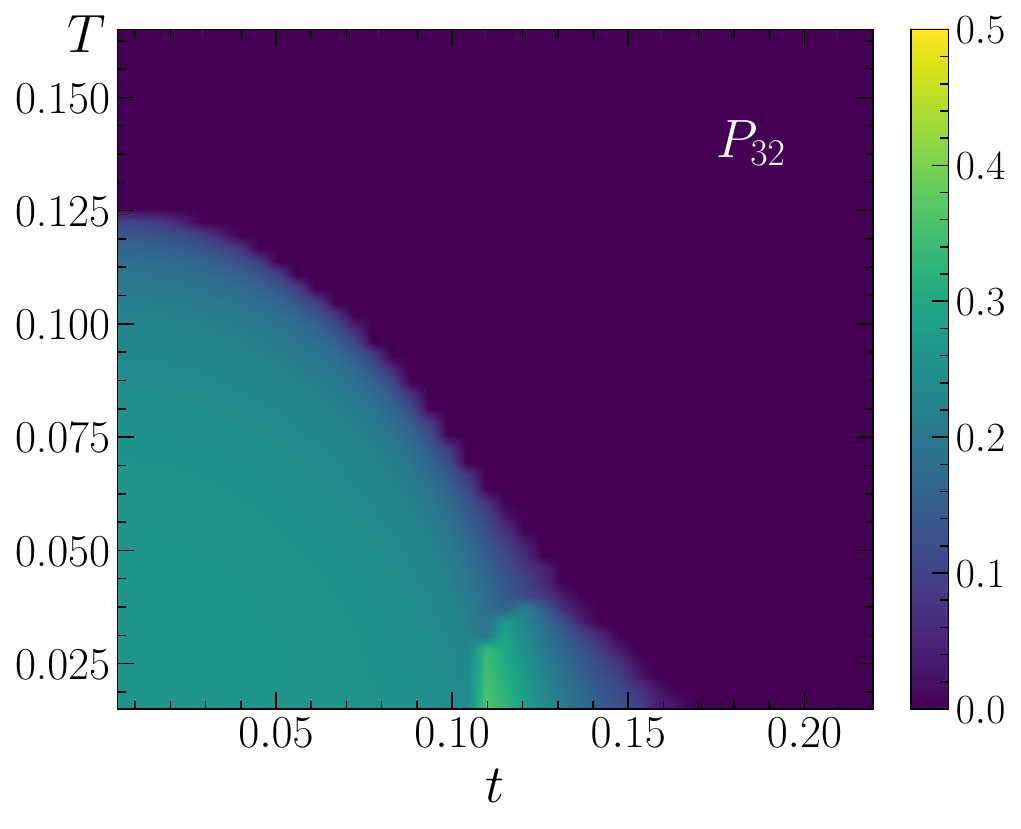} \label{Cubic lattice P32} }\hfill
	\subfloat[total density deviation from half-filling]{ \includegraphics[width=0.235\textwidth, height = 0.22\textwidth]{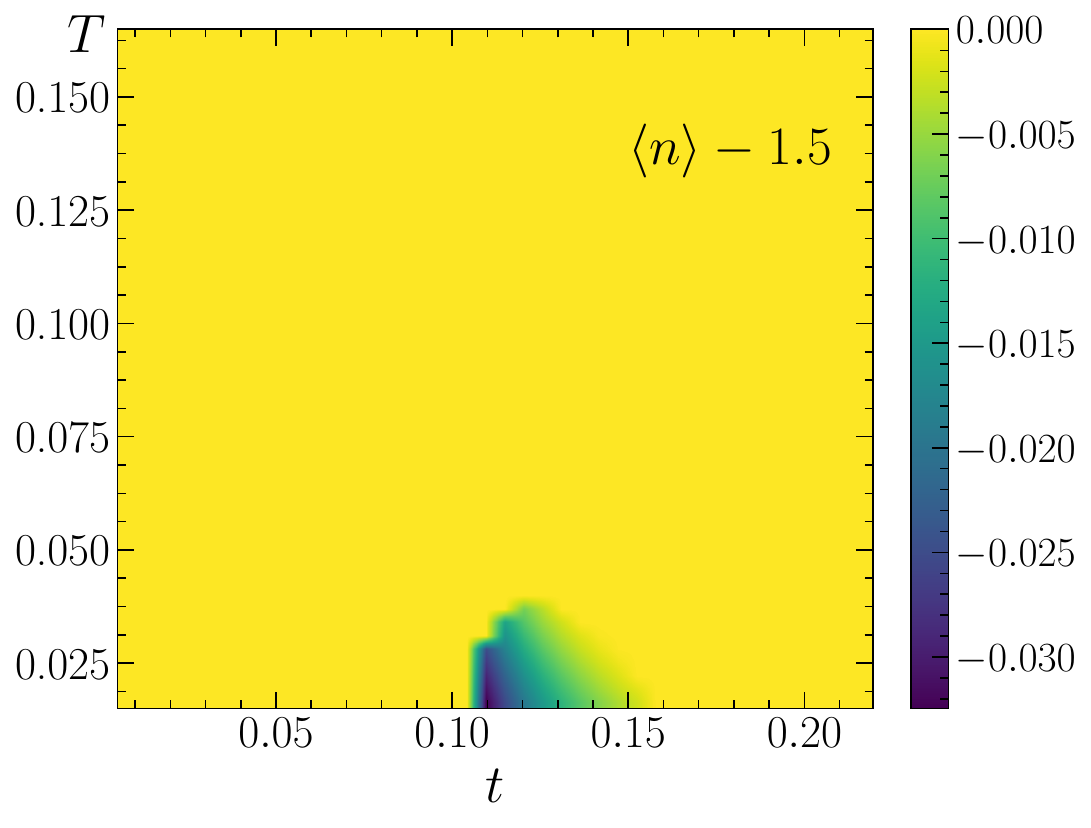} \label{Cubic lattice density deviation} }
        \subfloat[$\k(T)$]{ \includegraphics[width=0.235\textwidth, height = 0.22\textwidth]{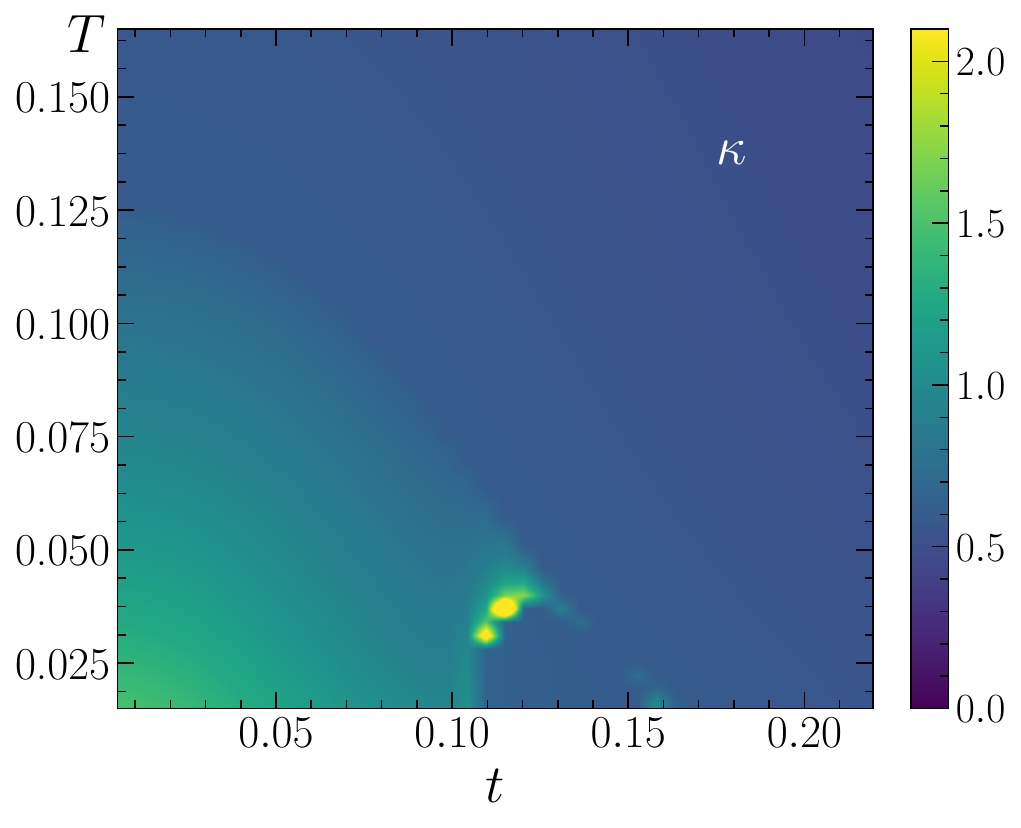} \label{Cubic lattice compressibility plot} } 
	\caption{Orbital polarizations, (a) $P_{21}\equiv n_2-n_1$ and (b) $P_{32}\equiv n_3-n_2$, for the cubic lattice model 
 with $V = 1.2$, $\delta t/t = 0.8$, and $\mu = 0$. (c) Total density $n = (n_1+n_2+n_3)$ deviation from $n=1.5$
 shows that the $[+,-,-]_{\rm M}$ phase spontaneously breaks PHS. (d) The $[+,-,-]_{\rm M}$ phase boundary is
 shows a strongly enhanced compressibility $\kappa$ symptomatic of the spontaneous
 density deviation from half-filling.}
	\label{orbital polarization example for cubic lattice model}
\end{figure}

To characterize each phase, we computed the orbital densities, specific heat, compressibility, and transport properties.
Depending on $t$ and $V$, the thermal transition from $[0,0,0]$ to $[+,0,-]$ can be either first-order or continuous as
marked by the color of the phase boundaries.
 Fig.~\ref{Cubic lattice P21} and Fig.~\ref{Cubic lattice P32} show the
 polarizations $P_{21} \equiv n_2 - n_1$ and $P_{32} \equiv n_3 - n_2$ across the phase diagram for $V = 1.2$. 
 In this symmetric $[0,0,0]$ phase, $P_{21} = P_{32} = 0$ since all orbitals have the same density.
 In the $[+,0,-]$ phase, upon ordering the orbital labels such that $n_1 \geq n_2 \geq n_3$, the polarizations are
 such that $P_{21} = P_{32} > 0$ while the average density is at half-filling. 
 By contrast, the $[+,-,-]_{\rm M}$ phase shows $P_{21} = 0, P_{32} > 0$. In this case, PHS is broken
 and this is consistent with the total density shifting away from half-filling; this density deviation is plotted in 
  Fig.~\ref{Cubic lattice density deviation}. Concomitant with this spontaneous density deviation,
  the phase boundary of the $[+,-,-]_{\rm M}$
  exhibits a strongly enhanced compressibility as shown in Fig.~\ref{Cubic lattice compressibility plot}.
 Further details of the phase diagrams, including additional thermodynamic quantities, and discussion on anisotropy, are provided in Appendix.\ref{Appendix: Further discussion on the cubic lattice model}.

\subsection{Orbital Resolved Spectral Functions} \label{section IV: C Spectral Functions}
\begin{figure}
	\centering
	\subfloat[$t = 0.082$]{\includegraphics[width=0.485\linewidth]{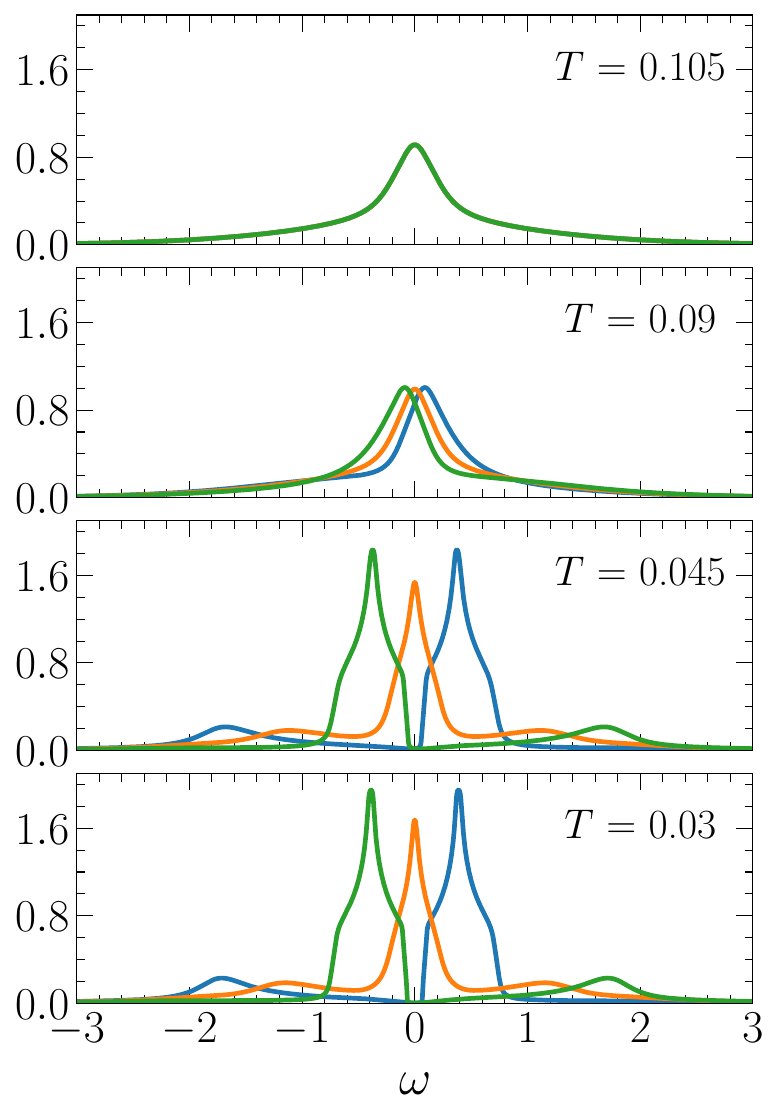}\label{V = 1.2 t = 0.082 Cubic Spectral}} \hfill
	\subfloat[$t = 0.115$]{ \includegraphics[width=0.485\linewidth]{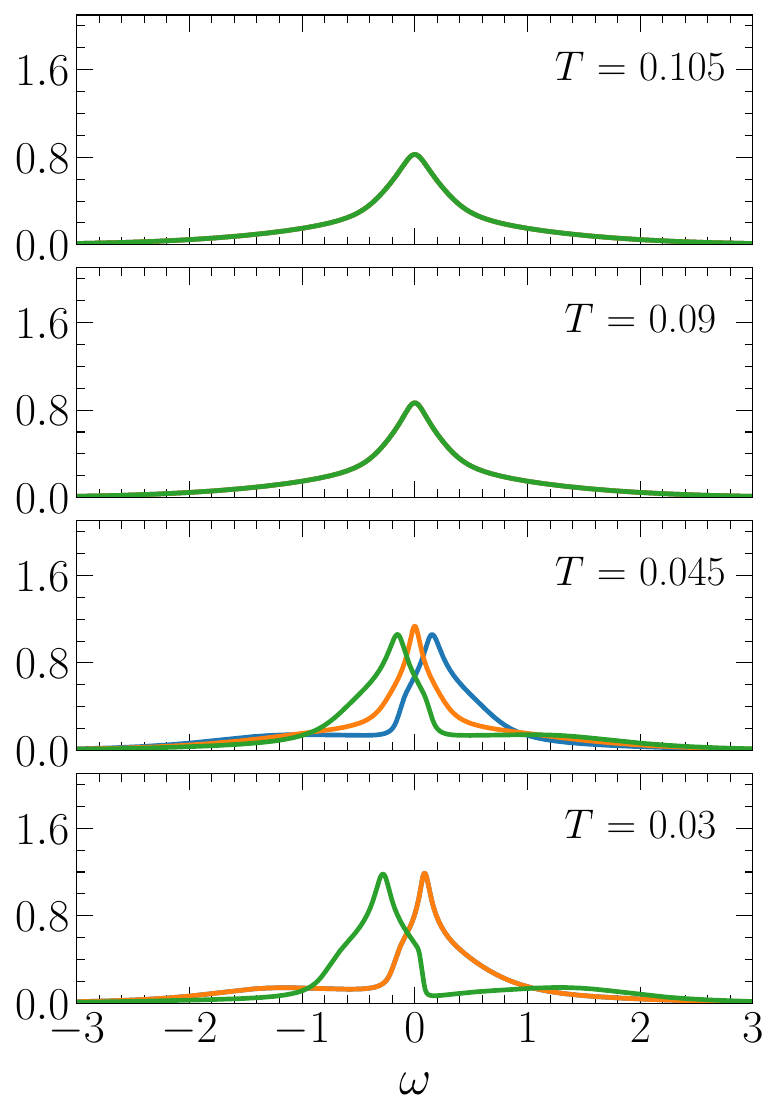} \label{V = 1.2 t = 0.115 Cubic Spectral} }\hfill
	\caption{Examples of spectral functions for cubic lattice model for $V = 1.2,\delta t/t = 0.8$, at two values of hopping, showing distinct transitions as illustrated in Fig.~\ref{3-orbital phase diagram on cubic lattice with V = 1.2}.
 Colors are used to distinguish the spectral function of the three orbitals. }
	\label{Examples of spectral functions for cubic lattice model with V = 1.2J.}
\end{figure}

Fig.~\ref{Examples of spectral functions for cubic lattice model with V = 1.2J.} shows the orbital resolved 
spectral functions for $V = 1.2$ and $ \delta t/t = 0.8$ for two vertical cuts through the phase diagram
at $t = 0.082$ and $t= 0.115$.
At lower hopping (Fig.~\ref{V = 1.2 t = 0.082 Cubic Spectral}), $A_s(\w)$ of the isotropic nFL phase $[0,0,0]$ at high temperature are orbital and particle-hole symmetric.
Upon cooling to the $[+,0,-]$ phase,  $A(\w)$ splits with $A_1(\w)$ and $A_3(\w)$  moving away from $\omega = 0 $ with $A_1(\w)= A_3(-\w)$.  $A_2(\w)$ remains symmetric around $\omega = 0$. 
The two insulating $A(\w)$ are less-strongly interacting and resemble the non-interacting density of states
$g(\varepsilon)$. Upon further cooling, $A_1(\w)$ and $A_3(\w)$ gap out, while $A_2(\w)$ remains gapless. 
We consider a spectral gap defined by $A_s(\w \rightarrow 0) \rightarrow 0$, indicative of an insulator. We elaborate on the precise nature of whether these gaps are exponential hard gaps or soft pseudogaps in Appendix \ref{Appendix:Gap}. \par
At larger hopping $t = 0.115$, the behavior differs as the low $T$ phase is $[+,-,-]_{\rm M}$.
Starting from the isotropic nFL phase at high $T$, the three spectral functions split at intermediate 
$T$ in the $[+,0,-]$ phase.
Upon further cooling into the $[+,-,-]_{\rm M}$ phase, the central unpolarized orbital shifts away
to positive energy. Despite the significant spectral asymmetry in this phase, none of the spectral functions
show gapped insulating behavior.

\subsection{Transport} \label{Section III D: Transport}

\begin{figure}[t]
	\centering
	{\includegraphics[width=0.98\linewidth]{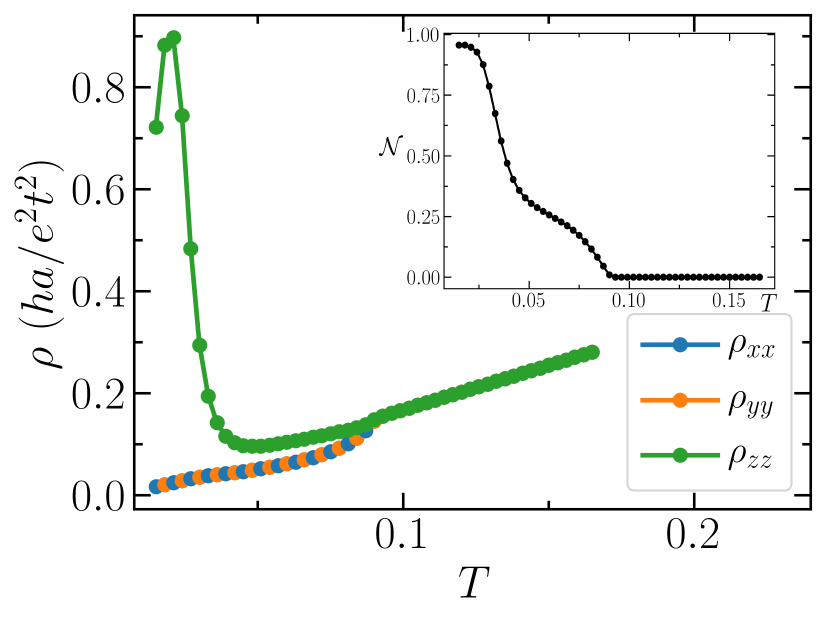}} \hfill
	\caption{The d.c.\ resistivity vs temperature $T$, for cubic lattice model, in Cartesian coordinate. The parameters match Fig.~\ref{V = 1.2 t = 0.082 Cubic Spectral}, with $t = 0.082$. $\rho_{xx}$ coincides with $\rho_{yy}$ for all $T$. The inset show the nematicity $\mathcal{N}(T)$.}
    \label{Examples of nematicity vs Temp at t = 0.082 Cubic model}
\end{figure}

Fig.~\ref{V = 1.2 t = 0.082 Cubic Spectral} show the
temperature dependent longitudinal d.c.\ resistivity $\rho(T)$ along the cubic axes.
The high-$T$ isotropic nFL phase displays a well-known $T$-linear resistivity which is the same along all directions. 
Upon cooling into the symmetry broken $[+,0,-]$ phase, the resistivity becomes highly anisotropic, signalling the nematic order. 
We have chosen the ordering pattern such that the $xy$ orbital remains metallic while the $xz,yz$ orbitals become gapped.
The resistivity along $x$- and $y$- axis  smoothly drops to zero, while the resistivity along the $z$-axis drops  gradually at intermediate $T$ and then becomes rapidly insulating for at $T \lesssim 0.05$ before exhibiting a peak and an
eventual downturn at low $T \lesssim 0.02$. 
This strong dichotomy, between a metallic in-plane $xy$ transport and a wide regime of insulating $z$-axis transport is reminiscent of observations in the pseudogap phase of the underdoped cuprates. 
In the cuprates, and in single band Hubbard models, the distinct behaviors of in-plane and out-of-plane transport stems from a momentum-resolved interlayer tunneling.  
Since the pseudogap  develops near the antinodes, precisely where the interlayer tunneling is maximal, it suppresses interlayer transport and leads to insulating behavior while leaving the in-plane conductivity metallic \cite{caxis_IoffeMillis_PRB1998,caxis_Tremblay_PRB2013}.
In our model, on the other hand, this dichotomy stems from the orbital selective gaps and the highly anisotropic transport properties of each orbital. 
To further quantify this anisotropy, we construct the transport nematicity parameter $\mathcal{N} \equiv (\rho_{zz} - \rho_{xx})/(\rho_{zz}+ \rho_{xx})$ which is plotted in the inset
of Fig.~\ref{Examples of nematicity vs Temp at t = 0.082 Cubic model}. 
$\mathcal{N}(T)$ vanishes in the isotropic phase at high $T$, and increases slowly upon entering the nematic phase as $T$ decreases.
Upon further cooling to $T \lesssim 0.05$, the nematicity starts to increase rapidly, indicative of an orbital selective transition where the transport properties are dominated by a single orbital. \par
\begin{figure*}
	\centering
	{\includegraphics[width=0.98\linewidth]{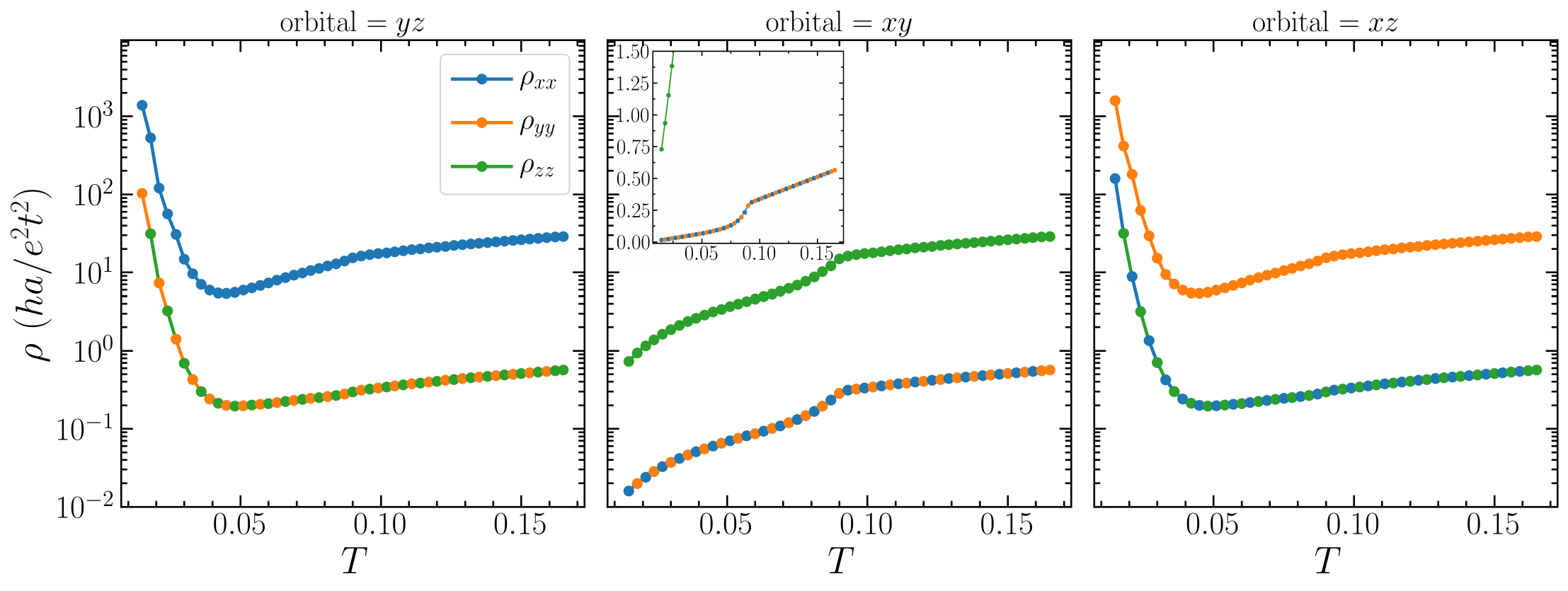}} \hfill
	\caption{Orbital-resolved d.c.\ resistivity as a function of temperature $T$ corresponding to the same parameters in Fig.~\ref{Examples of nematicity vs Temp at t = 0.082 Cubic model}, where the $y$-axis is displayed on a logarithmic scale. 
 The inset of the orbital $xy$ plot displays a zoom-in version of the resistivity within its easy-plane, with the $y$-axis shown on a linear scale. 
 For each orbital, the resistivity $\rho(T)$ within their hard-axis is a factor of $(t_\parallel / t_\perp)^2 = 49$ larger compared to the corresponding easy-plane.}
    \label{Examples of orbital resistivity vs Temp at t = 0.082 Cubic model}
\end{figure*}

The orbital selective nature of the transport is further highlighted in the orbital-resolved resistivity plot shown in
Fig.~\ref{Examples of orbital resistivity vs Temp at t = 0.082 Cubic model}.  At
$T \lesssim 0.05$, the second orbital ($xy$) 
becomes more conductive along all axes compared to the other two orbitals. This indicates 
that as two orbitals become gapped, the resistivity in all directions is reduced by the fully metallic $xy$ orbital . This
$T$-dependent orbital competition leads to a peak in the $z$-axis resistivity in 
Fig.~\ref{Examples of nematicity vs Temp at t = 0.082 Cubic model}.

\section{Triangular Lattice Model} \label{Section IV: Triangular lattice model}
\subsection{Lattice Model and Symmetries}
On the triangular lattice, we assume that each orbital has a distinct easy direction ($\hat{e}_1, \hat{e}_2, \hat{e}_3$) with a strong hopping $t_1 = t + \delta t$, and weak hoppings $t_2 = -(t - \delta t/2)$ on the other two directions, where $\hat{e}_1, \hat{e}_2, \hat{e}_3$ correspond to the three non-parallel direction vectors to the nearest-neighbour sites.
This lattice SYK model is symmetric under a $C_6$ rotation followed by a suitable orbital permutation. 
Unlike the cubic lattice SYK model, the triangular lattice model lacks  PHS at any $\mu$.
We investigate the model in the Canonical ensemble while fixing the total density of orbitals at half-filling. 

\subsection{Phase Diagram} \label{Section IV B: Phase diagram}

\begin{figure*}
	\centering
	\subfloat[phase diagram, $V = 0.8$]{\label{3-orbital phase diagram on triangular lattice with V = 0.8}\includegraphics[width=0.35\textwidth, height = 0.35\textwidth]{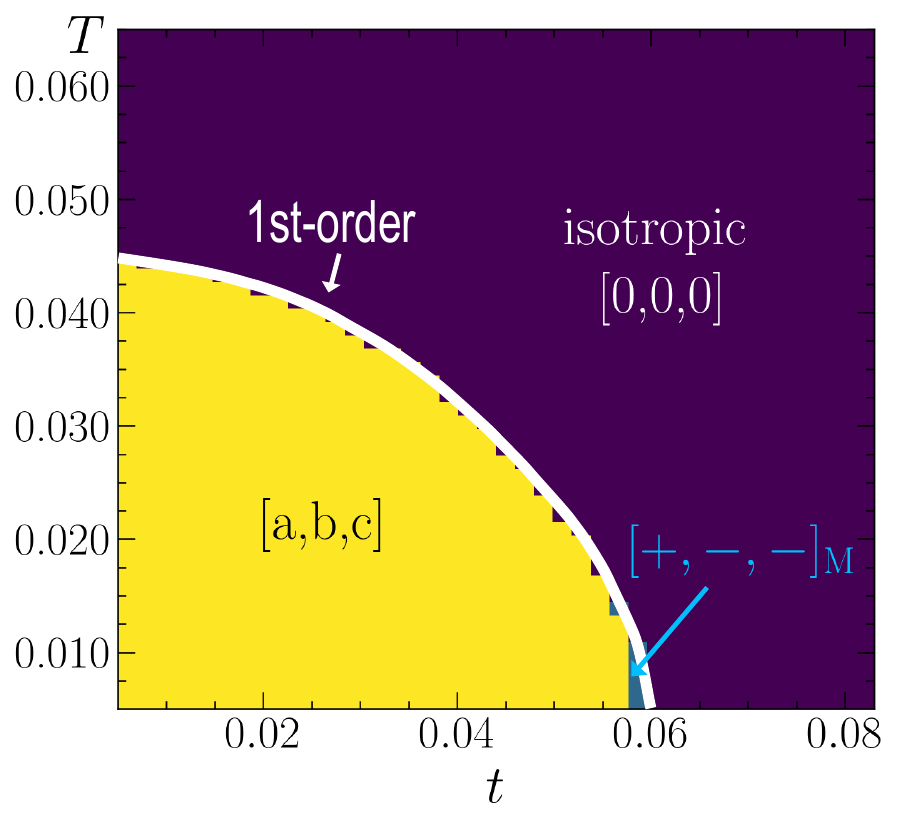}}
         \subfloat[phase diagram, $V = 1.2$]{\label{3-orbital phase diagram on triangular lattice with V = 1.2}\includegraphics[width=0.35\textwidth, height = 0.35\textwidth]{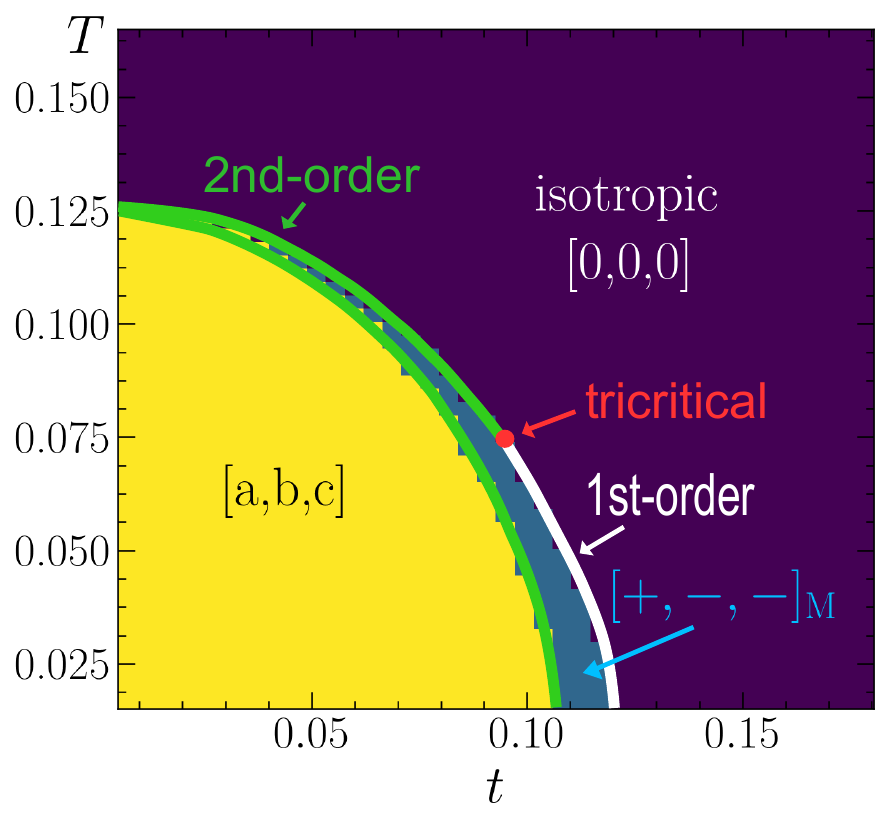}}
\caption{Phase diagram of the three-orbital triangular lattice SYK model for fixed anisotropy $\delta t /t = 0.8$ and fixed total density $n = 1.5$, with (a) $V = 0.8$ and (b) $V = 1.2$.
We find an orbital-symmetric nFL at high $T$, along with two nematic metal $[a,b,c]$ and $[+,-,-]_{\rm M}$ at low $T$, depending on the values of  $V$ and $t$.}

	\label{3-orbital phase diagram data on triangular lattice}
\end{figure*}

\begin{figure}
	\centering
	\subfloat[$P_{21}$]{\label{orbital polarization example for triangular lattice model (a)}\includegraphics[width=0.235\textwidth, height = 0.22\textwidth]{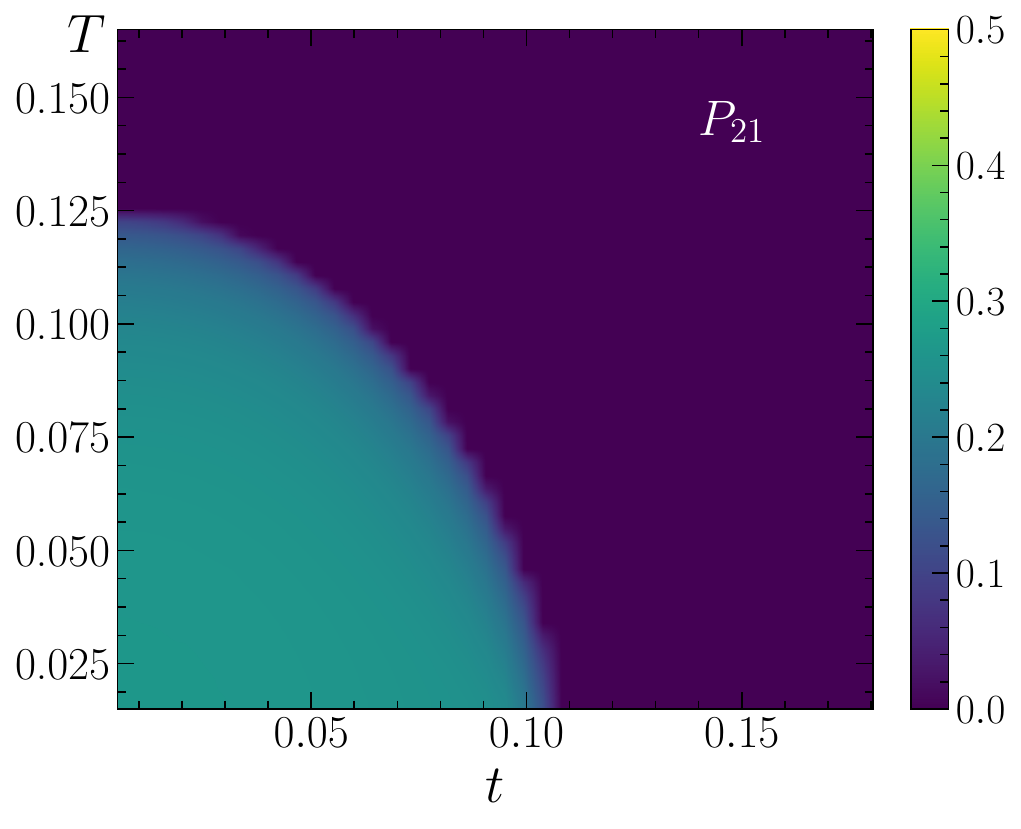}} \hfill
	\subfloat[$P_{32}$]{\label{orbital polarization example for triangular lattice model (b)}\includegraphics[width=0.235\textwidth, height = 0.22\textwidth]{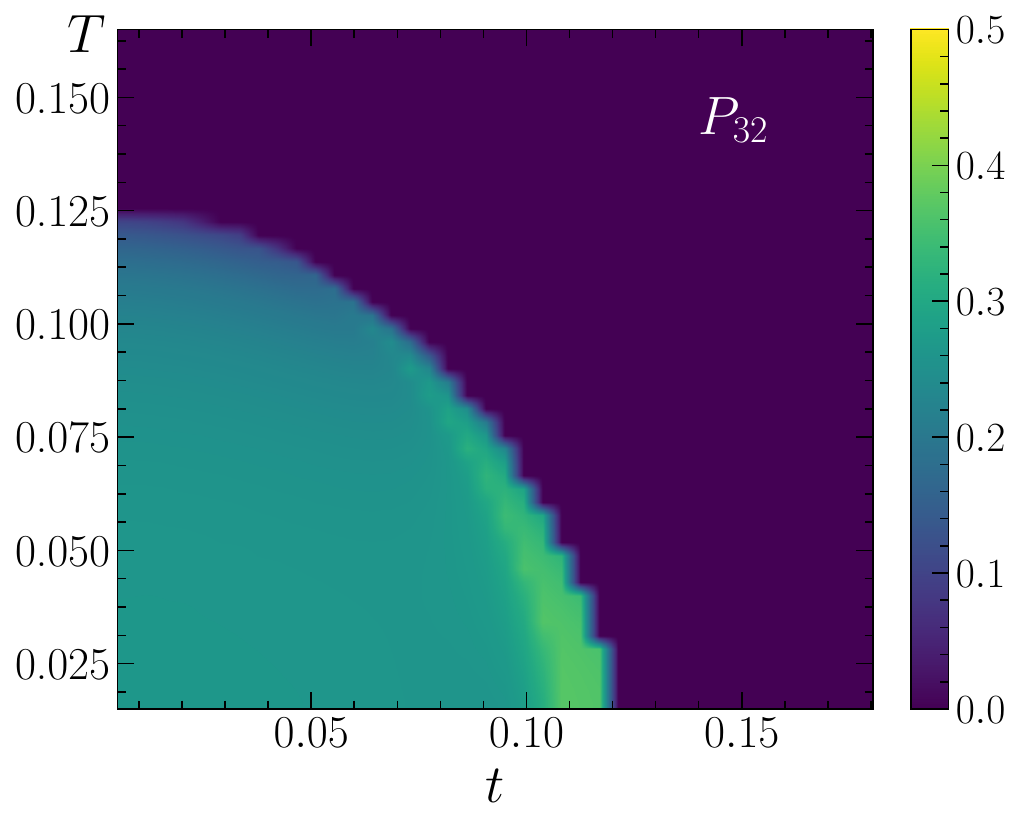} }
	\label{Orbital polarization for triangular model}
	\subfloat[The phase $\varphi$ in the order parameter.]{\label{orbital polarization example for triangular lattice model (c)}\includegraphics[width=0.235\textwidth, height = 0.22\textwidth]{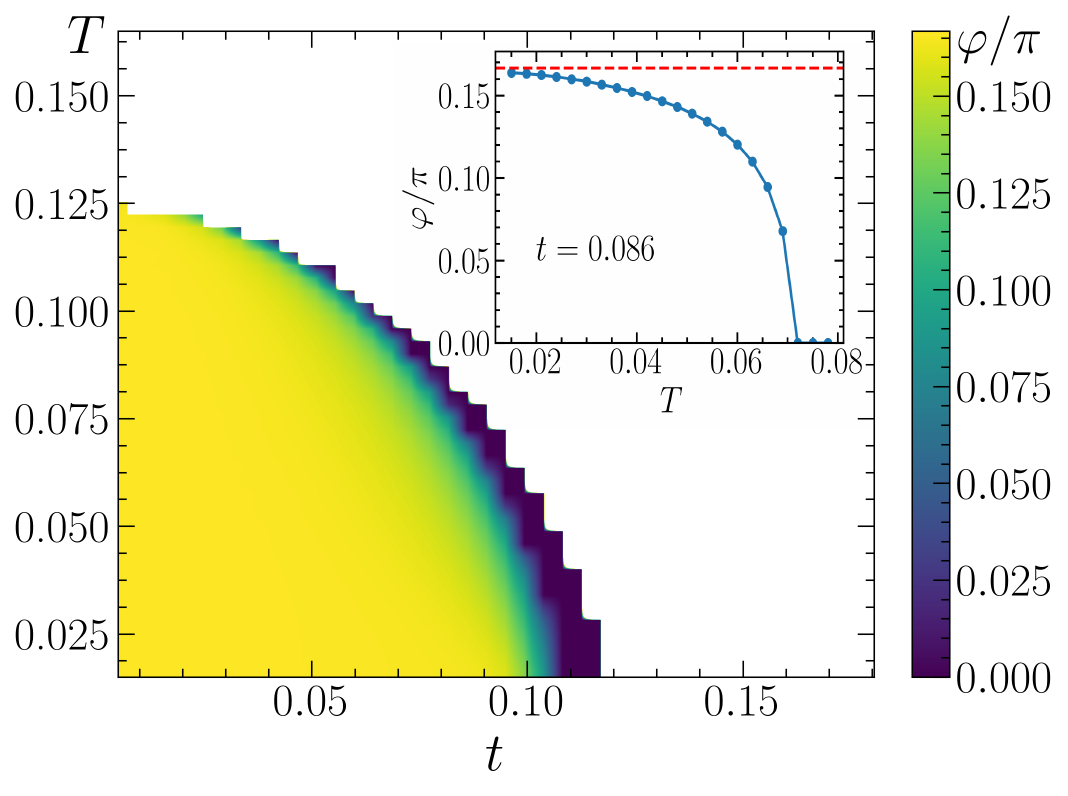} }
        \subfloat[$\k(T)$]{\label{compressibility triangular lattice model (d)}\includegraphics[width=0.235\textwidth, height = 0.22\textwidth]{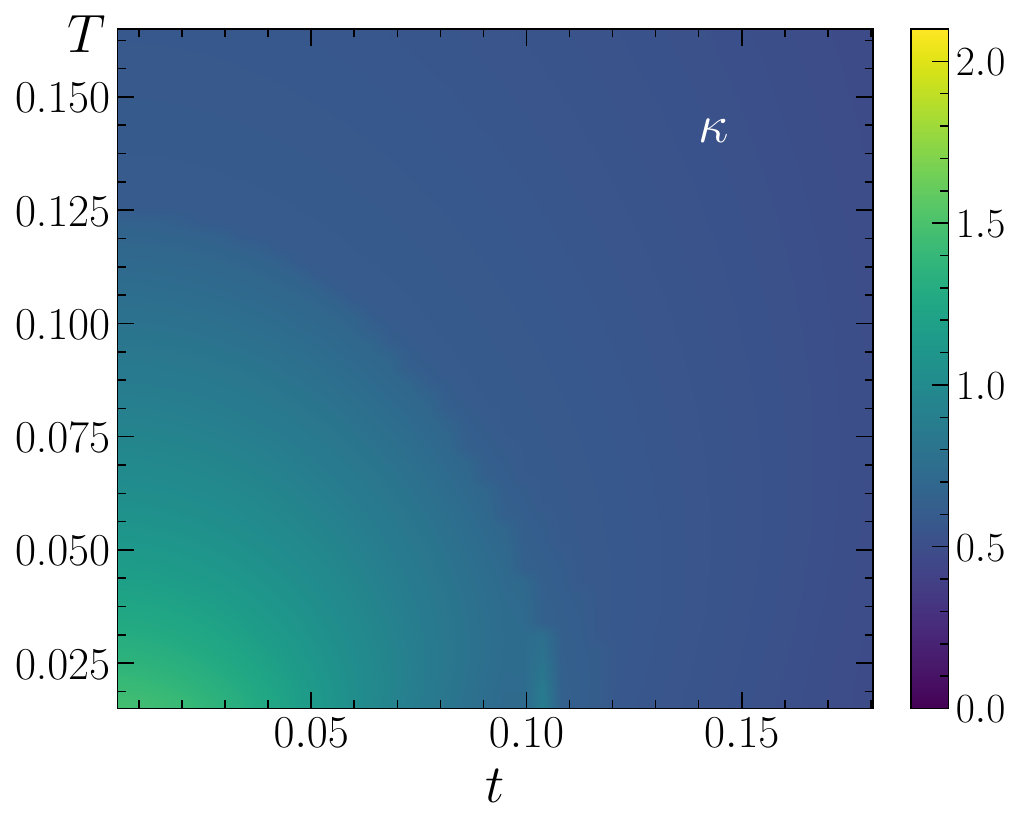} }
	\caption{Orbital polarizations for the three orbital model on triangular lattice with $V = 1.2, \delta t/t = 0.8$ in canonical ensemble with half-filling. (a) and (b) show the orbital polarizations corresponding to the phase diagram Fig.~\ref{3-orbital phase diagram on triangular lattice with V = 1.2}, demonstrating the three cases of orbital density imbalance discussed in Sec.~\ref{Section IV B: Phase diagram}. (c) shows the phase $\varphi$ defined in the order parameter, where $\varphi$ is not well-defined in the white region. 
 The inset shows a slice of the plot at the value of $t = 0.086$.}
	\label{orbital polarization example for triangular lattice model}
\end{figure}

Using the same convention as in previous discussions, we work in units where $J = 1$, with $V \equiv V/J, t\equiv t/J, T\equiv T/J$. 
As in the cubic lattice model, we focus on the regime of a large anisotropy $\delta t/t = 0.8$ and discuss numerical results for two values of $V$: $V = 0.8 < J$ and $V = 1.2 > J$. Fig.~\ref{3-orbital phase diagram on triangular lattice with V = 0.8} and Fig.~\ref{3-orbital phase diagram on triangular lattice with V = 1.2} show the $T-t$ phase diagrams obtained by minimizing the Helmholtz free energy $F = \Omega +\mu \langle n \rangle$, where $\Omega$ is evaluated using Eq.\eqref{grand potential functional}, and $\mu$ is found by fixing $\langle n \rangle = 1.5$ at each points in the phase diagrams.
Similar to the 0D model in Sec.~\ref{Section I: three orbital dot} and the cubic lattice model in Sec.~\ref{Section III: Cubic lattice model}, we uncover three phases in the phase diagram for both $V = 0.8$ and $V = 1.2$.

(i) Similar to previous models, at high $T$, we find an orbital-symmetric nFL state $[0,0,0]$ with linear-in-$T$ resistivity. 

(ii) At low $T$, we find a $[+,-,-]_{\rm M}$ phase which breaks the $C_3$ lattice symmetry but preserve the mirror symmetry, with two orbitals having equal densities, which is lower from that of the third orbital.  This nematic metal is distinct from the insulating $[+,-,-]_{\rm I}$ phase found in the SYK dot. This phase occurs only within a small window of large hopping $t$ for $V = 0.8$, but emerges at infinitesimal $t$ for $V = 1.2$. We note that at $V = 1.2$, the $T$ range where the $[+,-,-]_{\rm M}$ state is stable increases with $t$.

(iii) At low $T$, we also find a nematic metal $[a,b,c]$ which breaks all lattice symmetries. This phase is reminiscent of the $[+,0,-]$ phase in the SYK dot and the cubic lattice model. The $[a,b,c]$ phase asymptotically approaches the $[+,0,-]$ phase in the $T \to 0$ limit as indicated in Fig.~\ref{orbital polarization example for triangular lattice model (c)}, where all three orbitals have an infinitesimal deviation from half-filling at finite $T$.

To characterize each phase, we computed the
orbital densities, $n_s$, specific heat $C_V$,  compressibility $\k$, and
transport $\s_{\a \b}$. The nature of the transition depends on the values of $V$. At $V = 0.8$, the phase diagram unveils two types of isotropic-to-nematic transition, namely $[0,0,0]$-to-$[a,b,c]$ and $[0,0,0]$-to-$[+,-,-]_{\rm M}$, both of which are first-order. At $V = 1.2$, only the $[0,0,0]$-to-$[+,-,-]_{\rm M}$ transition occurs. Depending on $t$, this transition can be either first or second-order as marked by the color of the phase boundaries. Additionally, a second-order transition from the $[+,-,-]_{\rm M}$ to  the $[a,b,c]$ phase also occurs at small to moderate $t$ which is not presence at $V = 0.8$. Further details of the phase diagrams, including additional thermodynamic quantities, and discussion on anisotropy, are provided in Appendix.\ref{Appendix: Further discussion on the triangular lattice model}.

Fig.~\ref{orbital polarization example for triangular lattice model (a)} and \ref{orbital polarization example for triangular lattice model (b)} show the polarizations $P_{21} \equiv n_2 - n_1$ and $P_{32} \equiv n_3 - n_2$ across the phase diagram for $V = 1.2$. 
The phase $\varphi$ of the complex $\mathbb{Z}_3$ nematic order parameter $\psi = |\psi|e^{i\varphi}$ is shown in Fig.~\ref{orbital polarization example for triangular lattice model (c)}, which will be discussed in details in Sec.~\ref{Section V: Landau Theory of Z3 nematic order}. In the isotropic $[0,0,0]$ phase, $P_{21} = P_{32} = 0$, $|\psi| = 0$ and $\varphi$ is not a well-defined quantity. The $[+,-,-]_{\rm M}$ phase shows $P_{21} = 0, P_{32}> 0$ and $\varphi = 0$. The $[a,b,c]$ phase shows $P_{21} \neq P_{32} > 0$, although it approaches the $[+,0,-]$ phase with $\varphi = \pi/6$ upon cooling. 
Unlike the cubic lattice model, the compressibility only gradually enhanced at the phase boundary of the $[+,-,-]_{\rm M}$ as shown in Fig.~\ref{compressibility triangular lattice model (d)}. \par 

\subsection{Orbital Resolved Spectral Functions} \label{section V: C Spectral Functions}

\begin{figure}[b]
	\centering
	\subfloat[$t = 0.095$]{\includegraphics[width=0.48\linewidth]{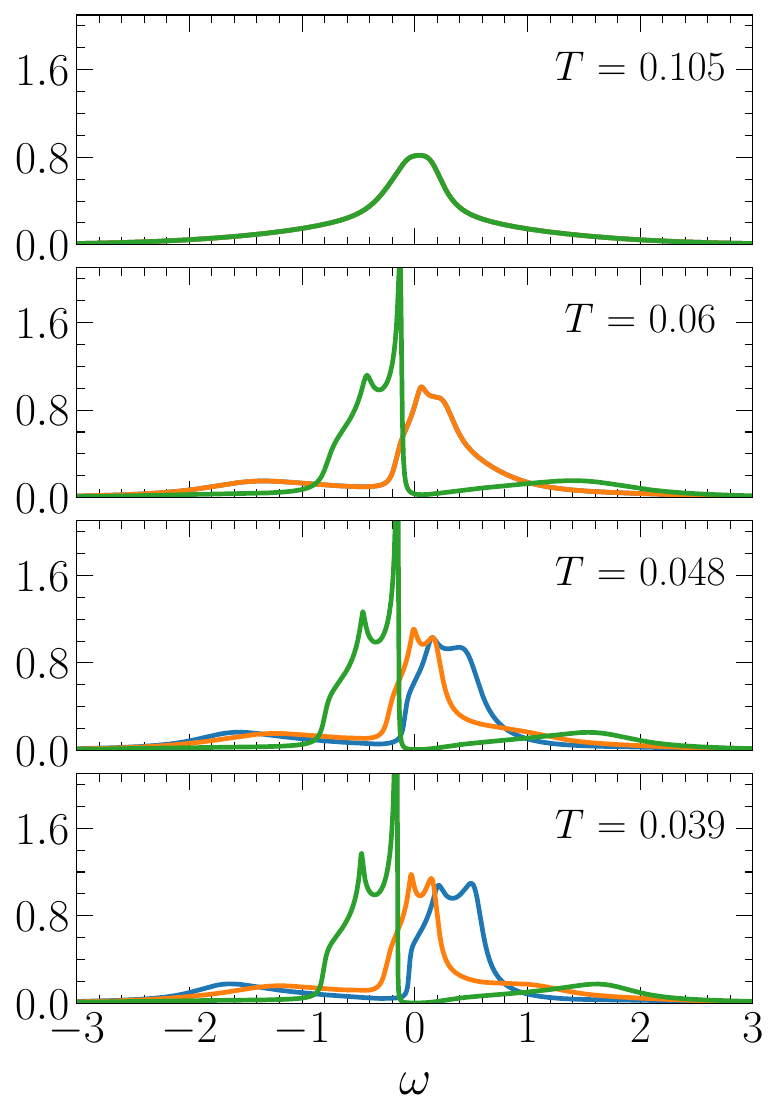} \label{V = 1.2 t = 0.095 Triangular Spectral} } \hfill
	\subfloat[$t = 0.109$]{ \includegraphics[width=0.48\linewidth]{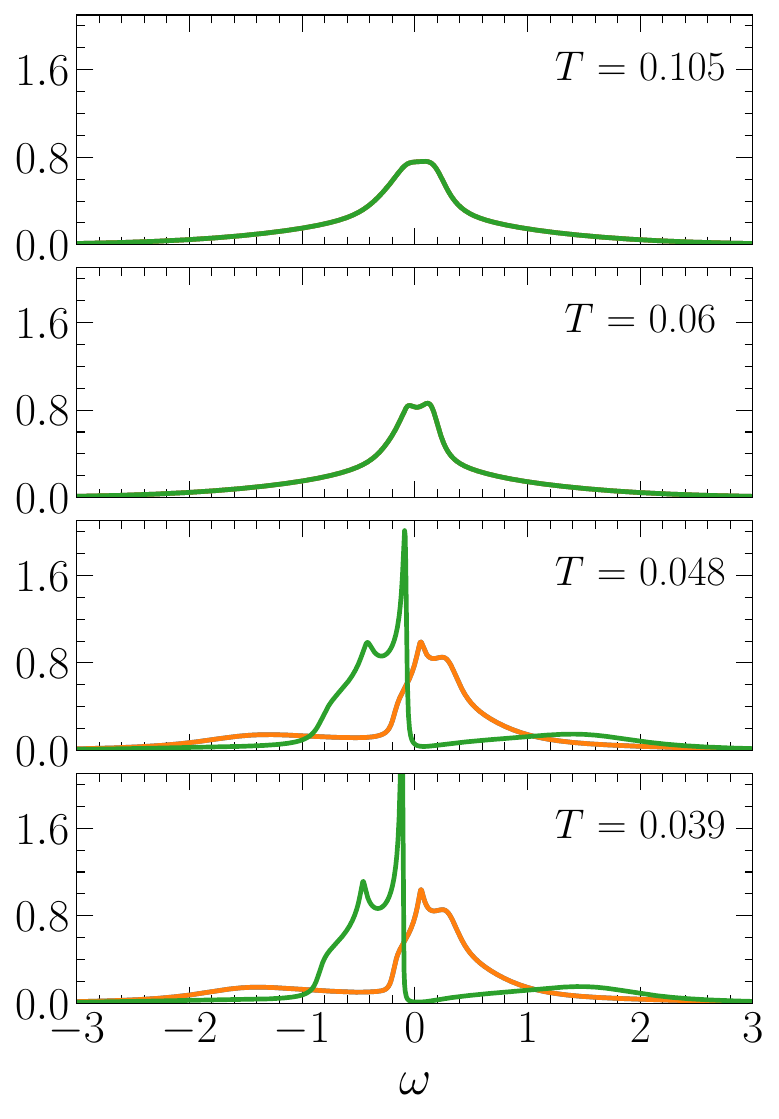} \label{V = 1.2 t = 0.1085 Triangular Spectral} }\hfill
	\caption{Spectral functions for triangular lattice model for $V = 1.2, \delta t/t = 0.8$, at two values of hopping, showing distinct transitions as illustrated in Fig.~\ref{3-orbital phase diagram on triangular lattice with V = 1.2}. Colors are used to distinguish the spectral function of the three orbitals. }
	\label{Examples of spectral functions for triangular lattice model with V = 1.2J.}
\end{figure}

Similar to Sec.~\ref{section IV: C Spectral Functions}, the spectral behavior of the model depends on the nature of the ordering. 
In order to quantify these effects, Fig.~\ref{Examples of spectral functions for triangular lattice model with V = 1.2J.} shows the orbital-resolved spectral functions for $V = 1.2$ and $\delta t/t = 0.8$ for two vertical cuts through the phase diagram at $t = 0.095$ and $t = 0.109$.
At lower hopping $t = 0.095$ (Fig.~\ref{V = 1.2 t = 0.095 Triangular Spectral}), $A_s(\w)$ of the isotropic nFL phase at high $T$ are orbital symmetric.
Upon cooling to the nematic $[+,-,-]_{\rm M}$ phase at intermediate $T$, the spectral weight $A_s(\w \rightarrow 0)$ reduces for all orbitals, while  $A_2(\w) = A_3(\w)$ remains degenerate, and $A_1(\w \rightarrow 0) \rightarrow 0$, indicative of an insulating orbital. Entering the $[a,b,c]$ phase at lower $T$, the two orbital degeneracy is broken  $A_2(\w) \neq A_3(\w)$, although both orbitals remain metallic. 
Upon further cooling, $A_3(\w \rightarrow 0) \rightarrow 0$ as the orbital becomes insulating. At larger hopping $t = 0.109$ (Fig.~\ref{V = 1.2 t = 0.1085 Triangular Spectral}), the behavior is very different, since the low $T$ phase is $[+,-,-]_{\rm M}$. 
Starting from the isotropic nFL phase at high $T$, the spectral functions stay in the $[+,-,-]_{\rm M}$ phase.  Two orbital spectral functions, $A_2(\w) = A_3(\w)$ remains degenerate and the systems metallic at the lowest $T$ available.
The form of $A_s(\w)$ indicates the reduced role of interactions and $\Sigma(\w)$ as they more closely resemble $g(\e)$ of the non-interacting trianglular lattice.

\subsection{Transport} \label{Section IV D: Transport}

\begin{figure}[]
	\centering
	\subfloat[$t = 0.095$]{\label{nematicity for t = 0.095 triangular model}\includegraphics[width=0.9\linewidth]{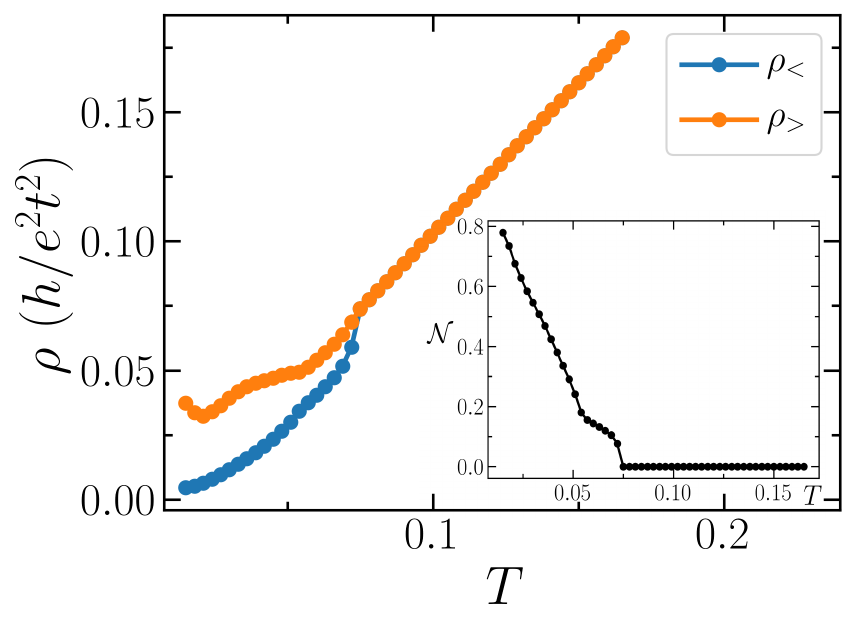}} \hfill
	\subfloat[$t = 0.109$]{\label{nematicity for t = 0.1085 triangular model} \includegraphics[width=0.9\linewidth]{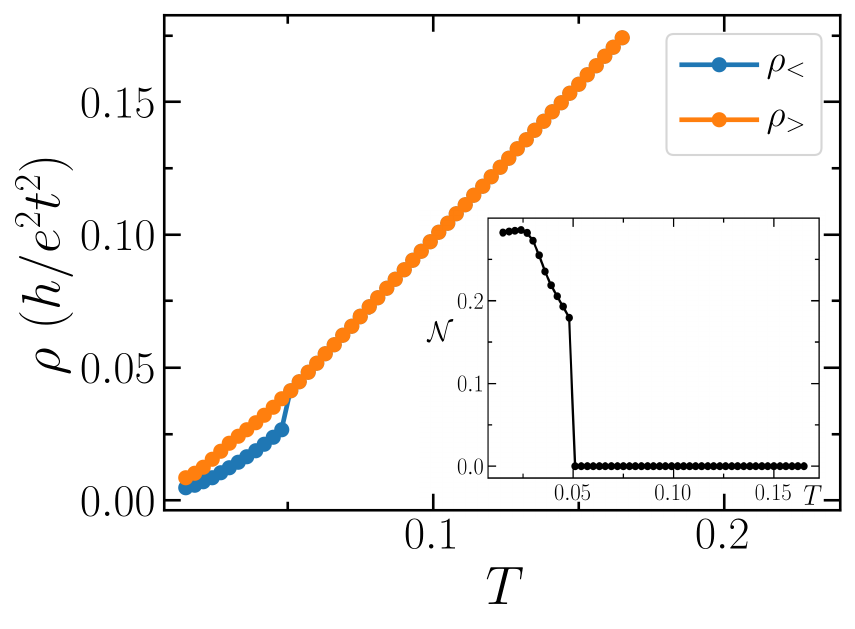} } \hfill
	\caption{The d.c.\ resistivity $\rho$ vs temperature $T$ for the triangular lattice models, with respect to the two principal axes. The parameters match Fig.~\ref{Examples of spectral functions for triangular lattice model with V = 1.2J.}, with (a) $t = 0.095$ and (b) $t = 0.109$.  The inset show the nematicity $\mathcal{N}(T)$. }
    \label{nematicity of triangular lattice model}
\end{figure}

We study the transport properties in each of the three phases by examining the parameter points in Fig.~\ref{Examples of spectral functions for triangular lattice model with V = 1.2J.}.
Fig.~\ref{nematicity of triangular lattice model} shows the temperature dependent d.c.\ resistivity $\rho(T)$. Here $\rho_{<}, \rho_{>}$ correspond to the resistivity along the two principal axes. 
The principal axes are chosen to be along the $x$- and $y$-axes in both the isotropic nFL $[0,0,0]$ phase and the nematic metal $[+,-,-]_{\rm M}$ phase due to the presence of a mirror symmetry. 
The directions of the principal axes depend on the details of $A_s(\w)$ once entering the $[a,b,c]$ phase.

As in the cubic lattice model discussed in Sec.~\ref{Section III D: Transport}, the high-$T$ isotropic nFL phase displays an isotropic $T$-linear resistivity. 
Upon entering the two phases $[+,-,-]_{\rm M}$ and $[a,b,c]$, the resistivity becomes highly anisotropic, signaling the nematic order. 

In order to quantify this anisotropy, the resistive anisotropy, or nematicity, $\mathcal{N} \equiv (\rho_{>} - \rho_{<})/(\rho_{>} + \rho_{<})$ is plotted in the inset of Fig.~\ref{nematicity for t = 0.095 triangular model} and \ref{nematicity for t = 0.1085 triangular model}.
$\mathcal{N}(T)$ vanishes in the isotropic phase at high $T$, and increases continuously (discontinuously), in correspondence with the transition being second (first)-order. 
The second-order transition has a small increase in the intermediary $[+,-,-]_{\rm M}$ state, followed by a rapid increase in $\mathcal{N}$ upon entering the $[a,b,c]$ phase.
The first-order transition to the $[+,-,-]_{\rm M}$ phase has a saturation of $\mathcal{N}$, less than the $[a,b,c]$ state's $\mathcal{N}$. 
This behavior is reminiscent of the transport properties in the cubic lattice model discussed in Sec.~\ref{Section III D: Transport}, indicative of an orbital selective transition where the transport properties are dominated by a single orbital.

\begin{figure*}
	\centering
	\subfloat[$t = 0.095$]{\label{Examples of orbital resistivity vs Temp at t = 0.095 Triangular model}\includegraphics[width=0.9\textwidth]{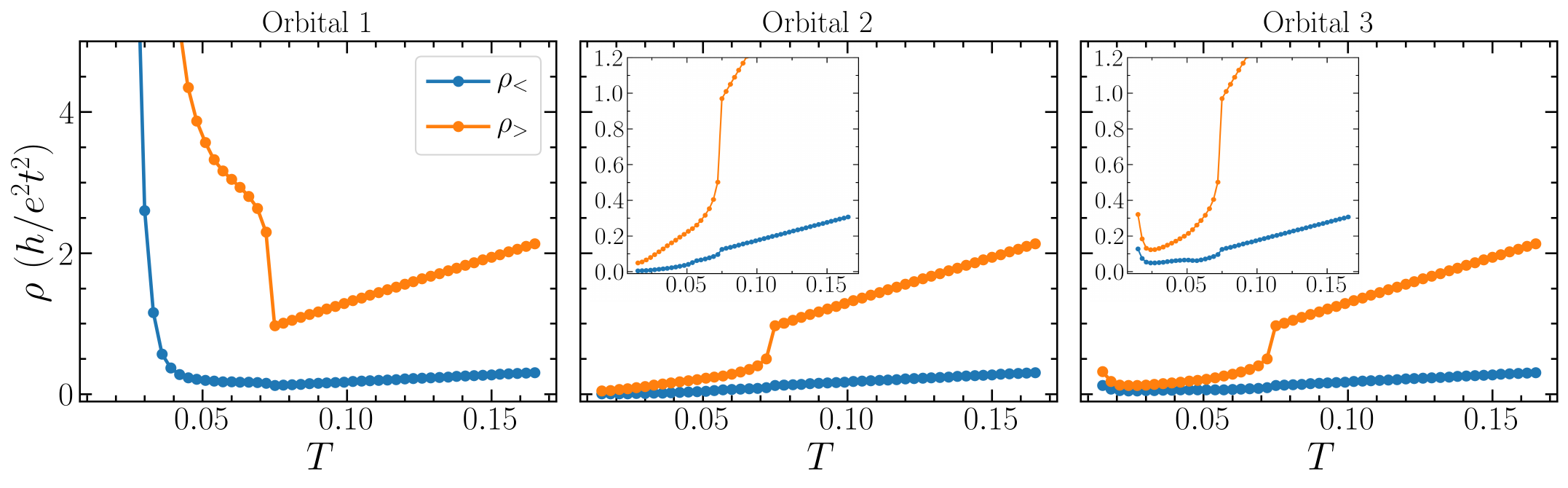}} \hfill
	\subfloat[$t = 0.109$]{\label{Examples of orbital resistivity vs Temp at t = 0.1085 Triangular model} \includegraphics[width=0.9\textwidth]{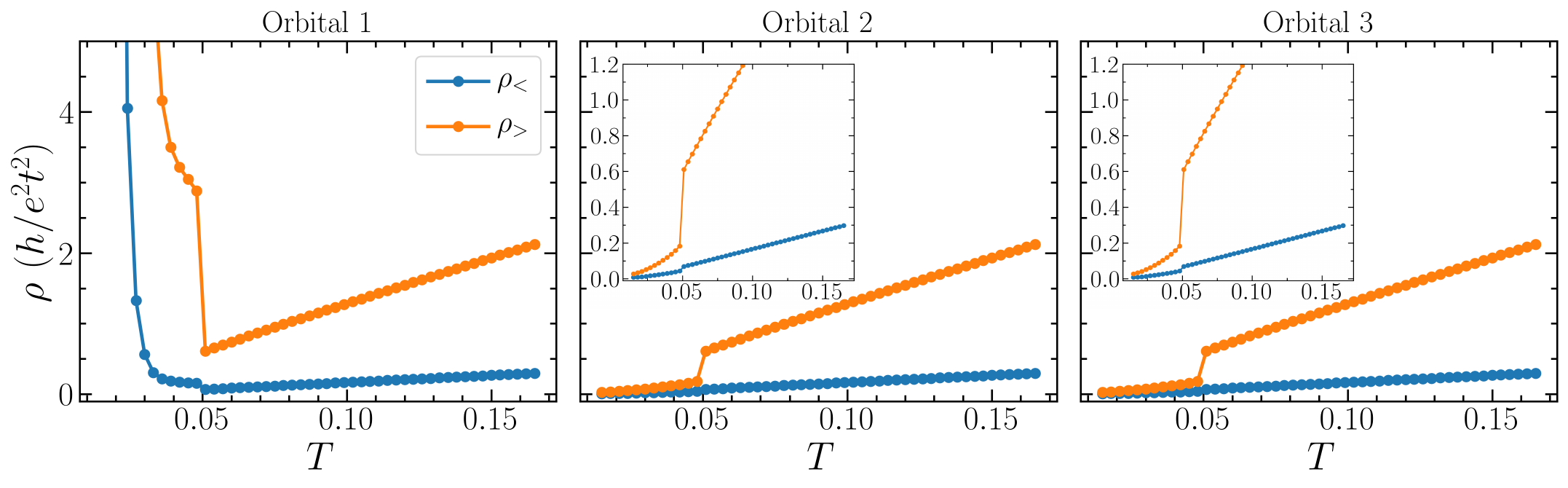} }\hfill
	\caption{Orbital-resolved resistivity as a function of temperature $T$ corresponding to transport properties shown in Fig.~\ref{nematicity for t = 0.1085 triangular model} with (a) $t = 0.095$, and (b) $t = 0.109$. The inset plots emphasize the low $T$ regions.}
\end{figure*}

The orbital selective nature of the transport is further highlighted in the orbital-resolved resistivity shown in Fig.~\ref{Examples of orbital resistivity vs Temp at t = 0.095 Triangular model} and \ref{Examples of orbital resistivity vs Temp at t = 0.1085 Triangular model} for $t = 0.095$ and $t = 0.109$ respectively. At $t = 0.109$ and $T \lesssim 0.05$, the second and third orbital becomes more conductive as $T$ decreases, down to the lowest $T$ available, while the first orbital becomes insulating, exhibiting large and rapidly increasing resistivity along all directions. At $t = 0.095$ and $T \lesssim 0.05$, the first orbital also shows increasing resistivity, similar to $t = 0.109$. However, the resistivity of the third orbital stops decreasing at $T \lesssim 0.05$, and starts increasing at $T \lesssim 0.025$. It exhibits significantly higher resistivity than that of the second orbital at the lowest few temperatures shown.

\section{Landau theory of $\mathbb{Z}_3$ nematic order} \label{Section V: Landau Theory of Z3 nematic order}

We next turn to a Landau theory, which captures aspects of the broken symmetries in the
phase diagrams of these models. 
The theory is formulated in terms of the three orbital densities $n_i$ \cite{Boudjada2018}, 
which we can recast in terms of two fields: a complex field $\psi$ and a real density field $\delta n$. Here,
$\psi = \sum_{i=1}^{3} \omega^{i-1} n_i$
with $\omega =e^{i\frac{2\pi}{3}}$, and
$\delta n = \sum_{i=1}^{3} (n_i - 1/2)$
is the deviation of the uniform density from half-filling. 

With $\psi = |\psi| {\rm e}^{i\varphi}$, the
different phases correspond to: 

\noindent (i) Symmetric $[0,0,0]$: $\psi=0$, $\delta n =0$;

\noindent (ii) $[+,0,-]$: $|\psi| \!\neq\! 0$, $\varphi\!=\!(2 p\! + \! 1)\frac{\pi}{6}$ ($p\!=\!0,\ldots 5$), $\delta n \!=\! 0$;

\noindent (iii) $[+,-,-]$: $|\psi|\! \neq\! 0$, $\varphi\!=\!2 p \frac{\pi}{6}$ ($p\!=\!0,\ldots 5$), $\delta n \!\neq\! 0$;

\noindent (iv) $[a,b,c]$: $|\psi| \!\neq\! 0$, generic $\varphi$, $\delta n \!\neq\! 0$.

\subsection{Model with Particle-Hole Symmetry}

\begin{figure}[b]
\centering
\subfloat[Phase Diagram]{\label{GL_3d_phase_diagram}\includegraphics[width=0.22\textwidth, height = 0.22\textwidth]{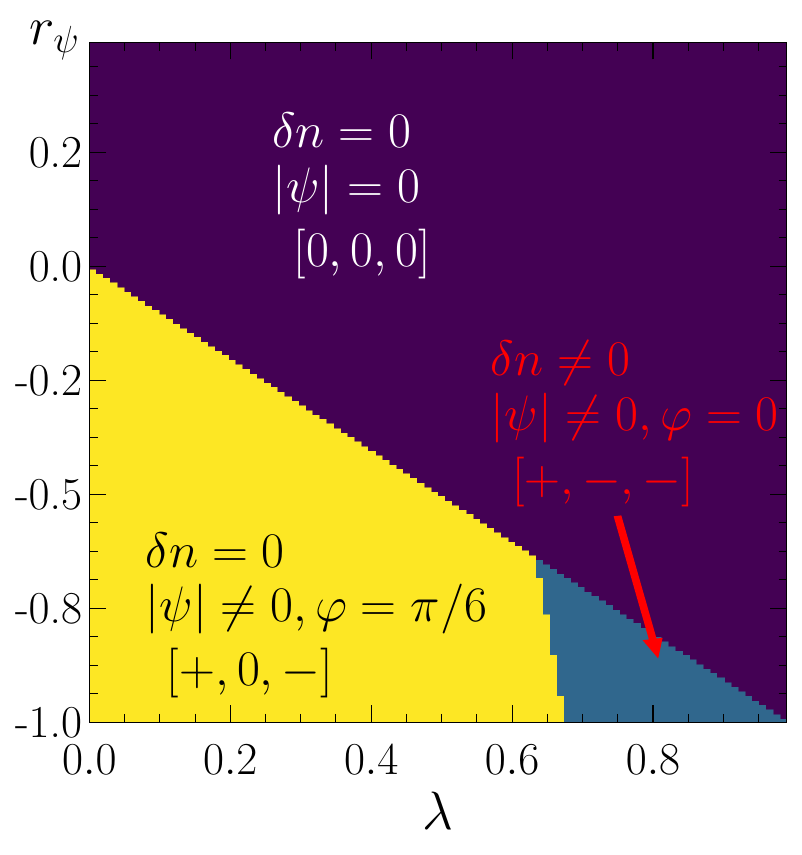}}
\subfloat[$\delta n$]{\label{GL_3d_dens}\includegraphics[width=0.235\textwidth, height = 0.22\textwidth]{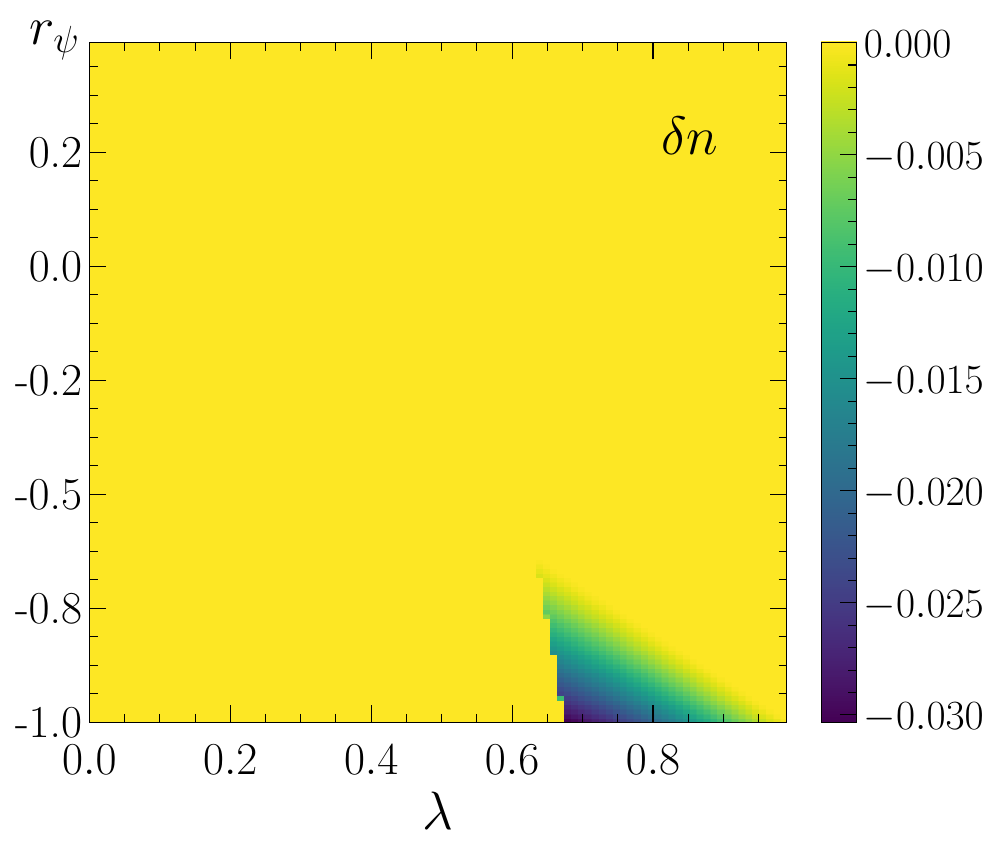}}
\caption{Landau theory phase diagram for the cubic lattice model. We assume $r_\psi \sim T$ and $\lambda \sim t$ to connect with
the microscopic SYK phase diagram in  Fig.~\ref{3-orbital phase diagram on cubic lattice with V = 1.2} and 
Fig.~\ref{Cubic lattice density deviation}.}
\label{GL phase diagram for cubic lattice model}
\end{figure}

For the three-orbital SYK dot and the 3D cubic
lattice, we work at a fixed chemical potential $\mu=0$. Spontaneous PHS breaking can lead to $\delta n \neq 0$.
PHS and orbital permutation or crystalline symmetries requires the Landau free energy functional to be invariant
under $(\psi,\delta n)  \to (-\psi, -\delta n)$, $\psi \to \omega \psi$, $\psi \to \omega \psi^*$. This 
fixes the form of the Landau free energy functional as
\begin{eqnarray}
\!\!	F_{\rm PHS} \!&=&\! (r_\psi \!+\! \lambda) |\psi|^2 \!+\! \lambda \delta n (\psi^3 \!+\! \psi^{*3}) + \lambda_6 (\psi^6 \!+\! \psi^{*6}) \nonumber \\
 \!&+&\! u_4 |\psi|^4 \!+\! u_6 |\psi|^6 \!+\! r_n \delta n^2 \!+\! w_4 \delta n^4 \!+\! g \delta n^2 |\psi|^2,
\end{eqnarray}
where we have dropped higher order terms. To simplify the discussion, we set $u_4=u_6=r_n=w_4=g=1$; 
their precise values are unimportant for our purpose.
We fix $\lambda_6=0.1$, and use the mass parameters 
$r_\psi$ and $\lambda$ as tuning parameters for the Landau theory phase diagram. The choice of sign of
$\lambda_6$ is dictated by the fact that we want to reproduce the $[+,0,-]$ phase and hence favor $\varphi=(2p+1)\pi/6$.
We construct the mass of
the $\psi$ field using two parameters, $r_\psi \sim (T-T_{\rm orb})$
and $\lambda \sim t$, since the nematic order is lost by increasing the temperature or the
hopping amplitude. 
At this point, we do not have a microscopic basis for setting the coefficient of  $\delta n (\psi^3 + \psi^{*3})$ to also scale as $\lambda$.
Fig.~\ref{GL_3d_phase_diagram} shows
the phase diagram in the $r_\psi$-$\lambda$ plane which captures the $T$-$t$ phase diagram of the cubic lattice SYK model.
Fig.~\ref{GL_3d_dens} shows the density deviation from half-filling in the $[+,-,-]$ phase,
in qualitative agreement with the SYK model phase diagram in Fig.~\ref{orbital polarization example for cubic lattice model}.
\subsection{Model with Broken Particle-Hole Symmetry}

\begin{figure}[t]
\centering
\subfloat[Phase Diagram]{\label{GL_2d_phase_diagram}\includegraphics[width=0.22\textwidth, height = 0.22\textwidth]{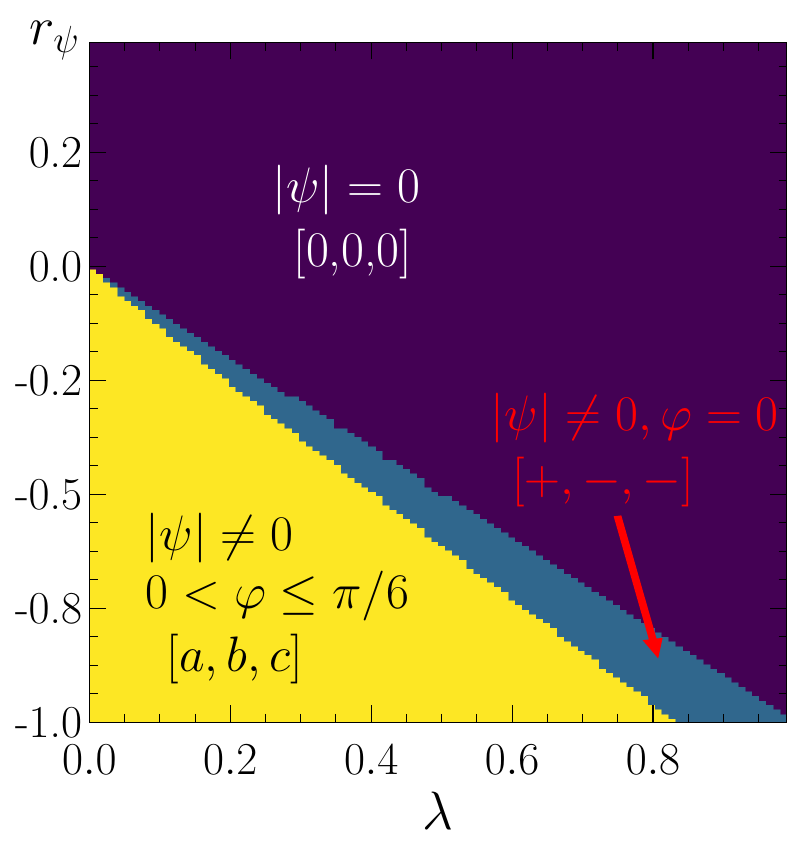}}
\subfloat[$\varphi/\pi$]{\label{GL_2d_phi}\includegraphics[width=0.235\textwidth, height = 0.22\textwidth]{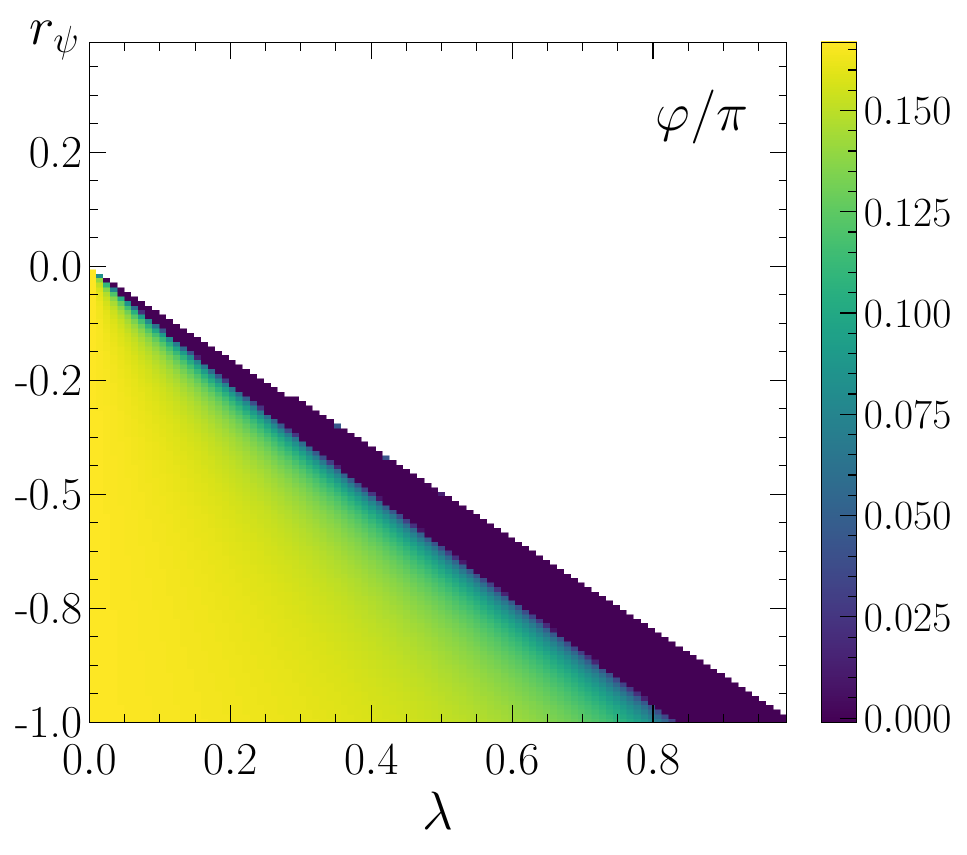}}
\caption{Landau theory phase diagram for triangular lattice model. We assume $r_\psi \sim T$ and $\lambda \sim t$ to connect with
the microscopic SYK phase diagram in  Fig.~\ref{3-orbital phase diagram data on triangular lattice} and
	Fig.~\ref{orbital polarization example for triangular lattice model (c)}.}
\label{GL phase diagram for triangular lattice model}
\end{figure}

On the triangular lattice, the dispersion breaks PHS.
The Landau free energy functional can, therefore, allow terms odd in $\psi$.
Furthermore, since PHS is broken, we tune the chemical potential to fix $\delta n=0$ in our SYK computations, so the density field does not play a role in the Landau theory.
This leads to the following Landau free energy functional for $\psi$:
\begin{eqnarray}
\!\!	F_{\triangle} \!&=&\! (r_\psi \!+\! \lambda) |\psi|^2 \!+\! \alpha \lambda (\psi^3 \!+\! \psi^{*3}) + 
\lambda_6 (\psi^6 \!+\! \psi^{*6}) \nonumber \\
 \!&+&\! u_4 |\psi|^4 \!+\! u_6 |\psi|^6,
\end{eqnarray}
We pick coefficients in $F_{\triangle}$ to be identical to the PHS case, fixing $\lambda_6=0.1$ and assuming, as before,
that $r_\psi \sim (T-T_{\rm orb})$
and $\lambda \sim t$. Given the observed
small density deviation ($\delta n < 0$) in the spontaneous PHS broken state in the 
Landau theory phase diagram for the cubic lattice(
Fig.~\ref{GL_3d_dens}), we also set the cubic term here to have a small fixed coefficient ($\alpha=-0.01$). The resulting phase diagram for the
triangular lattice is shown in Fig.~\ref{GL_2d_phase_diagram}. We see that the phase
diagram shows $[a,b,c]$ order over a wide range of parameters, so $\varphi$ is not generically pinned to 
multiples of $\pi/6$ in this case as seen
from Fig.~\ref{GL_2d_phi}. In addition, a growing sliver of $[+,-,-]$ phase
intervenes between the symmetric $[0,0,0]$ phase and the $[a,b,c]$ phase, corresponding to the SYK model phase diagram
in Fig.~\ref{3-orbital phase diagram data on triangular lattice}.

\section{Orbital Charge Susceptibility} \label{Section VI: Susceptibility}
Throughout this paper, we have considered ferro-orbital ordering, with $\mathbf{q} = 0$ 
ordering wavevector.
In order to consider other potential instabilities, we compute the static orbital charge susceptibility 
$\chi(\bq)$ (at zero frequency) to one-loop order in the nFL regime using the exact spectral function.
As compared to $\sigma(\w)$, $\chi(\bq)$ does not have vanishing vertex corrections in local theories \cite{Georges1996}.  
Despite this, our one-loop calculation provides qualitative understanding of potential alternative orderings. We compute the orbital-dependent charge susceptibilities as the product of two interacting Green's function, and is given by
\begin{eqnarray}
		\!\!\!\!\! \chi_{ab}(\mathbf{q}) \!&=&\!  \!\! \int_{\mathbf{k}} \int_{-\infty}^{\infty} \! \int_{-\infty}^{\infty}
  \!\!\!\! d\omega_\a d\omega_\b A_{ab}(\mathbf{k}\!+\! \mathbf{q}, \omega_\a) A_{ba}(\mathbf{k}, \omega_\b) \nonumber \\
		& \times & \left(  \frac{n_F(\omega_\b) - n_F(\omega_\a)}{\omega_\b - \omega_\a + i0^+}  \right),
\end{eqnarray}
where $a,b$ denotes orbital index, $A_{ab}(\mathbf{k},\omega)= -\frac{1}{\pi} \text{Im}G_{ab}(\mathbf{k},\omega)$ is the momentum-dependent spectral weight and $n_F(\cdots)$ is the Fermi-Dirac function. Since the Green's function is diagonal in orbital basis, $\chi_{ab}(\mathbf{q})$ has a vanishing off-diagonal part. \par
The orbital-dependent charge susceptibilities $\chi_{11}, \chi_{22}, \chi_{33}$ along a high-symmetry path are shown in Fig.~\ref{Charge susceptibility along path for V = 1.2, t = 0.115 cubic lattice at T = 0.096} for the cubic lattice model
and Fig.~\ref{Charge susceptibility along path for V = 1.2, t = 0.095 triangular lattice at T = 0.135} for the triangular lattice model.
As $\pdv{\Sigma(\w)}{\bk} = 0$, $\chi(\bq) $ retains the same form as the noninteracting Lindhard susceptibility $\chi_0(\bq)$ but
with a reduction of amplitude due to self-energy effects.
In both models, $\chi(\bq)$ shows weak $\bq$ modulations with the largest suscepbitility at $\mathbf{q}\neq 0$,
indicative of possible tendency to non-uniform orbital orders. Stabilizing uniform $\bq=0$ order might, therefore, need non-local 
interactions which favor small momentum exchange, leading to a $\bq$-dependence which is peaked at $\bq=0$. Alternatively,
weak uniaxial strain might enhance the tendency to $\bq=0$ ordering leading to a metanematic transition into the ferro-orbital state.

\begin{figure}[t]
\centering
{\includegraphics[width=0.48\textwidth]{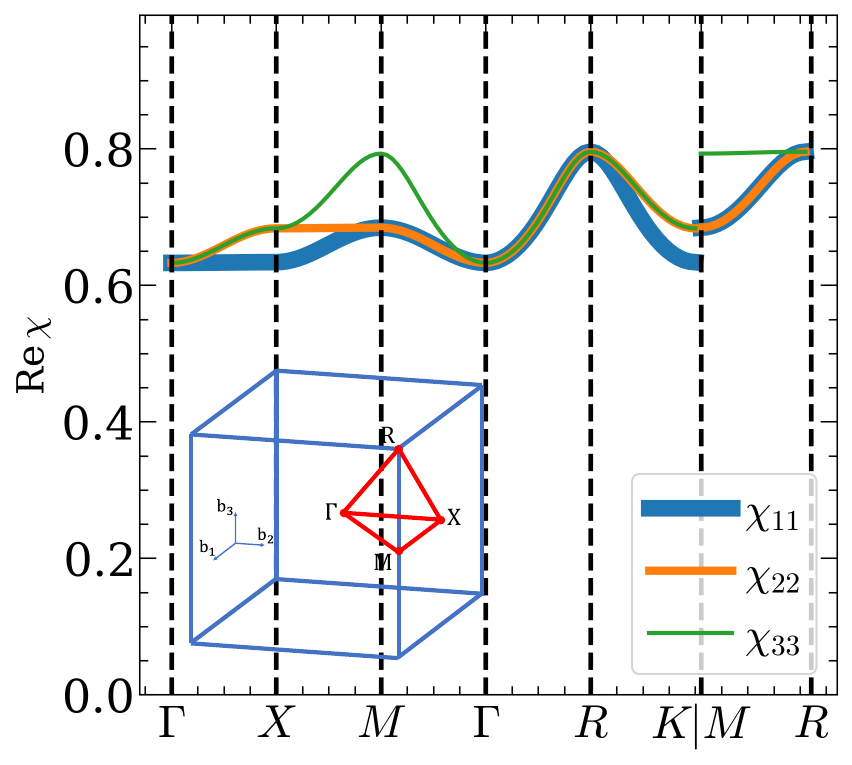}}
\caption{Orbital charge susceptibility $\chi$ along the path $\Gamma \to X \to M \to \Gamma \to R \to X$ and $M \to R$ for the cubic lattice model with $V = 1.2, t = 0.115, \delta t/t = 0.8$ in the orbital
symmetric phase at $T = 0.096$.}
	\label{Charge susceptibility along path for V = 1.2, t = 0.115 cubic lattice at T = 0.096}
\end{figure} 

\begin{figure}[h]
\centering
{\includegraphics[width=0.48\textwidth]{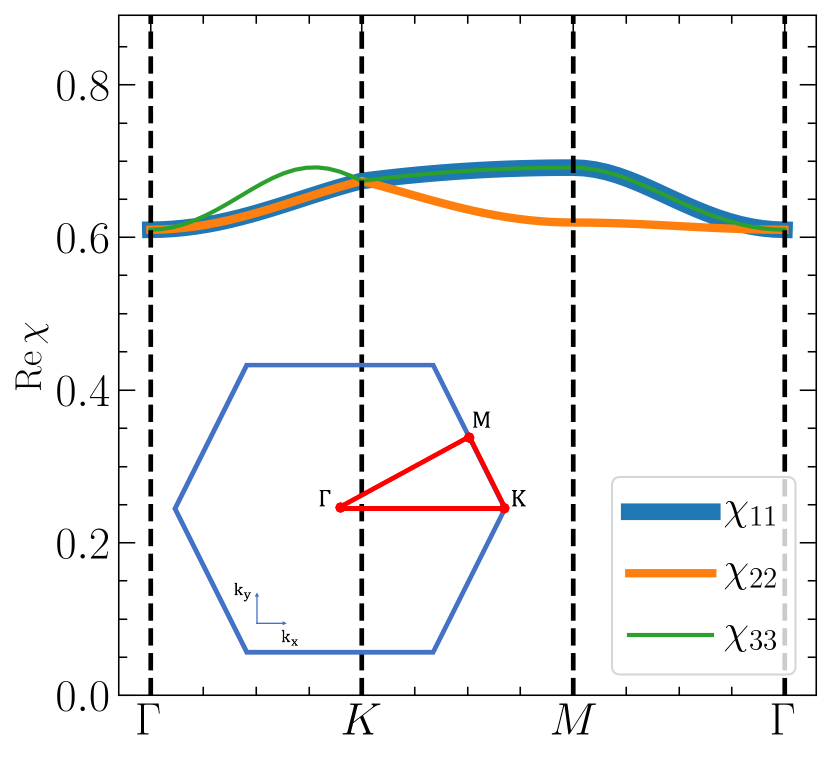}}
\caption{Orbital charge susceptibility ($\chi$ along the path $\Gamma \to K \to M \to \Gamma$ for triangular lattice model with $V = 1.2, t = 0.095, \delta t/t = 0.8$ in the orbital
symmetric phase at $T = 0.135$.} 
\label{Charge susceptibility along path for V = 1.2, t = 0.095 triangular lattice at T = 0.135}
\end{figure} 

\section{Discussion} \label{Section VII: Outlook}
In this paper, we have studied an exactly solvable three-orbital
Sachdev-Ye-Kitaev (SYK) model. 
Our thermodynamic results in the single-site limit exhibit a spontaneous orbital-selective polarization transition, which preserves average particle-hole symmetry. The nature of the transition depends on the magnitude of the ratio of the intraorbital and interorbital SYK interactions $V/J$. We find that the $[0,0,0]$-to-$[+,0,-]$ transition is first-(second-) order for small(large) values of $V/J$. \par 

We have also examined two lattice extensions of this model, in both $2$ and $3$ dimensions, demonstrating  a natural Potts-nematic $\mathbb{Z}_3$ symmetry. For the 3D cubic lattice model, we have considered the case of fixed $\mu=0$, where the Hamiltonian has
particle-hole symmetry.
 At both values of $V/J$ that we studied, an nFL-to-$[+,0,-]$ and a $[+,0,-]$-to-$[+,-,-]_{\rm M}$ transition are observed. 
The $[+,0,-]$-to-$[+,-,-]_{\rm M}$ second-order transition is  accompanied by a diverging compressibility $\k$. The nature of the nFL-to-$[+,0,-]$ transition depends on the value of $V/J$, where it is second-order for $V/J = 1.2$ but is dependent on $t$ at $V/J = 0.8$, with the presence of a tricritical point beyond which the transition becomes second-order.  The $\mathbb{Z}_3$ nematic state forms a layered metal which spontaneously breaks particle-hole symmetry at large $t$.
The $[+,-,-]_{\rm M}$ phase indicates an orbital-selective transition where the transport properties at low $T$ are dominated by a single orbital - this phase exhibits strongly anisotropic transport.
We have also investigated the 2D triangular lattice model at half-filling.
We found three stable phases throughout the phase diagram. At $V = 1.2$, an isotropic nFL-to-$[+,-,-]_{\rm M}$ transition and a $[+,-,-]_{\rm M}$-to-$[a,b,c]$ transition are found. 
The nature of the nFL-to-$[+,-,-]_{\rm M}$ transition depends on the hopping, with the presence of a tricritical point beyond which the transition becomes first-order, while the $[+,-,-]_{\rm M}$-to-$[a,b,c]$ transition is always second-order. 
At $V/J = 0.8$, we observed a nFL-to-$[a,b,c]$ transition and a nFL-to-$[+,-,-]_{\rm M}$ transition, both of which are first-order.
The nature of the phase diagram strongly depends on the values of $V/J$.
The correlated $\mathbb{Z}_3$ nematic metal indicates orbital selective ordering and exhibit strongly anisotropic transport, as discussed in Sec.~\ref{Section IV D: Transport}. 
We have corroborated our results using a Landau theory with clock anisotropy terms for describing the nematic 
order in both models. This Landau theory demonstrates that the $N \rightarrow \infty$ limit is not essential for the nematic ordering. This large $N$ limit is however required to study the relevant nFL parent state. We take the perspective however that the large-$N$ limit emerges as a universality class of critical systems. One manner of obtaining this is considering the large $N$ limit as a coarse-grained average over $N$ disordered systems \cite{Chowdhury2018, Allocca2024}. The large $N$ limit behaviour also appears for SU$(2)$ spins in SY and related models at criticality  \cite{cha_linear_2020, Dumitrescu2022, Hardy2024}.   \par

The original motivation for this study was the observation of  $\mathbb{Z}_3$ symmetry breaking and nFL behavior in twisted Moire systems \cite{Nematic_Moire_Abhay_Nature2019, jiangChargeOrderBroken2019,cao_nematicity_2021, rubio-verdu_moire_2022, Nematic_Moire_NadjPerge_NatPhys2019,jinStripePhasesWSe22021,xu_moireZ3_prb2020}. Connecting the effective three-orbital model to the original lattice spin-valley degrees of freedom of these experiments remains an intriguing question. The degree to which these large $N$ studies relate to physical materials is also of considerable interest. This systems have considerable disorder in the relative twist angle due lattice relaxation effects. Rather than considering a connection between $N$ and a microscopic degree of freedom, we again take the perspective that $N$ corresponds to a coarse-grained average of disordered sites within real samples \cite{Chowdhury2018, Allocca2024}. 
This lattice mismatch also induces residual strain in samples. Extensions to incorporate the response of $\mathbb{Z}_3$ orders
to strain are important directions to explore in future studies.

\bigskip

\begin{acknowledgments}
We thank D. Chowdhury and F. Kugler for insightful discussions.
This research was supported by the Natural Sciences and Engineering Research Council (NSERC) of Canada.
AH acknowledges an NSERC Graduate Fellowship (PGS-D). YZX acknowledges support from an Ontario Graduate Scholarship (OGS). 
Numerical computations were performed on the Niagara supercomputer at the SciNet HPC Consortium and the Digital Research Alliance of Canada. 
AH also acknowledges support from a predoctoral fellowship at the Flatiron Institute while part of this work was being completed. The Flatiron Institute is a division of the Simons Foundation.
\end{acknowledgments}

\bibliography{apssamp}% Produces the bibliography via BibTeX.
\appendix
\onecolumngrid

\section{Effective Action} \label{Appendix A: Effective action}
The Hamiltonian of our model is
\begin{gather}
	\begin{aligned}
		H = & \left(- \sum_{\mathbf{r},\vec{\delta}} \sum_{s,i} t_{s, \vec{\delta}} c_{\mathbf{r},s,i}^\dag c_{\mathbf{r}+\vec{\delta},s,i}  + h.c. \right) - \mu \sum_{\mathbf{r},s} \sum_{i} c_{\mathbf{r},s,i}^\dag c_{\mathbf{r},s,i} +  \sum_{\mathbf{r},s,(ijkl)} J^{(s)}_{ij;kl}(\mathbf{r}) c_{\mathbf{r},s,i}^\dag c_{\mathbf{r},s,j}^\dag c_{\mathbf{r},s,k} c_{\mathbf{r},s,l}  \\
		& + \left( \sum_{s < s'} \sum_{\mathbf{r},(ijkl)} V_{ij;kl}^{(s,s')}(\mathbf{r}) c_{\mathbf{r},s,i}^\dag c_{\mathbf{r},s,j}^\dag c_{\mathbf{r},s',k} c_{\mathbf{r},s',l} +\text{h.c} \right) .
	\end{aligned}
\end{gather}
 which is also given by Eq.\eqref{SYK Model Hamiltonian} in the main text. 
We disorder average over $\{J_{ij;kl}^{(s)}, V_{ij;kl}^{(s,s')} \}$ through the replica trick, $\overline{\ln Z} = \lim_{n\to 0} (\overline{Z}^n - 1)/n$, where $n$ is the number of replicas of the system \cite{Chowdhury2021b}. After disorder averaging over $\{J_{ij;kl}^{(s)}, V_{ij;kl}^{(s,s')} \}$, we obtain the averaged partition function given by
\begin{gather}\label{Appendix replicated partition function}
	\overline{Z^n} = \int \mathcal{D}(\bar{c},c) e^{-S[\bar{c},c]} ,
\end{gather}
where
\begin{gather}
	\begin{aligned}
		S = &\int d\tau d\tau' \left[ \sum_{\alpha = 1}^{n} \sum_{\mathbf{r},\mathbf{r}',s,i} \bar{c}_{\mathbf{r},s,i,\alpha}(\tau) \bigg[ (\partial_{\tau} - \mu) \delta (\tau - \tau')  \delta_{\mathbf{r},\mathbf{r}'} - t_{s,\mathbf{r}-\mathbf{r}'} \bigg] c_{\mathbf{r}',s,i,\alpha}(\tau') \right ] \\
		&+ \frac{N}{4} \int d\tau d\tau' \sum_{\alpha,\gamma = 1}^{n} \sum_{\mathbf{r},s} (J^{(s)})^2 \left( \sum_{i,j} \frac{1}{N^2} \bar{c}_{\mathbf{r},s,i,\gamma}(\tau) c_{\mathbf{r},s,i,\alpha}(\tau') \bar{c}_{\mathbf{r},s,j,\alpha}(\tau') c_{\mathbf{r},s,j,\gamma} (\tau) \right)^{2} \\
		&+ \frac{N}{2} \int d\tau d\tau' \sum_{\alpha,\gamma = 1}^{n} \sum_{s<s'} \sum_{\mathbf{r}} (V^{(ss')})^2 \left( \sum_{i,j} \frac{1}{N^2}   \bar{c}_{\mathbf{r},s,i,\gamma}(\tau) c_{\mathbf{r},s,i,\alpha}(\tau') \bar{c}_{\mathbf{r},s',j,\alpha}(\tau') c_{\mathbf{r},s',j,\gamma} (\tau) \right)^{2} ,
	\end{aligned}
\end{gather}
where $\alpha,\gamma = 1,\cdots, n$ denote the replica index. We introduce the following Lagrange multiplier fields
\begin{gather}
	G_{s,\mathbf{r},\alpha\gamma}(\tau, \tau') \equiv -\frac{1}{N}\sum_{i} \bar{c}_{\mathbf{r},s,i,\gamma}(\tau') c_{\mathbf{r},s,i,\alpha}(\tau) ,
\end{gather}
by inserting the following identity into the replicated partition function in Eq.\eqref{Appendix replicated partition function},
\begin{gather}
	\begin{aligned}
		1 &= \prod_{s,\mathbf{r},\alpha \gamma} \int \mathcal{D}G_{s,\mathbf{r},\alpha \gamma}(\tau,\tau') \delta \bigg( N G_{s,\mathbf{r},\alpha \gamma}(\tau',\tau) + \sum_{i} \bar{c}_{\mathbf{r},s,i,\gamma}(\tau) c_{\mathbf{r},s,i,\alpha}(\tau')  \bigg) \\[5pt]
		&= \int \left( \prod_{s,\mathbf{r},\alpha \gamma} \mathcal{D} G_{s,\mathbf{r},\alpha \gamma} \mathcal{D}\Sigma_{s,\mathbf{r},\alpha \gamma} \right) \exp \left[ \sum_{s,\mathbf{r},\alpha \gamma} N \int d\tau d\tau' \Sigma_{s,\mathbf{r},\alpha \gamma}(\tau, \tau') \left(  G_{s,\mathbf{r},\alpha \gamma}(\tau',\tau) + \frac{1}{N}\sum_{i} \bar{c}_{\mathbf{r},s,i,\gamma}(\tau) c_{\mathbf{r},s,i,\alpha}(\tau ') \right) \right] .
	\end{aligned}
\end{gather}
The action then becomes
\begin{gather}
	\begin{aligned}
		S = &\int d\tau d\tau' \Bigg\{ \left[ \sum_{\alpha = 1}^{n} \sum_{\mathbf{r},\mathbf{r}',s,i} \bar{c}_{\mathbf{r},s,i,\alpha}(\tau) \bigg[ (\partial_{\tau} - \mu) \delta (\tau - \tau')  \delta_{\mathbf{r},\mathbf{r}'} - t_{s,\mathbf{r}-\mathbf{r}'} \bigg] c_{\mathbf{r}',s,i,\alpha}(\tau') \right ] \\
      &+ \sum_{s,\mathbf{r},\alpha \gamma} \Sigma_{s,\mathbf{r},\alpha \gamma}(\tau,\tau') \sum_{i} \bar{c}_{\mathbf{r},s,i,\gamma}(\tau) c_{\mathbf{r},s,i,\alpha}(\tau ') + \frac{N}{4} \sum_{\alpha,\gamma = 1}^{n} \sum_{\mathbf{r},s} (J^{(s)})^2 \bigg( G_{s,\mathbf{r},\alpha \gamma}(\tau',\tau) G_{s,\mathbf{r},\gamma \alpha}(\tau,\tau') \bigg)^{2} \\
		&+ \frac{N}{2} \sum_{\alpha,\gamma = 1}^{n} \sum_{s<s'} \sum_{\mathbf{r}} (V^{(ss')})^2  \bigg( G_{s,\mathbf{r},\alpha \gamma}(\tau',\tau) G_{s,\mathbf{r},\gamma \alpha}(\tau,\tau') \bigg)^{2} + N \sum_{s,\mathbf{r},\alpha,\gamma} \Sigma_{s,\mathbf{r},\alpha \gamma}(\tau, \tau') G_{s,\mathbf{r},\alpha \gamma}(\tau', \tau) \Bigg\}.
	\end{aligned}
\end{gather}
Next, we consider a replica-diagonal ansatz, i.e, $\Sigma_{s,\mathbf{r},\alpha \gamma} = \Sigma_{s,\mathbf{r}} \delta_{\alpha,\gamma}$ and $G_{s,\mathbf{r},\alpha \gamma} = G_{s,\mathbf{r}} \delta_{\alpha \gamma}$. This implies $\overline{Z^n} = \overline{Z}^n$, so we can work directly with $\overline{Z}$. Furthermore, we consider an average lattice-translational invariance, such that $\Sigma_{s,\mathbf{r}} = \Sigma_{s}$ and $G_{s,\mathbf{r}} = G_{s}$. After a Fourier transform and integrating out the fermionic fields, we obtain
\begin{gather}
	\overline{Z} = \int \mathcal{D}[G,\Sigma] e^{-S_{eff}[G,\Sigma]} ,
\end{gather}
where the effective action (per fermionic mode $N$ per lattice site $N_L$) is given by
\begin{gather}
	\begin{aligned}
		S_{eff}/N N_L &= \int d\tau d\tau' \bigg[ \sum_{s} \nu \int \frac{d^{d}\mathbf{k}}{(2\pi)^d} \text{Tr}\ln \bigg( \delta(\tau - \tau') \big(\partial_{\tau} - \mu + \varepsilon_s(\mathbf{k}) \big) + \Sigma_s(\tau,\tau')  \bigg)  \\
		& - \int d\tau d\tau' \sum_{s}\left[ \frac{(J^{(s)})^2}{4} G_{s}(\tau',\tau)^2 G_{s}(\tau,\tau')^2 + \Sigma_s(\tau,\tau')G_s(\tau',\tau) \right ] \\
      &- \int d\tau d\tau' \sum_{s,s': s<s'} \frac{(V^{(s,s')})^2}{2} G_{s}(\tau',\tau)^2 G_{s'}(\tau,\tau')^2 .
	\end{aligned}
\end{gather}
Since the Hamiltonian is time-translational invariant, we have $\Sigma_s(\tau, \tau') = \Sigma_s(\tau - \tau')$ and $G_s(\tau, \tau') = G_s(\tau - \tau')$ so we can work with the fermionic Matsubara frequencies, $\omega_n = (2n+1)\pi/\beta$ with $\beta = T^{-1}$. The effective action can then be rewritten as
\begin{gather}
	\begin{aligned}
		S_{eff}/NN_L &= \sum_{s} \bigg[ - \sum_{i\omega_n} \int d\varepsilon g_s(\varepsilon) \ln \big[ i\omega_n + \mu - \varepsilon - \Sigma_s(i\omega_n) \big] + \beta \int_{0}^{\beta} d\tau \Sigma_s(\tau) G_s(\beta - \tau) \\
      &\hspace{4em}- \beta \frac{\big(J^{(s)}\big)^2}{4} \int_{0}^{\beta} d\tau G_s^2(\beta - \tau) G_s^2(\tau) \bigg] - \beta \sum_{s,s': s<s'} \frac{(V^{(s,s')})^2}{2} \int_{0}^{\beta} G_{s}^2 (\beta - \tau) G_{s'}^2(\tau) .
	\end{aligned}
\end{gather}
where $g_s(\varepsilon) \equiv \nu \int \frac{d^d \mathbf{k}}{(2\pi)^d} \delta(\varepsilon - \varepsilon(\mathbf{k}))$ is the orbital-dependent lattice density of state. For $N \to \infty$, the saddle point of the effective action leads to the following Dyson's equations,
\begin{gather}
	\begin{aligned}
		\Sigma_s(\tau) &= -\big(J^{(s)}\big)^2G_s^2(\tau)G_s(-\tau) - \sum_{s': s'\neq s} \big(V^{(ss')}\big)^2 G_{s'}^2(\tau)G_{s}(-\tau) , \\
		G_s(i\omega_n) &= \int d\varepsilon \, g_s(\varepsilon) \big[ i\omega_n + \mu - \varepsilon - \Sigma_s(i\omega_n) \big]^{-1} .
	\end{aligned}
\end{gather}
\section{Real Frequency Self-Consistent Solutions} \label{Appendix: Real Frequency Self-Consistent Solutions}

In this Appendix, we derive the self-consistent equations for the spectral function through analytic continuation  of the Dyson equation. We determine $\Sigma_s(i\omega_n)$ from Eq.\eqref{DS1} through
\begin{gather}
	\begin{aligned}
		\Sigma_s(i\omega_n) &= \int_{0}^{\beta} d\tau e^{i\omega_n \tau} \Sigma_s(\tau) = -\frac{(J^{(s)})^2}{\beta^2} \sum_{n_1, n_2} G_s(i\omega_{n_1}) \, G_s(i\omega_{n_2}) \, G_s(i\omega_{n_1} + i\omega_{n_2} - i\omega_n) \\[5pt]
		&\qquad - \sum_{s': s'\neq s} \frac{V^{(ss')} {}^2}{\beta^2} \sum_{n_1, n_2} G_s(i\omega_{n_1}) G_{s'}(i\omega_{n_2}) G_{s'}(i\omega_{n_1} + i\omega_{n_2} - i\omega_n) ,
	\end{aligned}
\end{gather}
which is obtained through the Fourier decomposition
\begin{gather}
	G(\tau) = \frac{1}{\beta} \sum_{n} G(i\omega_n) e^{-i\omega_n \tau} .
\end{gather}
Now we express the spectral representation $G_s(i\omega_n) = \int_{-\infty}^{\infty} d\omega \frac{A_s(\omega)}{i\omega_n - \omega}$. The $J$ term for example can be expressed as
\begin{gather}
	-\sum_{n_1, n_2 } \frac{(J^{(s)})^2}{\beta^2} \int_{-\infty}^{\infty} \dfrac{\left( \prod_{\alpha = 1}^{3} \mathrm{d}\omega_{\alpha} A_s(\omega_{\alpha})\right)}{(i\omega_{n_1} - \omega_1) (i\omega_{n_2} - \omega_{2})(i\omega_{n_1} + i\omega_{n_2} - i\omega_n - \omega_3)} .
\end{gather}
The Matsubara sums in the expression can be computed as,
\begin{gather}
	\begin{aligned}
		& \sum_{n_1} \sum_{n_2} \frac{1}{(i\omega_{n_1} - \omega_1) (i\omega_{n_2} - \omega_{2})(i\omega_{n_1} + i\omega_{n_2} - i\omega_n - \omega_3)} \\[5pt]
		&= \beta \sum_{n_2} \left( \frac{n_F(\omega_1)}{(i\omega_{n_2} - \omega_{2})(\omega_1 + i\omega_{n_2} - i\omega_n - \omega_3)} + \frac{n_F(-i\omega_{n_2} + i\omega_{n} + \omega_3)}{(-i\omega_{n_2} + i\omega_n + \omega_3 - \omega_1) (i\omega_{n_2} - \omega_{2})} \right) \\[5pt]
		&= \beta \sum_{n_2} \frac{n_F(\omega_1) - n_F(\omega_3)}{(i\omega_{n_2} - \omega_{2})(\omega_1 - \omega_3 + i\omega_{n_2} - i\omega_n )} \\[5pt]
		&= \beta^2 (n_F(\omega_1) - n_F(\omega_3)) \left( \frac{n_F(\omega_2)}{\omega_1 -\omega_3 + \omega_2 - i\omega_n} + \frac{n_F(\omega_3 - \omega_1 + i\omega_n)}{-\omega_1 + \omega_3 + i\omega_n - \omega_2} \right) \\[5pt]
		&= \beta^2 \frac{\big( n_F(\omega_1) - n_F(\omega_3) \big) \big( n_F(\omega_2) + n_B(\omega_3 - \omega_1) \big)}{\omega_1 + \omega_2 - \omega_3 - i\omega_n} \\[5pt]
		&=  \beta^2 \frac{n_F(-\omega_1)n_F(-\omega_2)n_F(\omega_3) + n_F(\omega_1)n_F(\omega_2)n_F(-\omega_3)}{\omega_1 + \omega_2 - \omega_3 - i\omega_n} ,
	\end{aligned}
\end{gather}
where $n_F(\cdots)$ is the Fermi-Dirac function. In the first equality, we make use of the fact that $i\omega_{n_2} - i\omega_{n}$ is a bosonic Matsubara frequency to write $n_F( -i\omega_{n_2} + i\omega_{n} + \omega_3)= n_F(\omega_3)$ without further analytic extension for $\omega_{n_2}$. A similar identity gives $n_F(\omega + i\omega_n) = -n_B(\omega)$ since $i\omega_n$ is a fermionic Matsubara frequency. The $J$ term after analytic continuation of $i\omega_n \to \omega + i\eta$ becomes
\begin{gather}
	\begin{aligned}
		 J^{(s)}{}^2 \int_{-\infty}^{\infty} \left( \prod_{\alpha =1}^{3} d\omega_\alpha A_s(\omega_\alpha) \right)  \dfrac{n_F(-\omega_1)n_F(-\omega_2)n_F(\omega_3) + n_F(\omega_1)n_F(\omega_2)n_F(-\omega_3)}{-\omega_1 - \omega_2 + \omega_3 + \omega  + i\eta} .
	\end{aligned}
\end{gather}
A similar expression for the $V$ terms can be derived. We note the identity $\frac{1}{\omega'  + i0^+} = (-i) \frac{1}{i\omega' - 0^+} = -i \int_{0}^{\infty} dt e^{i(\omega' + i0^+)t}$, thus combining $J$ and $V$ terms we have
\begin{gather}
	\begin{aligned}
		\Sigma_s^{R}(\omega) = \Sigma_s(\omega + i0^+) &= 
		-i\int_{0}^{\infty} e^{i\omega t} \bigg( J^{(s)}{}^2 \big\{ n_{1,s}^2(t) n_{2,s}(t) + n_{3,s}^2(t) n_{4,s}(t) \big\} \\
		&\hspace{5em} + \sum_{s': s'\neq s} V^{(ss')} \big\{ n_{1,s'}^2(t) n_{2,s}(t) + n_{3,s'}^2(t) n_{4,s}(t) \big\} \bigg) ,
	\end{aligned}
\end{gather}
where
\begin{gather}
	\begin{aligned}
		n_{1,s}(t) &\equiv \int_{-\infty}^{\infty} d\omega' A_s(\omega') n_F(-\omega ') e^{-i\omega' t} , \\
		n_{2,s}(t) &\equiv \int_{+\infty}^{\infty} d\omega' A_s(\omega') n_F(\omega') e^{i\omega't} , \\
		n_{3,s}(t) &\equiv \int_{-\infty}^{\infty} d\omega' A_s(\omega') n_F(\omega') e^{-i\omega't} , \\
		n_{4,s}(t) &\equiv \int_{-\infty}^{\infty} d\omega' A_s(\omega') n_F(-\omega') e^{i\omega't} .
	\end{aligned}
\end{gather}
The corresponding Dyson equation via analytic continuation is
\begin{equation}
	G_s^R(\omega) = \int d\varepsilon g(\varepsilon) \big[ \omega + \mu - \varepsilon - \Sigma_s^R(\omega)\big]^{-1} ,
\end{equation}
which provides the spectral function 
\begin{equation}
	A_s(\omega) = -\frac{1}{\pi} \text{Im}\big( G_s^R(\omega) \big) .
\end{equation}

\section{Thermodynamic Numerical Procedure}
\label{Appendix: Thermodynamic Numerical Procedure}
\begin{figure}[H]
	\centering
        \subfloat[]{ \includegraphics[width=0.32\linewidth]{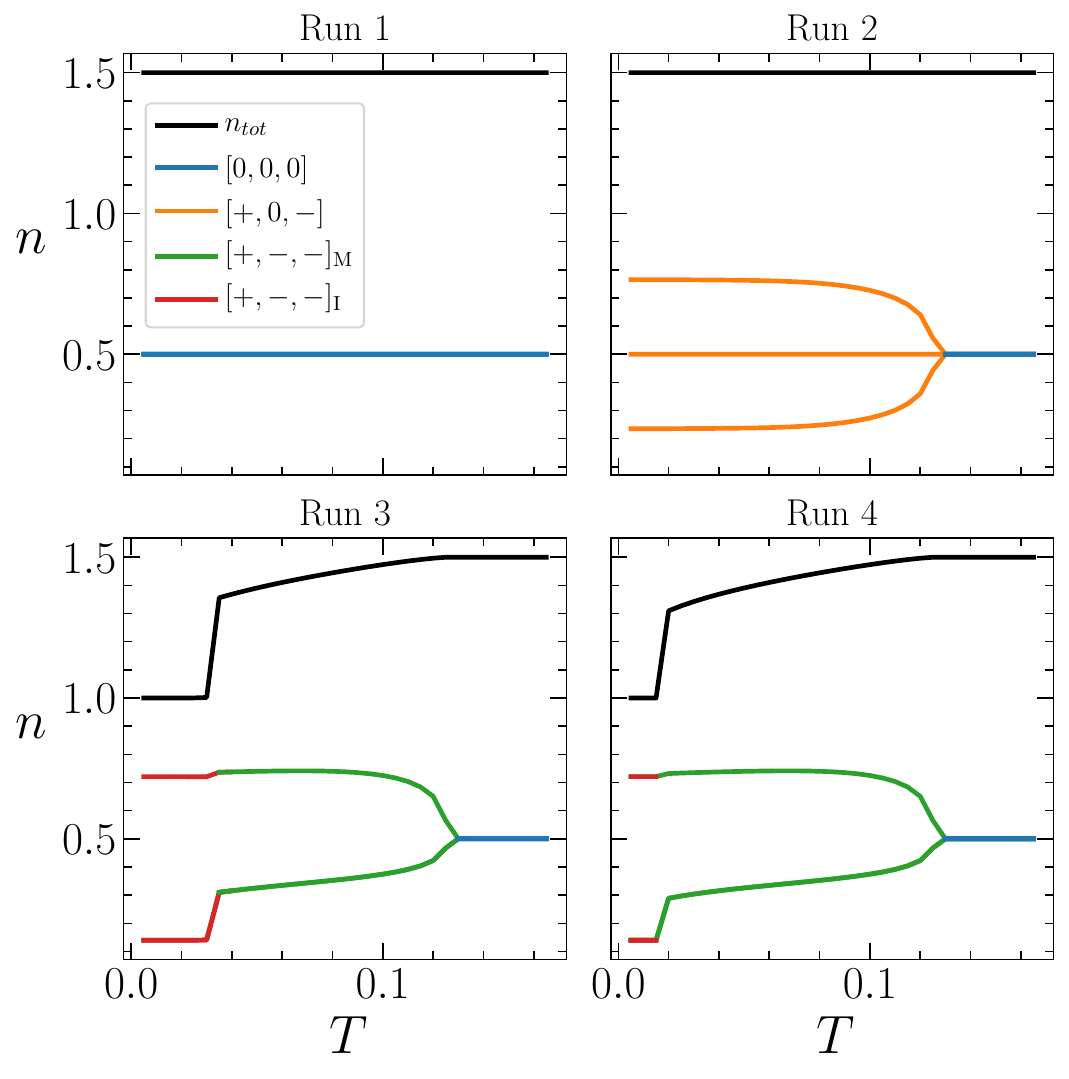}\label{V=1.2 density} } \hfill
	\subfloat[]{\includegraphics[width=0.32\linewidth]{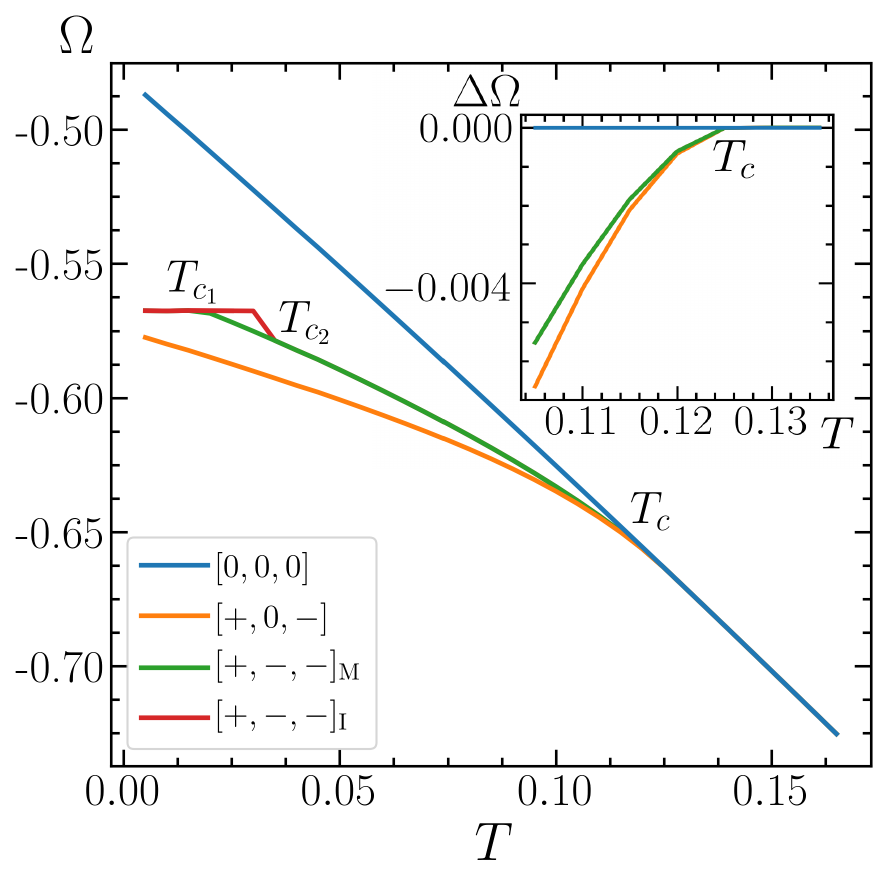}\label{V=1.2 free energy}} \hfill
        \subfloat[]{\includegraphics[width=0.32\linewidth]{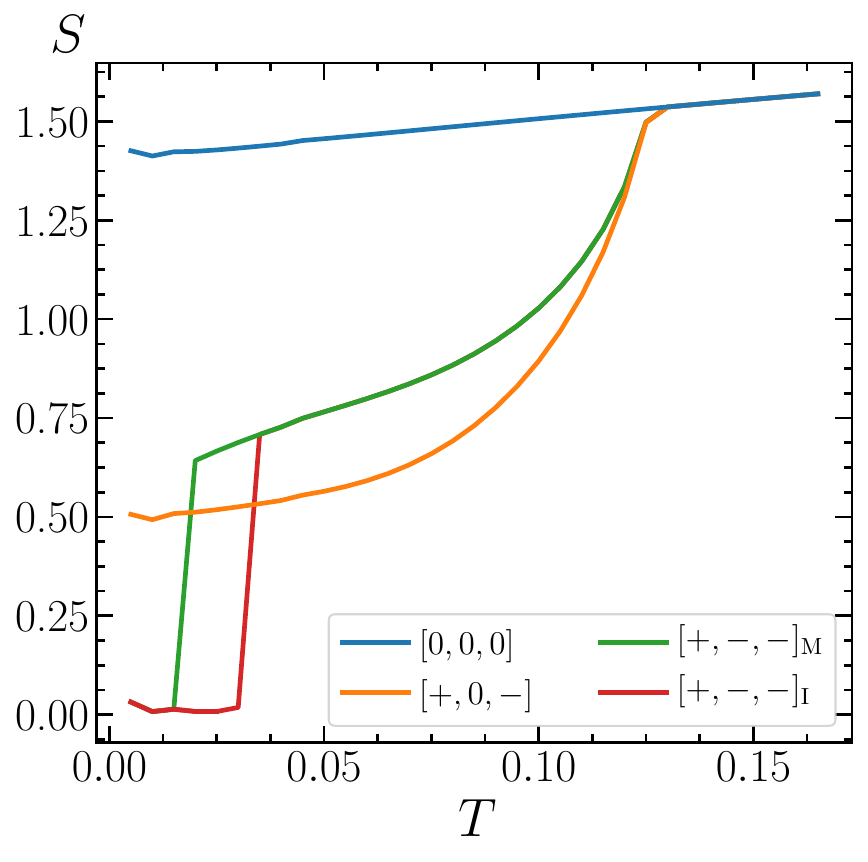}}\label{V=1.2 entropy} 
	\caption{(a) $n_s(T)$, (b) $\Omega(T)$, and (c)  $S(T)$ for $V = 1.2$ as a function of temperature, focusing on the low temperatures in Fig.~\ref{0D phase diagram}. Colors distinguish different phases,  denoted as: $[0,0,0]$ (blue), $[+,0,-]$ (yellow), $[+,-,-]_{\rm M}$ (green), and $[+,-,-]_{\rm I}$ (red).
 (a) shows results from four independent full-temperature runs, labelled as run $1$ to $4$, where each of them corresponds to a single curve in (b) and (c). (a) indicates that the $[0,0,0]$ and $[+,0,-]$ phase is particle-hole symmetric and remain half-filling at all temperature, while the $[+,-,-]_{\rm I}$ and $[+,-,-]_{\rm M}$ phase have spontaneous PHS breaking and the total filling deviate from half-filling.  (b) Four saddle point solutions coexist within the temperature region $T\in [T_{c1}, T_{c2}]$. 
 The inset of (b) shows the free energy difference of each phases relative to that of the $[0,0,0]$ phase near the phase transition at $T_c$, suggesting that the $[+,0,-]$ phase is stable for $T<T_c$ and $[0,0,0]$ phase is stable for $T>T_c$ in the Grand Canonical ensemble.  }
	\label{V = 1.2 figures for 0D 3-orbital SYK model}
\end{figure}

In this Appendix, we provide a detailed explanation of our procedure for obtaining the phase diagram (in Fig.~\ref{0D phase diagram}) of the three-orbital SYK dot. This procedure applies similarly to the lattice extension of the model, as discussed in Sec.~\ref{Section III B: Phase diagram} and \ref{Section IV B: Phase diagram}. We illustrate the details through an example of a cut through the phase diagram at $V = 1.2$.
As discussed in Sec.~\ref{Section I B: phase diagram} of the main text, four saddle point solutions are found across the phase diagram. To obtain each of these solutions, we conduct four independent full-temperature simulations, labeled as Runs 1 to 4 in Fig.~\ref{V=1.2 density}. Each of the four plots displays the orbital-resolved density $n_s(T)$ along with the total density $\sum_s n_s(T)$ as a function of temperature. 
For Runs 1 to 3, we start from the lowest temperature with three different initial conditions of $G_s(\tau)$ for the iterations, and apply the solution at each temperature as the initial condition for the subsequent temperature iteration while progressively increasing the temperature. For run 4, instead of starting from the lowest temperature, the algorithm starts from an intermediate temperature where the $[+,-,-]_{\rm M}$ phase exists as a saddle-point solution. The free energy density $\Omega$ at each of the saddle point solutions are then computed and is shown in Fig.~\ref{V=1.2 free energy}. We use these values to identify the thermodynamically stable phase by minimizing the free energy density. Subsequently, all thermodynamic quantities are updated corresponding to the thermodynamically stable phase.

Despite the fact that the $[+,-,-]_{\rm M}$ is not thermodynamically stable in the three-orbital SYK dot, here we provide further details on this phase, as it is discussed in the main text in Sec.~\ref{Section I B: phase diagram}. This phase shares two similarities with the PHS insulator $[+,-,-]_{\rm I}$: (1) Two of the three orbitals in both phases are degenerate, with their density being less than $1/2$. (2) Both phases have spontaneous PHS breaking, where the total filling deviates from half-filling. 
However, there are several features that distinguish the two phases: (1) The $[+,-,-]_{\rm M}$ phase has weak PHS breaking with a weak deviation of the total density from half-filling, where the deviation increases continuously as temperature decreases. In contrast, the $[+,-,-]_{\rm I}$ phase exhibits a strong deviation from half-filling where $\sum_s \langle n_s \rangle = 1.0$ at all temperature. 
(2) The $[+,-,-]_{\rm M}$ phase has larger entropy than the single SYK dot at all temperature, while the $[+,-,-]_{\rm I}$ phase has zero entropy whenever the solution exists. 
(3) The $[+,-,-]_{\rm I}$ phase has a vanishing compressibility and fully gapped spectral functions, while the $[+,-,-]_{\rm M}$ state has finite compressibility and is not fully gapped. \\

\section{Elaboration on the Three Orbital SYK-Dot} \label{Appendix: Further discussion on the three-orbital SYK dot}
\subsection{Thermodynamic Quantities}

\begin{figure}[!h] 
\centering
\subfloat[$P_{21}$]{\label{0D P21}\includegraphics[width=0.33\textwidth, height = 0.25\textwidth]{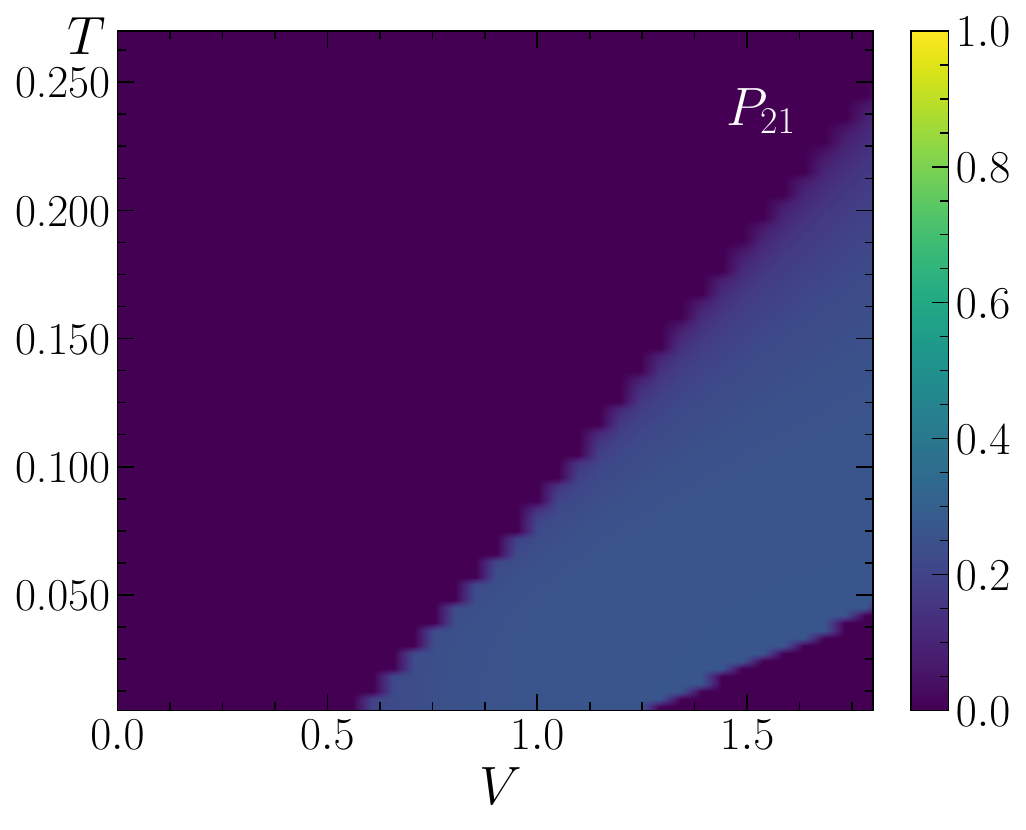}}
 \subfloat[$P_{32}$]{\label{0D P32}\includegraphics[width=0.33\textwidth, height = 0.25\textwidth]{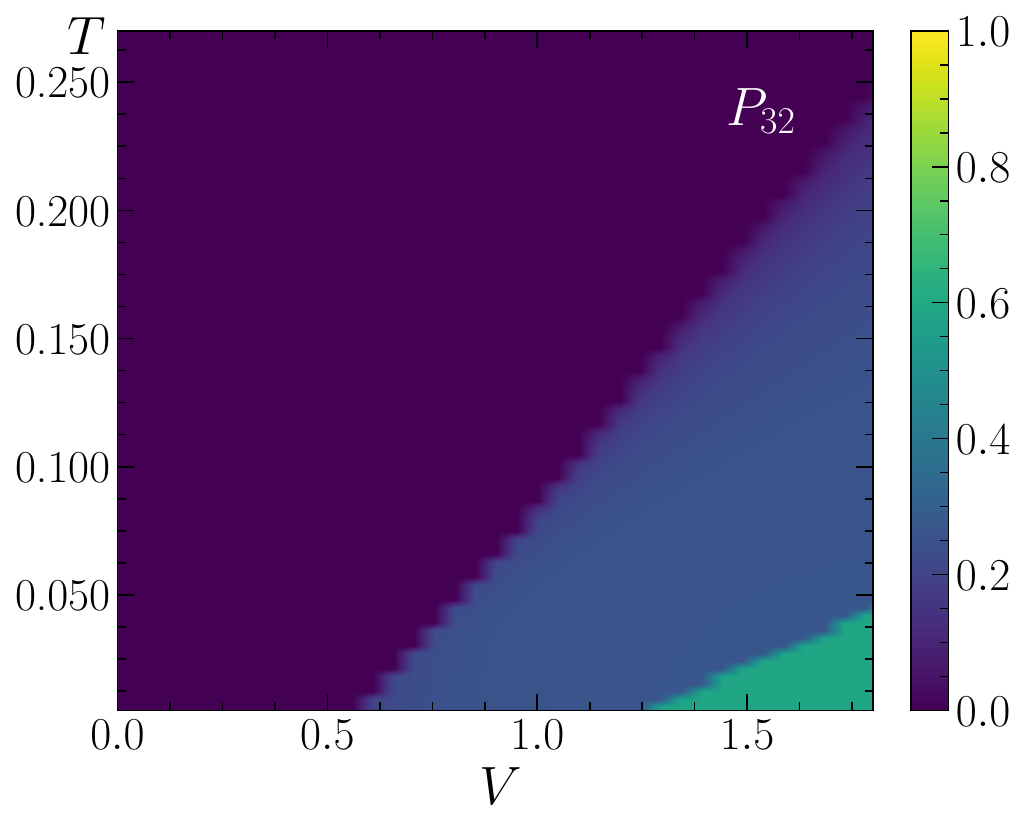}}
\subfloat[$n_{total} - 1.5$]{\label{0D density deviation}\includegraphics[width=0.33\textwidth, height = 0.25\textwidth]{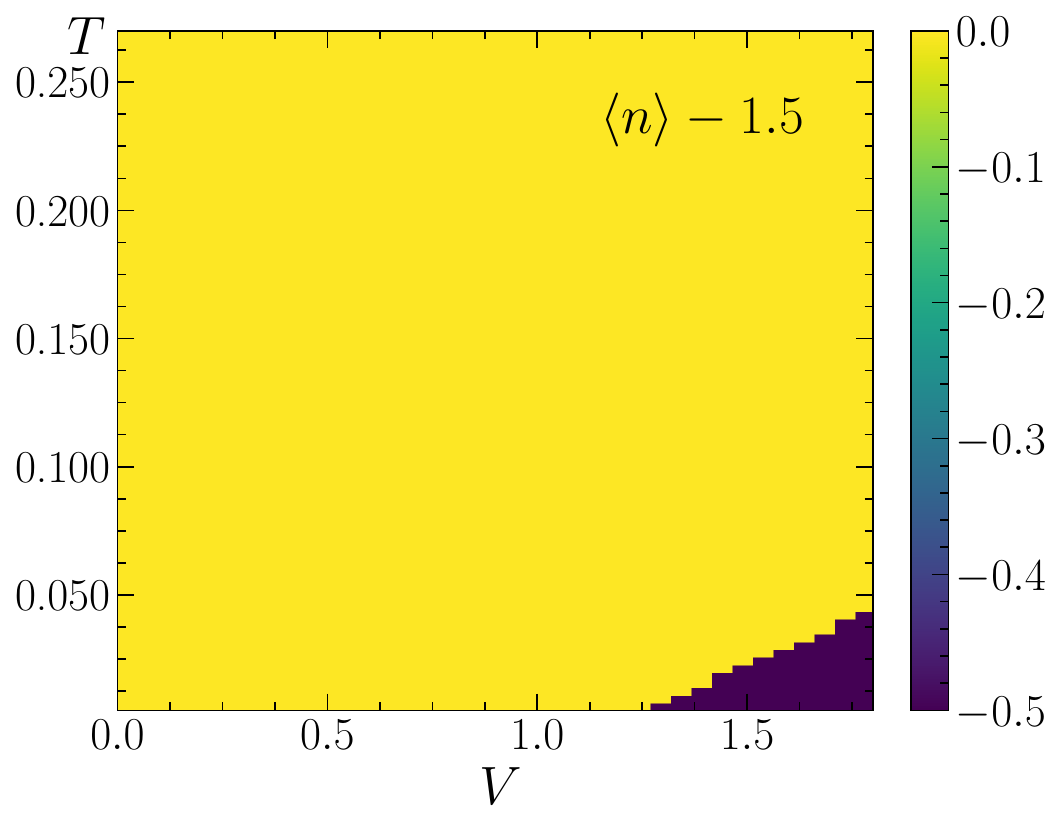}}
\caption{Polarization (a)-(b): $P_{21}, P_{32}$, (c) total density deviation from half-filling $n_{total} - 1.5$ for the three-orbital SYK dot. }
\label{0D Polarization}
\end{figure}

To classify each phase in the three-orbital SYK dot in Fig.~\ref{0D phase diagram}, we present additional details of the $V$-$T$ phase diagram. We illustrate the orbital polarizations $P_{21} = n_2 - n_1$ and  $P_{32} = n_3 - n_2$, along with the total density deviation from half-filling $\sum_s n_s - 1.5$ in Fig.~\ref{0D Polarization}. Specifically, $P_{21} = P_{32} = 0$ and $n = 1.5$ for the orbital-symmetric nFL phase $[0,0,0]$. The orbital-selective nFL phase $[+,0,-]$ satisfies $P_{21} = P_{32} \neq 0$ and $n = 1.5$. The PHS broken insulator $[+,-,-]_{\rm I}$ features a $P_{21} = 0, P_{32} \neq 0$, and a total density $n = 1.0$, strongly deviating from half-filling.

\begin{figure}[!h] 
\centering
 \subfloat[$S(T)$]{\label{0D entropy}\includegraphics[width=0.33\textwidth, height = 0.25\textwidth]{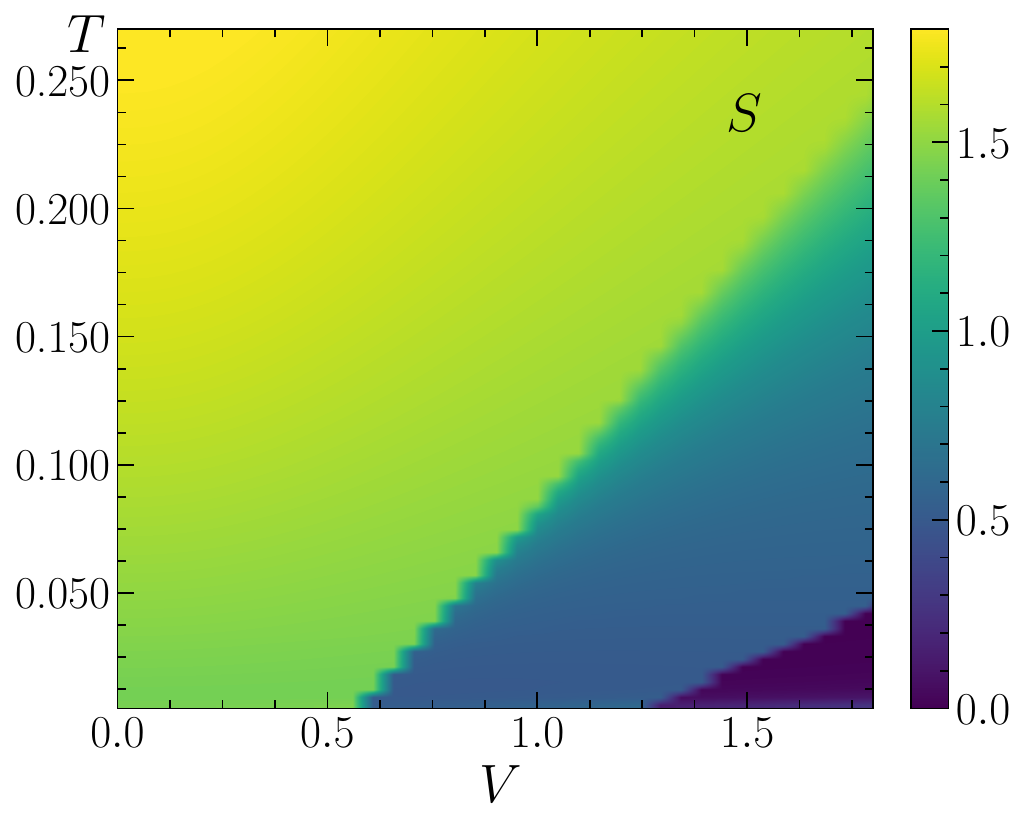}}
\subfloat[$C_V(T)$]{\label{0D specific heat}\includegraphics[width=0.33\textwidth, height = 0.25\textwidth]{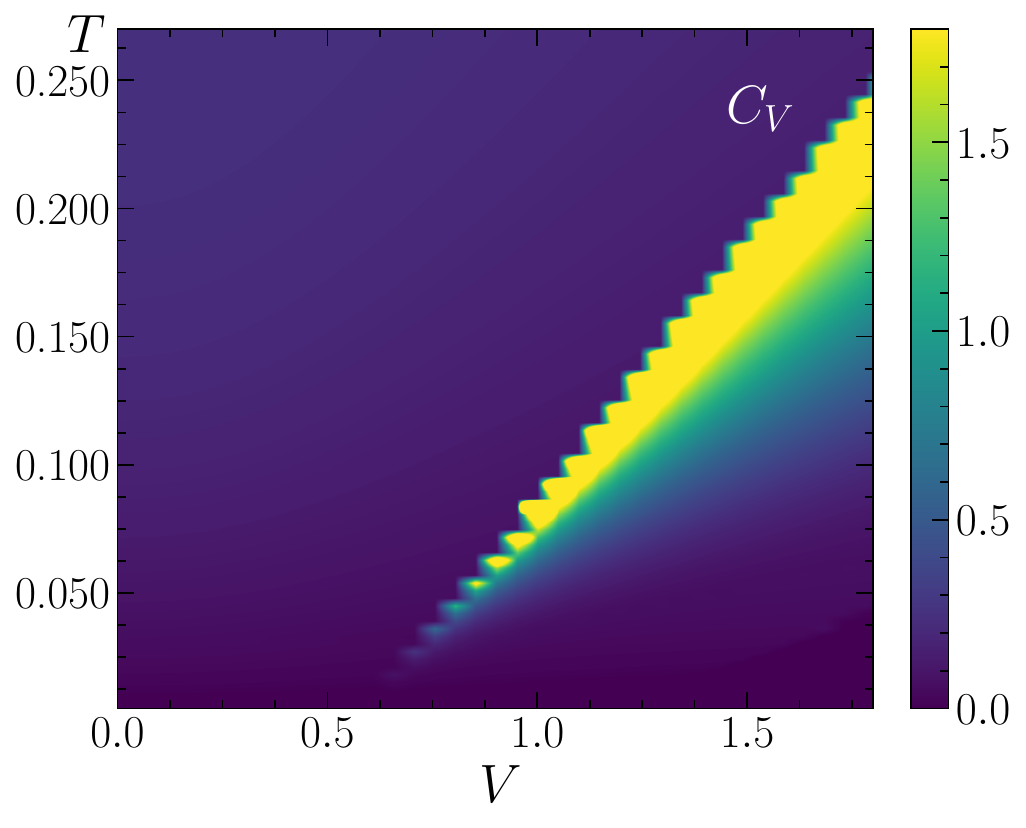}}
\subfloat[$\k(T)$]{\label{0D compress}\includegraphics[width=0.33\textwidth, height = 0.25\textwidth]{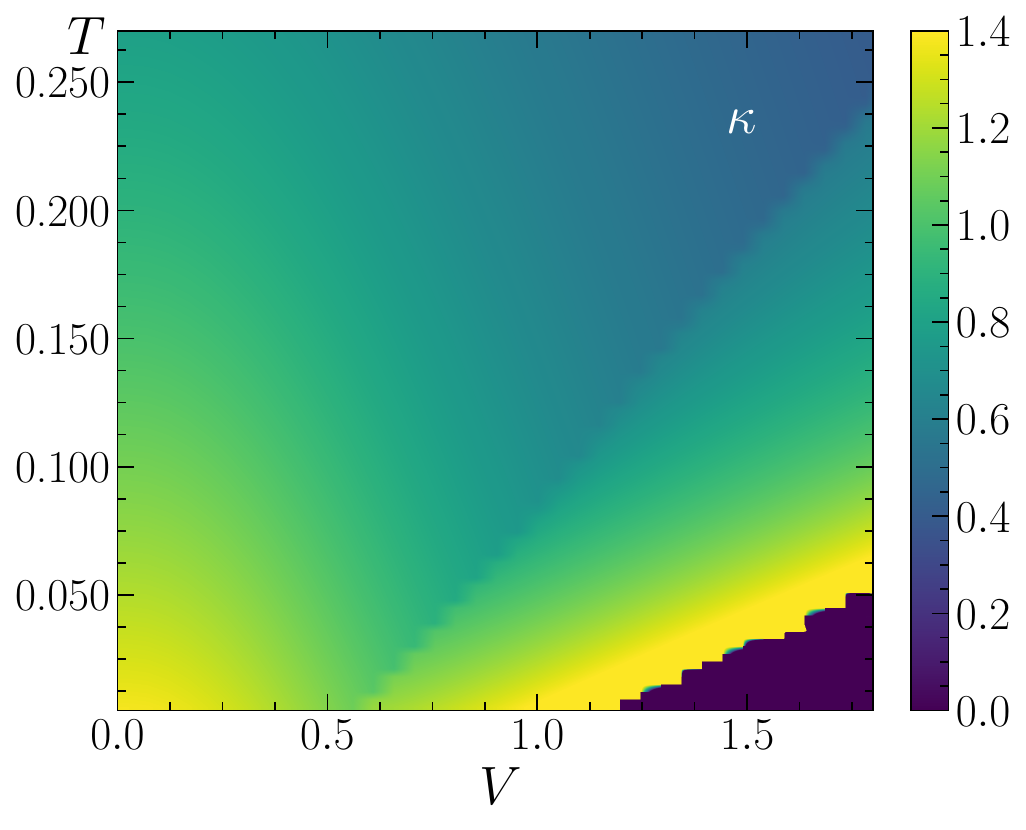}}
\caption{Thermodynamic quantities of the $V$-$T$ phase diagram for the three-orbital SYK dot.}
\label{thermodynamic quantities for the VT phase diagrams for 0D}
\end{figure}

\begin{figure}[H]
    \centering
    \includegraphics[width=0.5\textwidth, height = 0.4\textwidth]{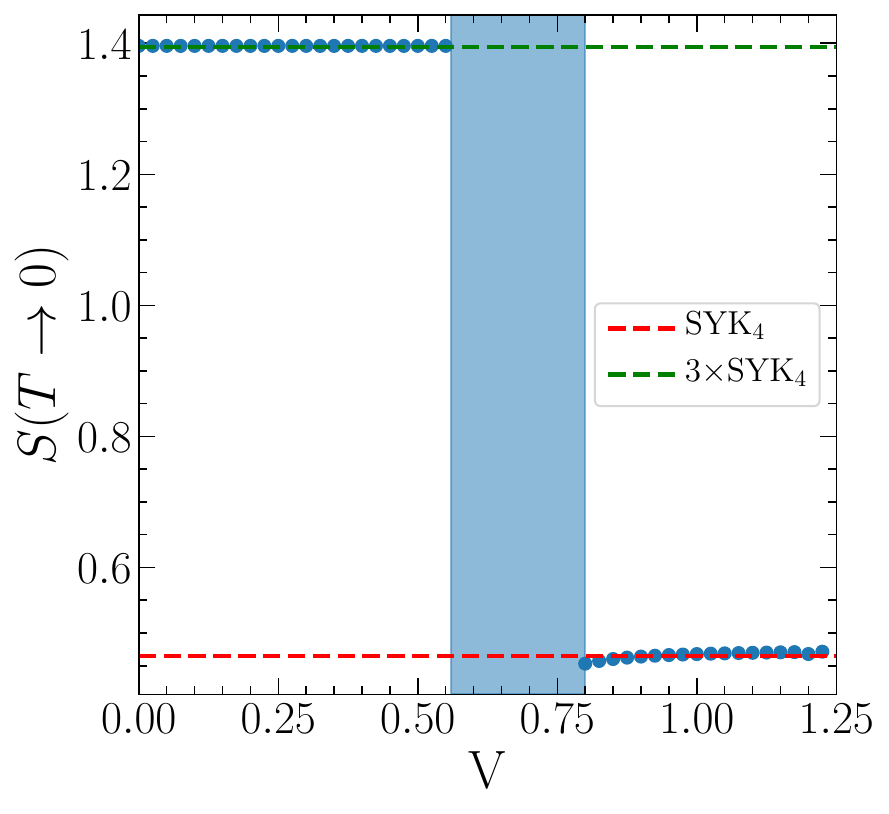}
    \caption{Residual entropy $S(T\to 0)$ of the lowest temperature phase as a function of the inter-orbital interaction $V$, extrapolated from the two lowest temperatures. Dash lines represent $1\times$ and $3\times$ the residual entropy of the original SYK model $S_{\text{SYK}_4} \approx 0.4648$ taken from \cite{Song2017a}, where all data points are within $2\%$ of the value. The blue region indicates the interval of $V$ where an accurate residual entropy cannot be computed from extrapolation due to the proximity of the lowest accessible temperature to the transition temperature, as illustrated in Fig.~\ref{0D phase diagram}.}
    \label{syk3dots entropy vs V}
\end{figure}

Additionally, we include several thermodynamic quantities to further characterize these phases and phase transitions. These includes the entropy $S(T) = -\left( \partial \Omega/\partial T \right)_{V,\mu}$, the specific heat $C_V(T) = T\left( \partial S/\partial T \right)_{V,\mu}$, and the compressibility $\kappa(T) = n^{-2}\left(\partial^2 \Omega/\partial \mu^2\right)_{V,T}$, where $n = \sum_s n_s$ is the total density. These quantities are shown in Fig.~\ref{thermodynamic quantities for the VT phase diagrams for 0D}.
In the phase diagram in Fig.~\ref{0D phase diagram}, the $[0,0,0]$-to-$[+,0,-]$ transition features a tricritical point at $V = 1$, beyond which the transition shifts from first-order to second-order. This transition features a divergence in the specific heat, and a jump in compressibility for $V < 1$. At large $V \geq 1.3$, we find a first-order transition from the $[+,0,-]$ phase to the PHS insulator $[+,-,-]_{\rm I}$, which features a divergence in compressibility. The $[0,0,0]$ phase has residual entropy $3\times \text{entropy}_{\text{SYK}_4}$, and the $[+,0,-]$ phase recover the residual entropy of the $\text{SYK}_4$ model, as indicated in Fig.~\ref{syk3dots entropy vs V}.

\section{Elaboration on the Cubic Lattice Model} \label{Appendix: Further discussion on the cubic lattice model}
\subsection{Thermodynamic Quantities}
\begin{figure}[H]
	\centering
	\subfloat[$S(T)$]{\label{entropy for 3-orbital phase diagram on cubic lattice  with V = 0.8}\includegraphics[width=0.33\textwidth, height = 0.25\textwidth]{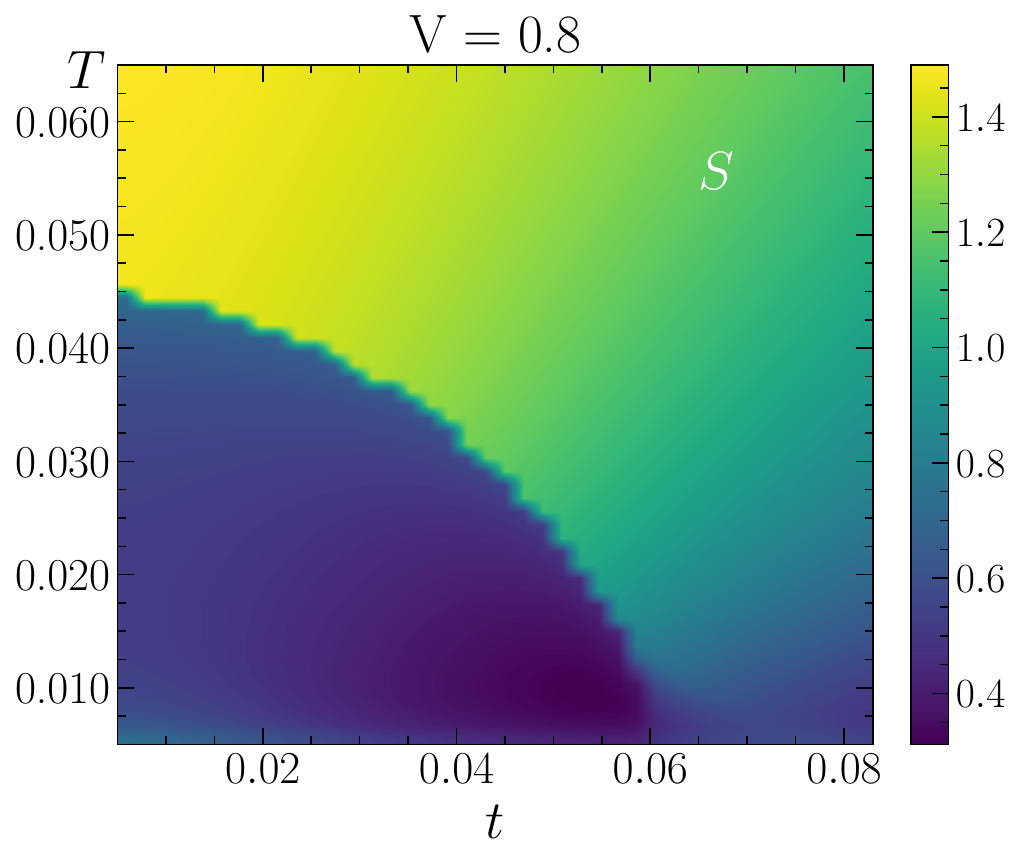}}
	\subfloat[$C_V(T)$]{\label{specific heat for 3-orbital phase diagram on cubic lattice  with V = 0.8}\includegraphics[width=0.33\textwidth, height = 0.25\textwidth]{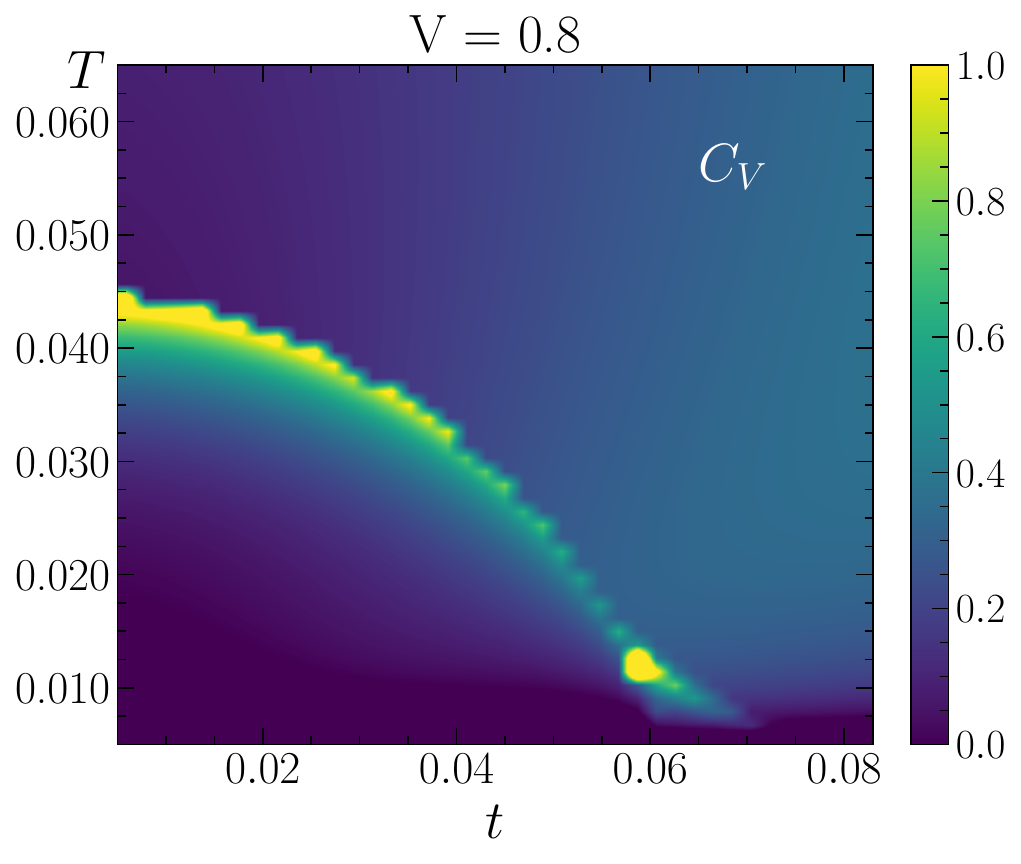}}
	\subfloat[$\k(T)$]{\label{compressibility for 3-orbital phase diagram on cubic lattice with V = 0.8}\includegraphics[width=0.33\textwidth, height = 0.25\textwidth]{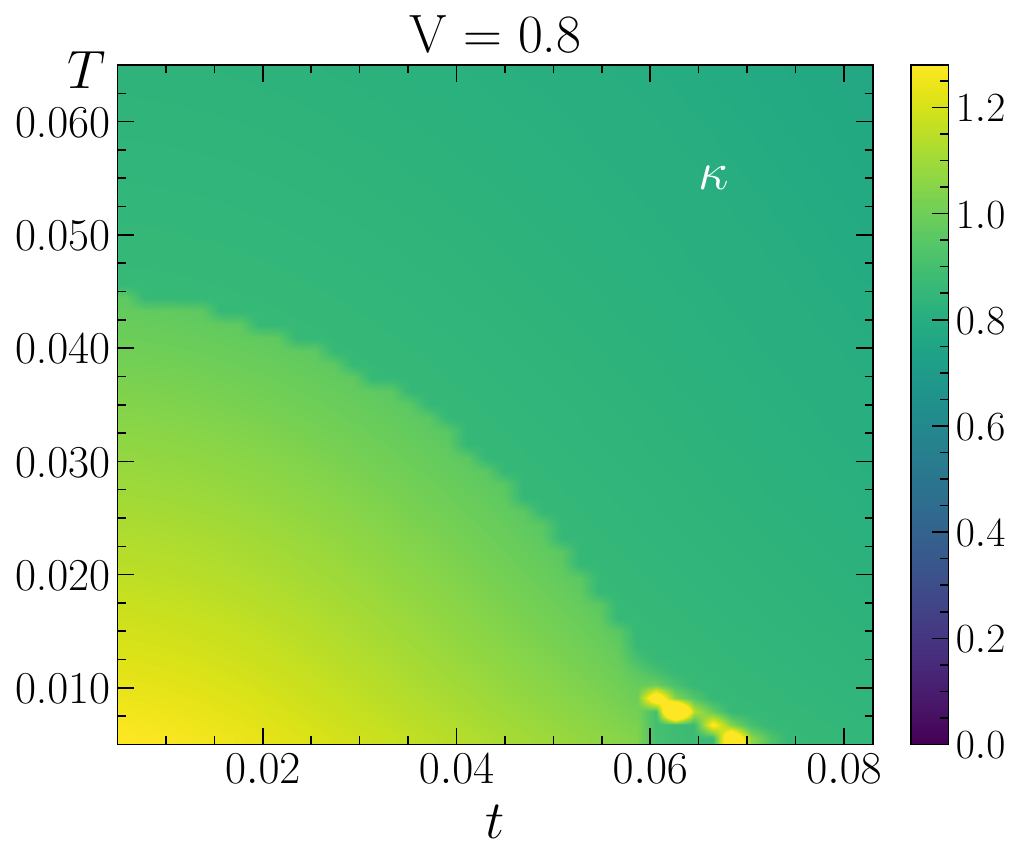}}

    \centering

	\subfloat[$S(T)$]{\label{entropy for 3-orbital phase diagram on cubic lattice with V = 1.2}\includegraphics[width=0.33\textwidth, height = 0.25\textwidth]{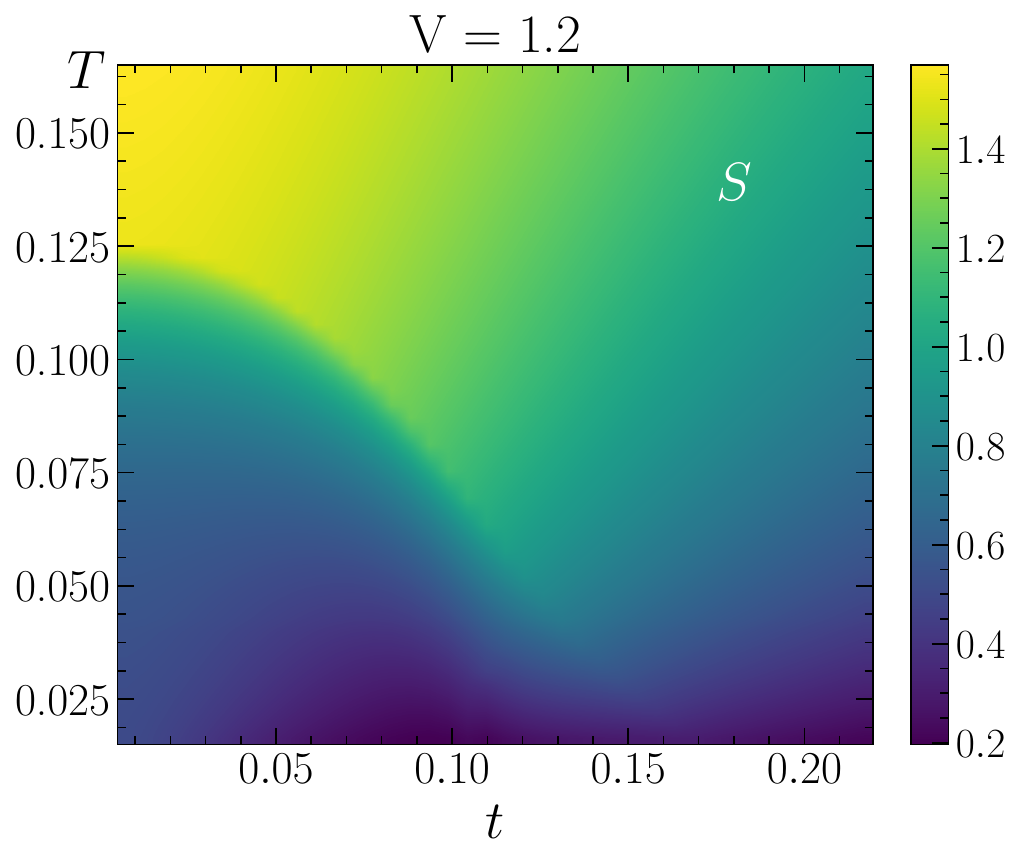}}
	\subfloat[$C_V(T)$]{\label{specific heat for 3-orbital phase diagram on cubic lattice with V = 1.2}\includegraphics[width=0.33\textwidth, height = 0.25\textwidth]{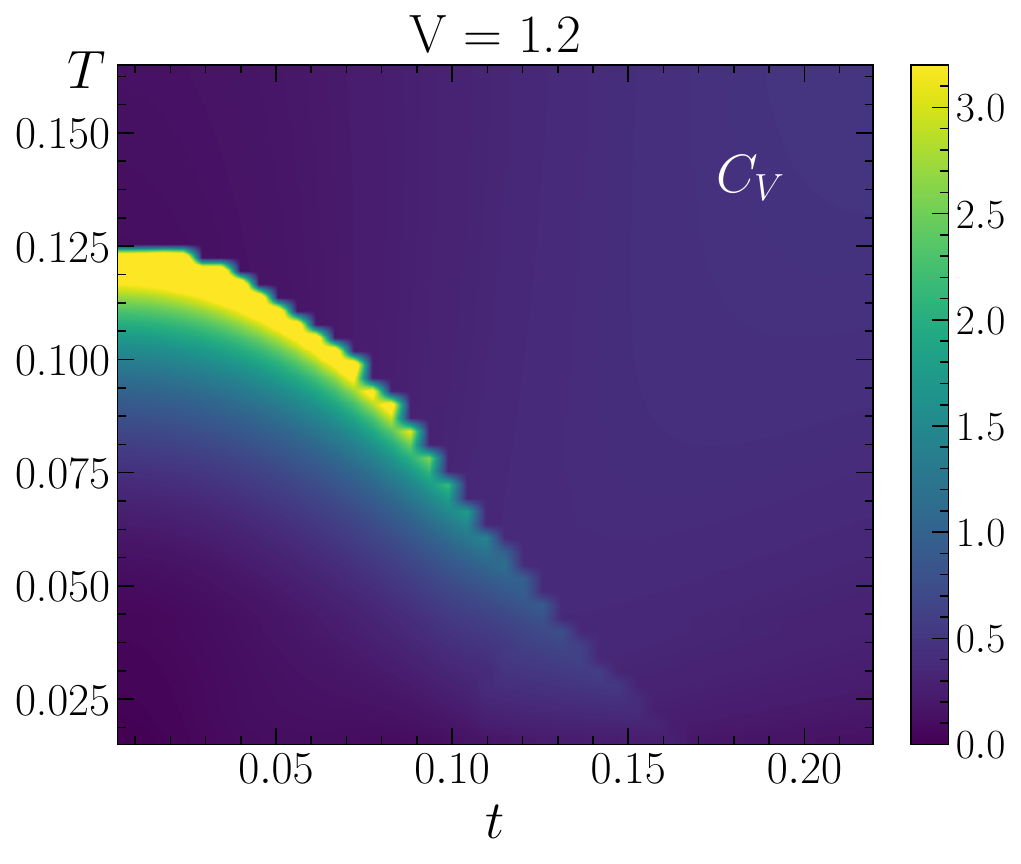}}
	\caption{Additional Thermodynamic quantities of the three orbital cubic lattice model: (a)-(c) $V = 0.8$, and (d)-(e) $V = 1.2$. }
	
	\label{3-orbital phase diagram data on cubic lattice}
\end{figure}

In this Appendix, we provide additional details on the phase diagrams of the cubic lattice model discussed in Sec.~\ref{Section III B: Phase diagram}. Fig.~\ref{3-orbital phase diagram data on cubic lattice} shows several thermodynamic quantities for $V = 0.8$ and $V = 1.2$, including  the entropy $S(T) = -\left( \partial \Omega/\partial T \right)_{V,\mu}$, the specific heat $C_V(T) = T\left( \partial S/\partial T \right)_{V,\mu}$, and the compressibility $\kappa(T) = n^{-2}\left(\partial^2 \Omega/\partial \mu^2\right)_{V,T}$, that are not shown in the main text. The entropy plots in Fig.~\ref{entropy for 3-orbital phase diagram on cubic lattice  with V = 0.8} and \ref{entropy for 3-orbital phase diagram on cubic lattice with V = 1.2} indicate that the isotropic phase at high $T$ and at small hopping $t$ retains a large entropy as in the three-orbital SYK dot. This is indicative of a nFL phase for temperatures exceeding the crossover scale $t^2/J$ to a Fermi liquid \cite{Song2017a}. Notably, since the $T_c$ for the isotropic-nematic transition is higher than the crossover scale, the $[0,0,0]$-to-$[+,0,-]$ transition is an example of a nFL-to-nematic transition. This transition features a divergence in specific heat as shown in Fig.~\ref{specific heat for 3-orbital phase diagram on cubic lattice  with V = 0.8} and \ref{specific heat for 3-orbital phase diagram on cubic lattice with V = 1.2}. Another interesting feature of the phase diagram is the strongly enhanced compressibility at the phase boundary of the $[+,-,-]_{\rm M}$ phase, which occurs at both $V = 0.8$ (Fig.~\ref{compressibility for 3-orbital phase diagram on cubic lattice with V = 0.8}) and $V = 1.2$ as discussed in the main text.

 \subsection{Discussion on the $[+,0,-]$ phase}
 \label{Appendix: NFL}
\begin{figure}[H]
	\centering
 \subfloat[]{ \includegraphics[width=0.49\linewidth,height = 0.4\linewidth]{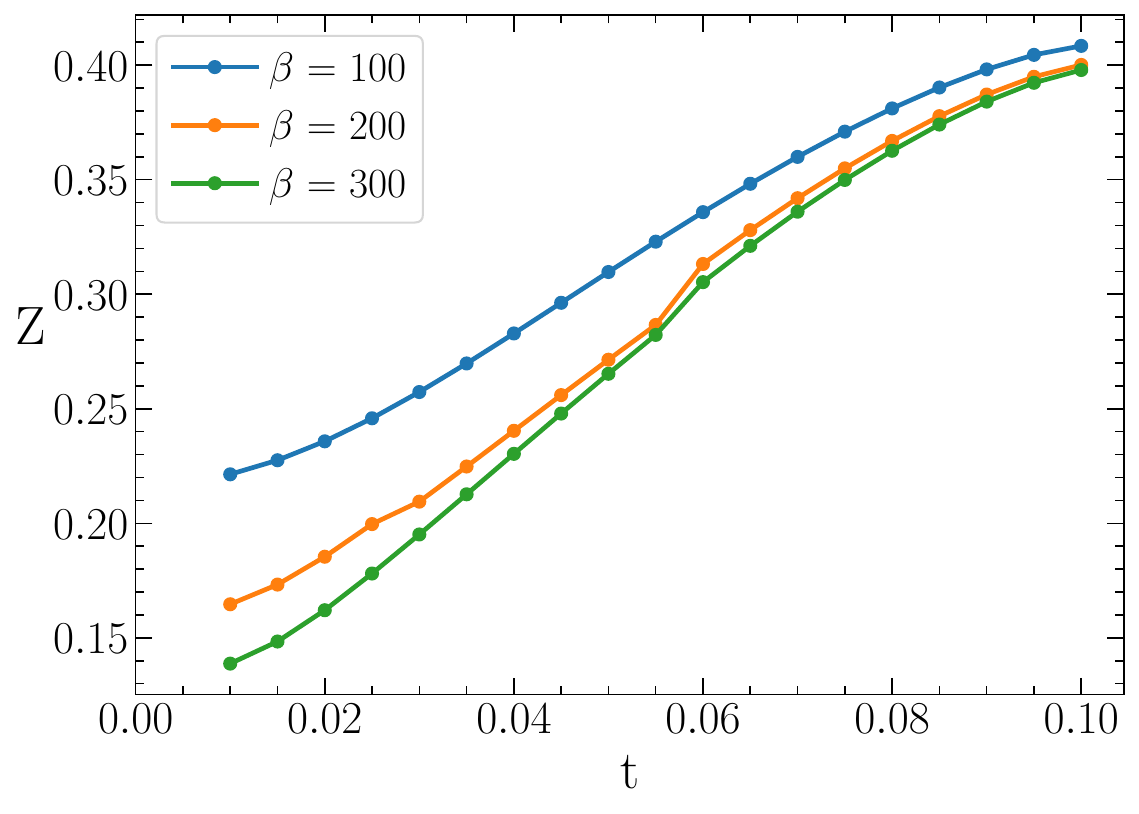}\label{Cubic_Z_vs_hopping} } \hfill
	\subfloat[]{\includegraphics[width=0.49\linewidth,height = 0.4\linewidth]{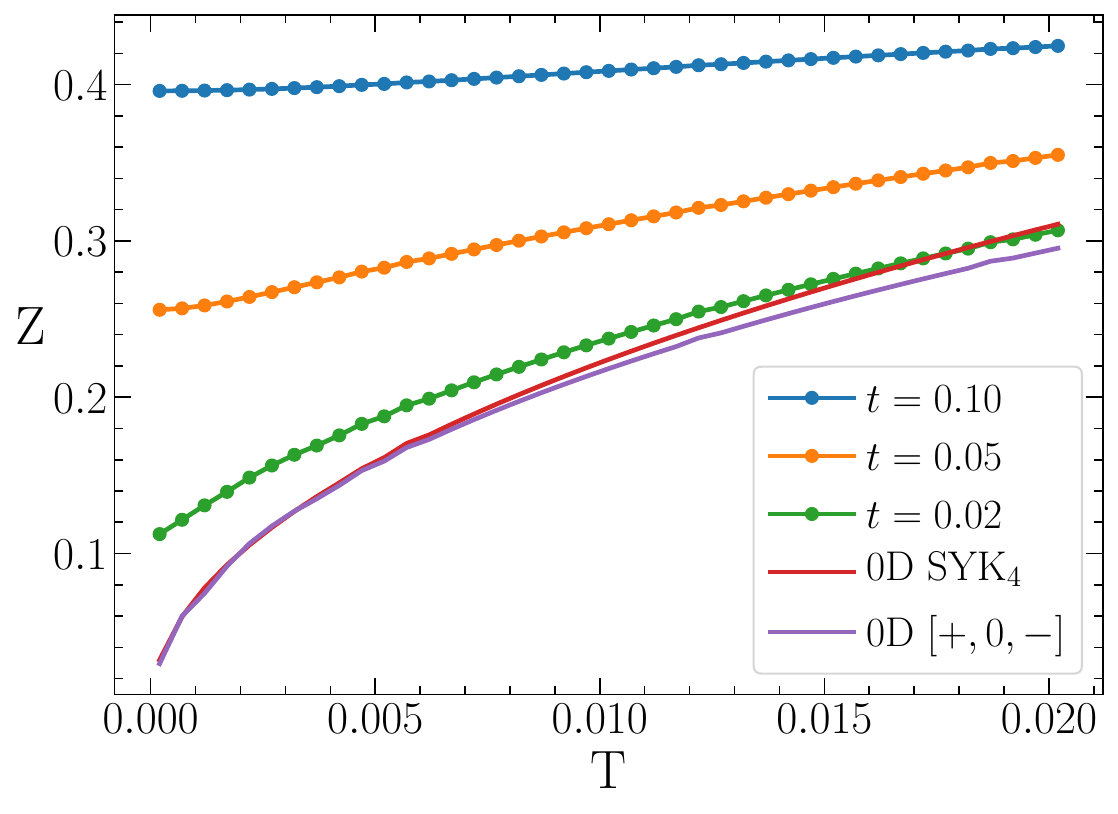}\label{Cubic_Z_vs_Temperature}}  
 \caption{The quasiparticle residue $Z$: (a) as a function of hopping amplitude at different temperatures, (b) as a function of temperature for the cubic lattice model at different hopping amplitudes $t$ (with markers), and the zero-dimensional models of one orbital(red) and of three orbitals(purple). Results shown correspond to the parameters $V = 1.2, \delta t/t = 0.8$. }
	\label{Cubic quasiparticle residue}
\end{figure}

 In order to explore the FL/NFL nature of the $[+,0,-]$ phase, we extract the quasiparticle residue $Z = \left[ 1 - \frac{\partial \text{Im}\Sigma(i\omega_n)}{\partial \omega}\big|_{\omega = 0} \right]^{-1}$ from the imaginary-frequency self-energy $\Sigma(i\omega_n)$ of the metallic orbital. 
Fig.~\ref{Cubic_Z_vs_hopping} shows that the temperature dependence of $Z$ decreases with increasing hopping $t$. 
 Fig.~\ref{Cubic_Z_vs_Temperature} shows a saturation of $Z$ at a temperature scale $T \lessapprox t^2/J (J = 1)$, which we found to have little $V$-dependence.  In the 0D limit, $Z$ drop substantially to zero at low temperatures. We also observe the convergence of $Z$ at low temperature between the original SYK model and the $[+,0,-]$ phase in the 0D three-orbital SYK model, indicating their shared low-frequency behaviour. In order to access low temperatures, we have employed the discrete Lehman representation \cite{kaye2022discrete} within the TRIQS (Toolbox for Research on Interacting Quantum Systems) Library \cite{parcollet2015triqs}.

 \subsection{Discussion on Anisotropy}
\begin{figure}[H]
	\centering
	\subfloat[$\delta t /t = 0.0$]{\includegraphics[width=0.32\linewidth]{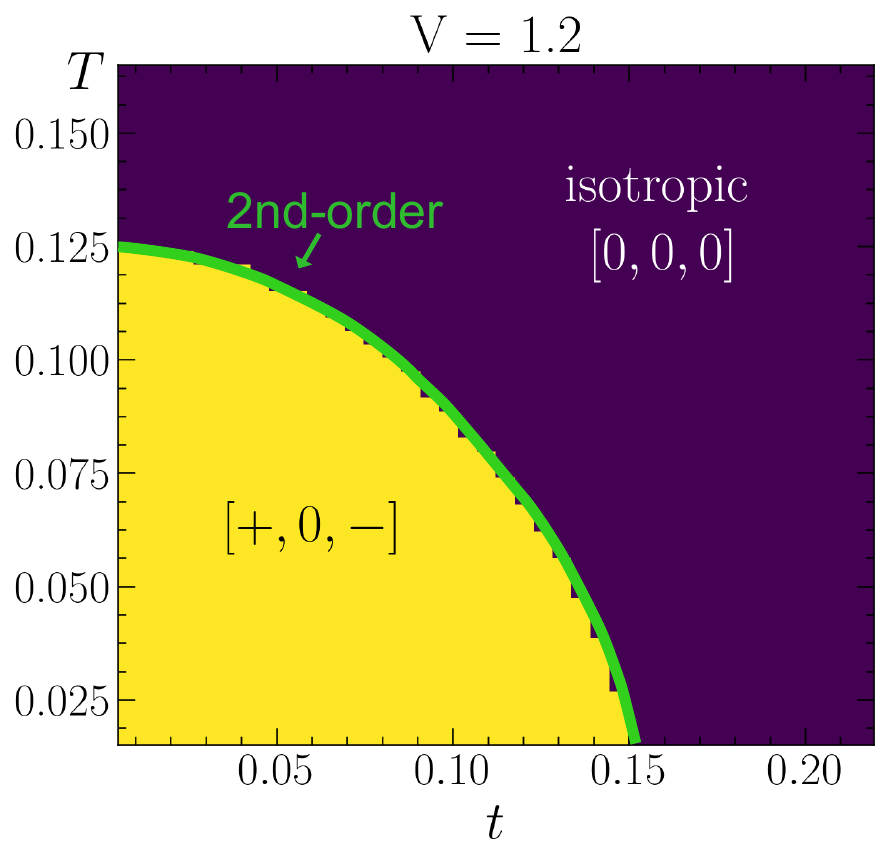}} \hfill
	\subfloat[$\delta t/t = 0.4$]{ \includegraphics[width=0.32\linewidth]{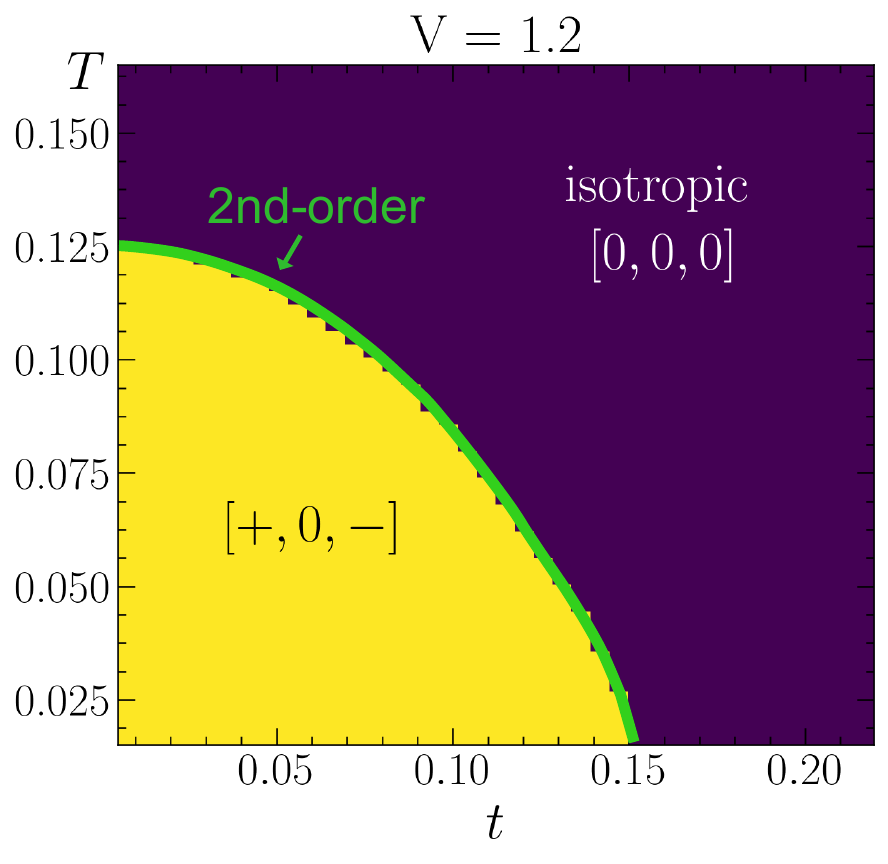} }\hfill
	\subfloat[$\delta t/t = 0.8$]{ \includegraphics[width=0.32\linewidth]{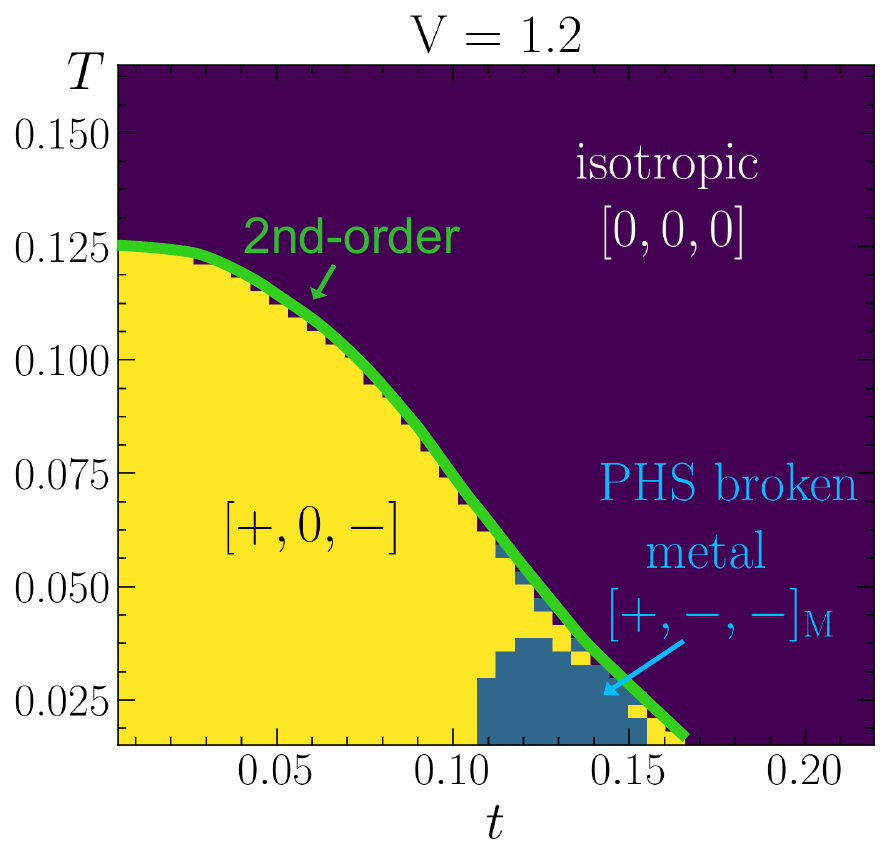} }\hfill
	\caption{Phase diagrams corresponding to different values of anisotropy $\d t$ in the cubic lattice model. }
	\label{Phase diagrams with different anisotropy for Cubic lattice model}
\end{figure}

  In the main text, we present results for a single value of anisotropy $\delta t/t = 0.8$. Here, we discuss the effect of anisotropy on the phase diagram. Fig.~\ref{Phase diagrams with different anisotropy for Cubic lattice model} shows phase diagrams with three different values of anisotropy at $V = 1.2$. The phase diagrams exhibit similar features qualitatively, except that the PHS broken phase $[+,-,-]_{\rm M}$ is suppressed by decreasing anisotropy. We attribute this suppression to the disappearance of the peak of $g(\mu)$ at small anisotropies. Further numerical results suggest that the window of the PHS broken metal $[+,-,-]_{\rm M}$ occurs when $\delta t / t > 0.5$.
  
\section{Elaboration on the Triangular Lattice Model} \label{Appendix: Further discussion on the triangular lattice model}
\subsection{Thermodynamic quantities}

\begin{figure}[H]
	\centering
	\subfloat[$S(T)$]{\label{entropy for 3-orbital phase diagram on triangular lattice with V = 0.8}\includegraphics[width=0.33\textwidth, height = 0.25\textwidth]{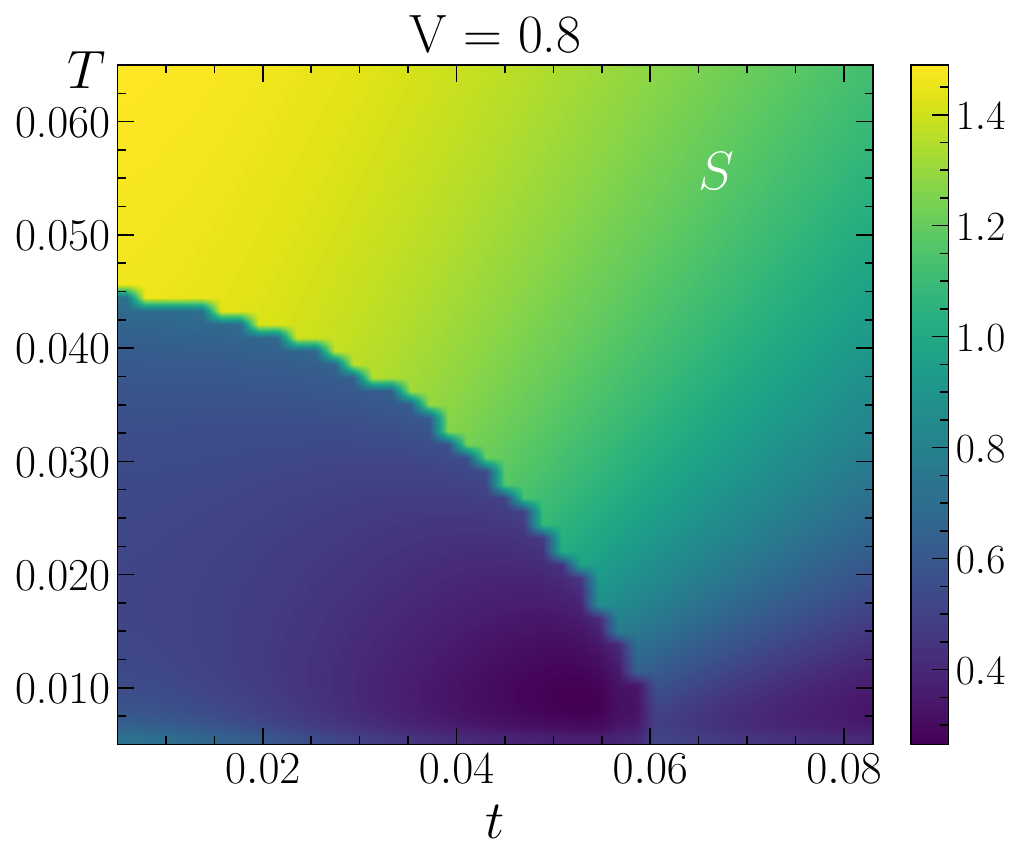}}
	\subfloat[$C_V(T)$]{\label{specific heat for 3-orbital phase diagram on triangular lattice with V = 0.8}\includegraphics[width=0.33\textwidth, height = 0.25\textwidth]{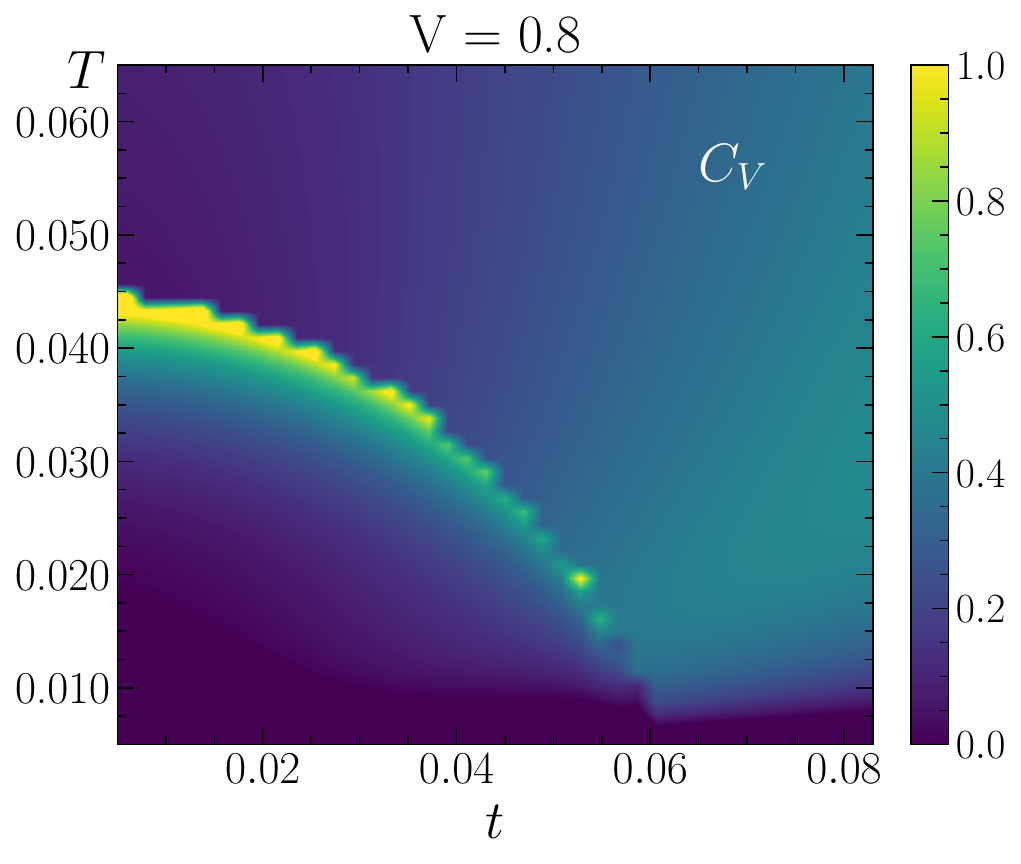}}
	\subfloat[$\k(T)$]{\label{compressibility for 3-orbital phase diagram on triangular lattice with V = 0.8}\includegraphics[width=0.33\textwidth, height = 0.25\textwidth]{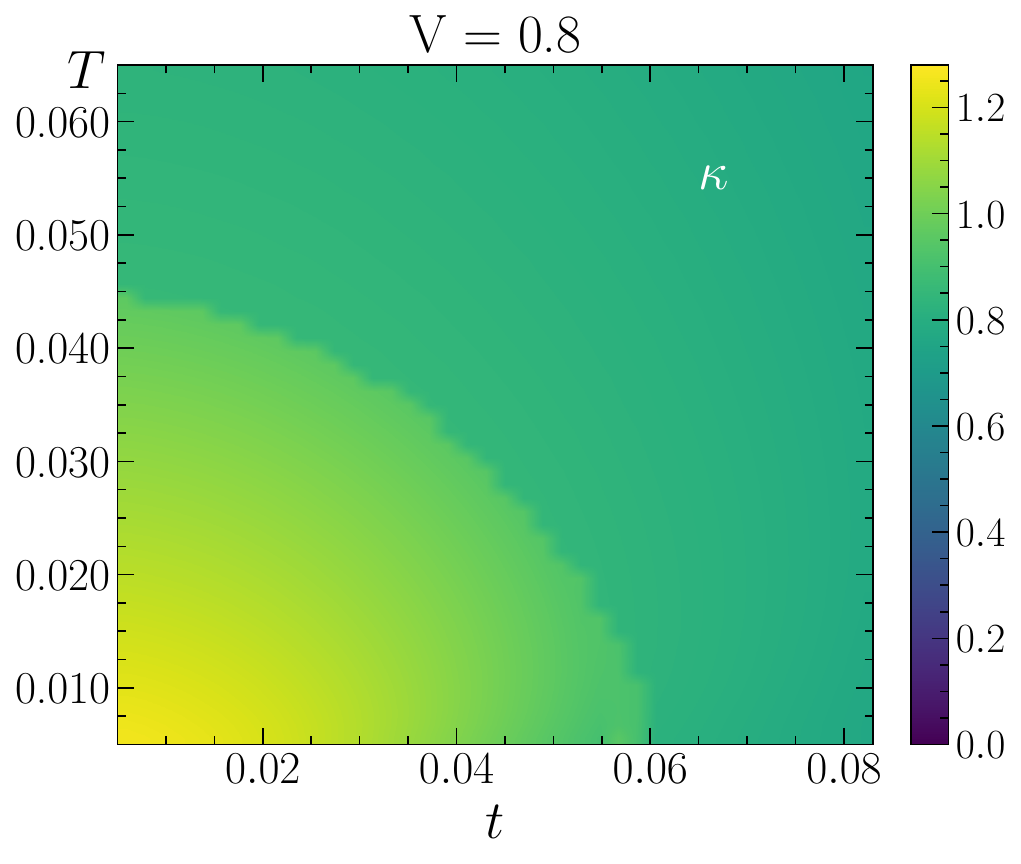}}

	\subfloat[$S(T)$]{\label{entropy for 3-orbital phase diagram on triangular lattice with V = 1.2}\includegraphics[width=0.33\textwidth, height = 0.25\textwidth]{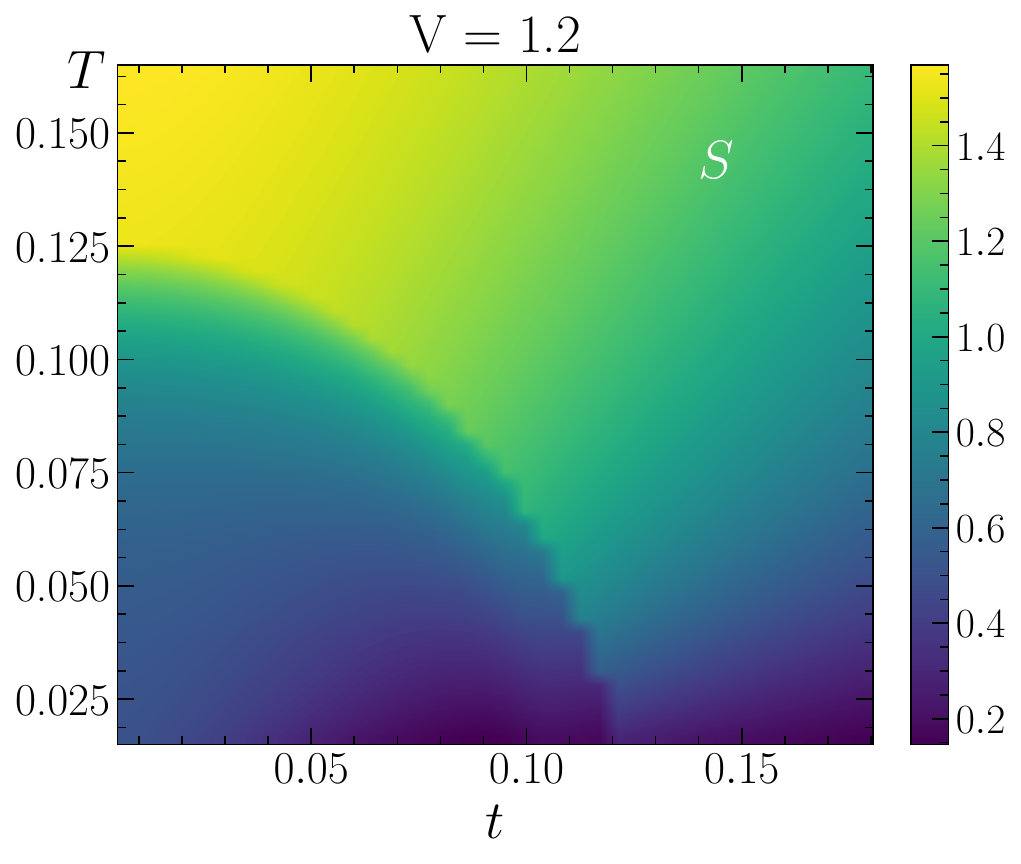}}
	\subfloat[$C_V(T)$]{\label{specific heat for 3-orbital phase diagram on triangular lattice with V = 1.2}\includegraphics[width=0.33\textwidth, height = 0.25\textwidth]{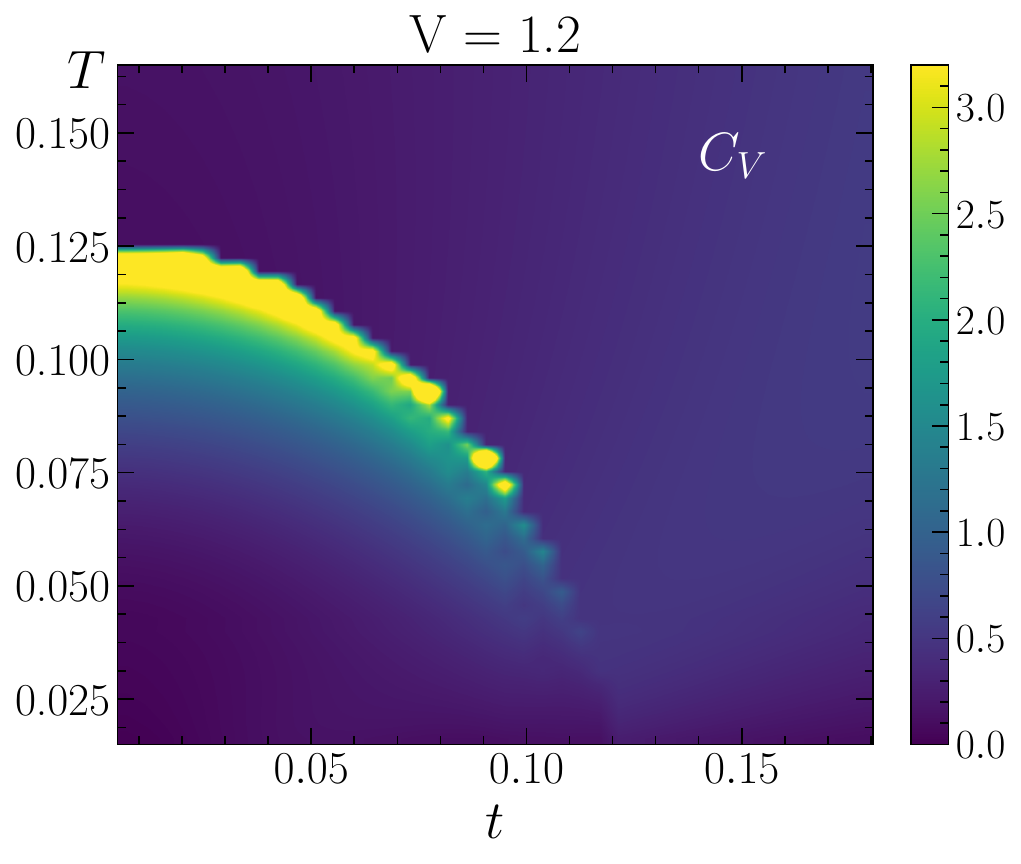}}
 
	\caption{Additional Thermodynamic quantities of the three orbital triangular lattice model: (a)-(c) $V = 0.8$, and (d)-(e) $V = 1.2$.}

	\label{appendix 3-orbital phase diagram data on triangular lattice}
\end{figure}

In this Appendix, we provide additional details on the phase diagrams of the triangular lattice model discussed in Sec.~\ref{Section IV B: Phase diagram}. Fig.~\ref{3-orbital phase diagram data on triangular lattice} shows several thermodynamic quantities for $V = 0.8$ and $V = 1.2$, including  the entropy $S(T) = -\left( \partial \Omega/\partial T \right)_{V,\mu}$, the specific heat $C_V(T) = T\left( \partial S/\partial T \right)_{V,\mu}$, and the compressibility $\kappa(T) = n^{-2}\left(\partial^2 \Omega/\partial \mu^2\right)_{V,T}$, that are not shown in the main text. As discussed in Sec.~\ref{Section IV B: Phase diagram} of the main text, depending on $V$ and $t$, the phase diagram reveal two types of isotropic-to-nematic phase transition: $[0,0,0]$-to-$[+,-,-]_{\rm M}$ and $[0,0,0]$-$[a,b,c]$. Both transitions feature a divergence in specific heat, as shown in Fig.~\ref{specific heat for 3-orbital phase diagram on triangular lattice with V = 0.8} and \ref{specific heat for 3-orbital phase diagram on triangular lattice with V = 1.2}. The $[+,-,-]_{\rm M}$-to-$[a,b,c]$ transition feature a finite peak at specific heat. The compressibility plot for $V = 0.8$ show similar feature as $V = 1.2$ discussed in Fig.~\ref{Cubic lattice compressibility plot}, which only gradually enhanced at the phase boundary of the $[+,-,-]_{\rm M}$. This qualitatively differs from that seen in the cubic lattice model.

\subsection{Discussion on Anisotropy}
\begin{figure}[H]
	\centering
	\subfloat[$\delta t /t = 0.0$]{\includegraphics[width=0.32\linewidth]{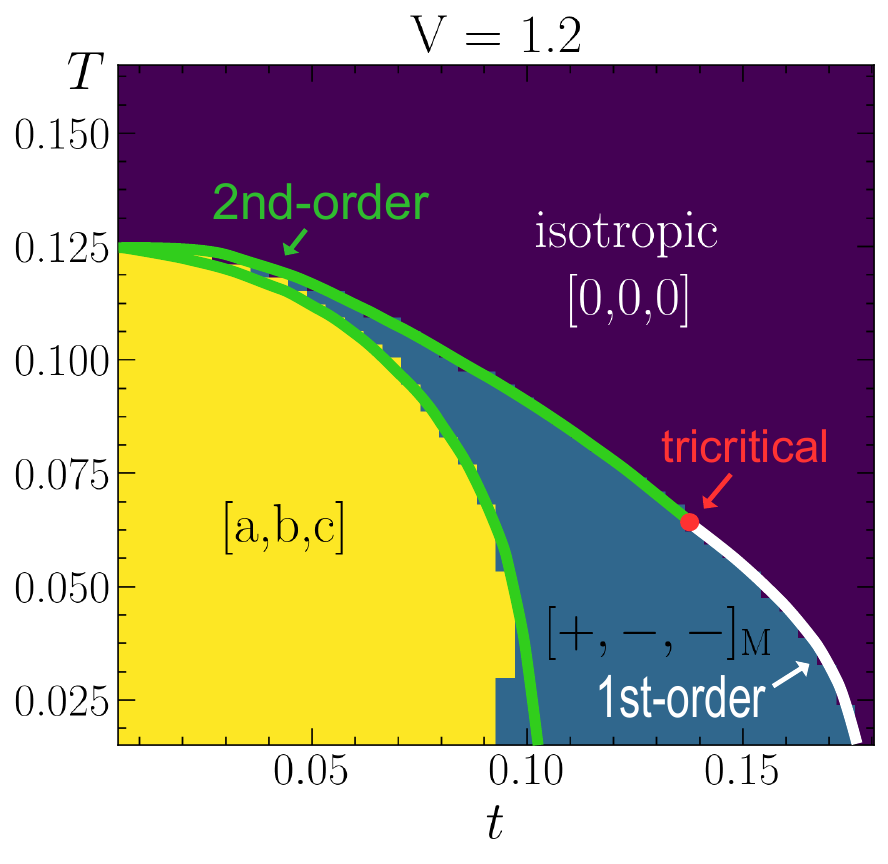}} \hfill
	\subfloat[$\delta t/t = 0.4$]{ \includegraphics[width=0.32\linewidth]{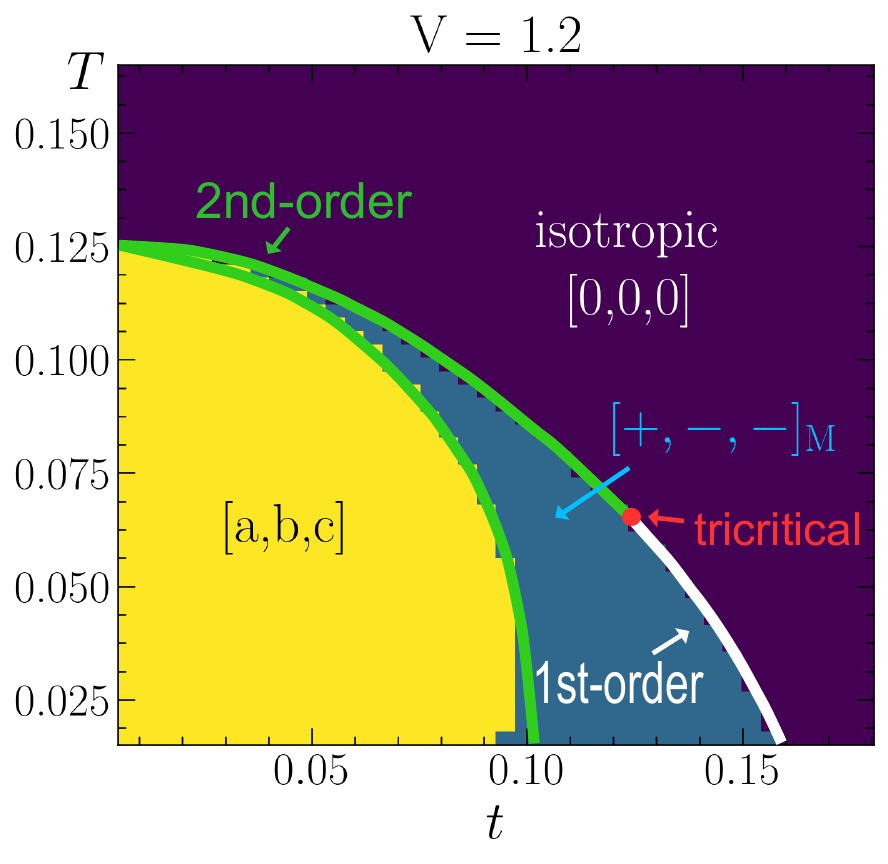} }\hfill
	\subfloat[$\delta t/t = 0.8$]{ \includegraphics[width=0.32\linewidth]{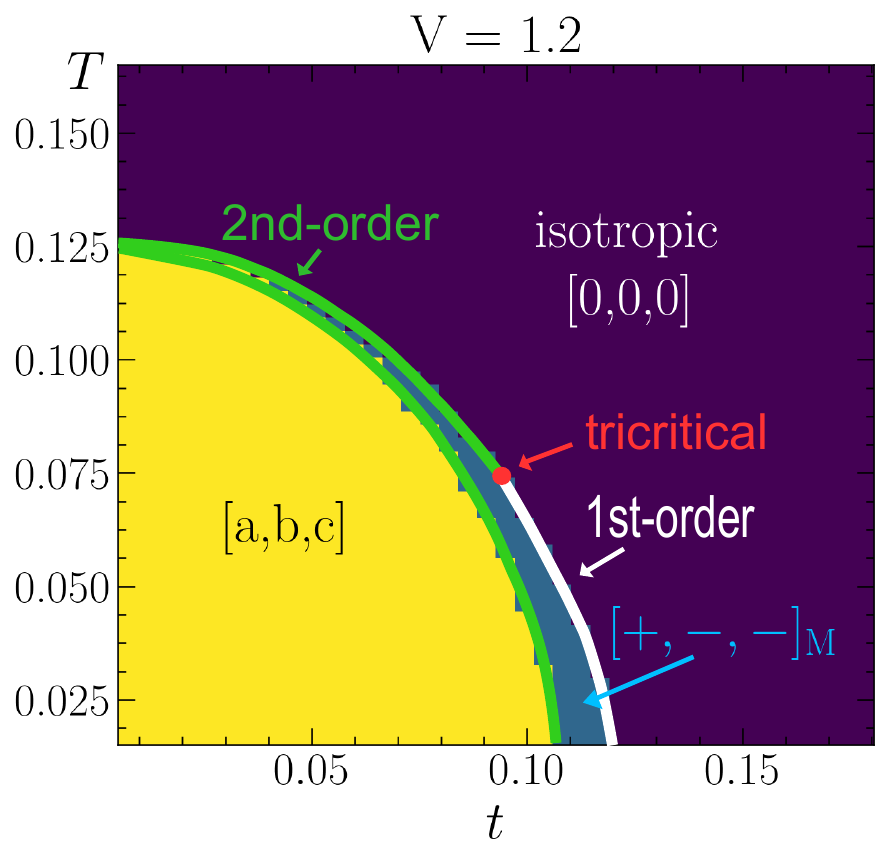} }\hfill
	\caption{Phase diagrams corresponding to different values of anisotropy $\d t$ in the triangular lattice model.  }
	\label{Phase diagrams with different anisotropy for triangular lattice model}
\end{figure}

In the main text, we present results for a single value of anisotropy $\delta t/t = 0.8$. Here, we discuss the effect of anisotropy on the phase diagram. Fig.~\ref{Phase diagrams with different anisotropy for triangular lattice model} shows phase diagrams with three different values of anisotropy at $V = 1.2$.  In the triangular lattice model, the phase diagrams show qualitatively similar results except that at low anisotropy, the $[+,-,-]_{\rm M}$ phase are stable for a wider range of hopping. The phase boundary at $t \sim 0.95$ has nearly degenerate states that make the precise boundary difficult to distinguish.

\section{Elaboration on the Spectral Gap}
\label{Appendix:Gap}
\begin{figure}[H]
	\centering
	\subfloat[gap $\Delta(T)$]{\includegraphics[width=0.49\linewidth,height = 0.4\linewidth]{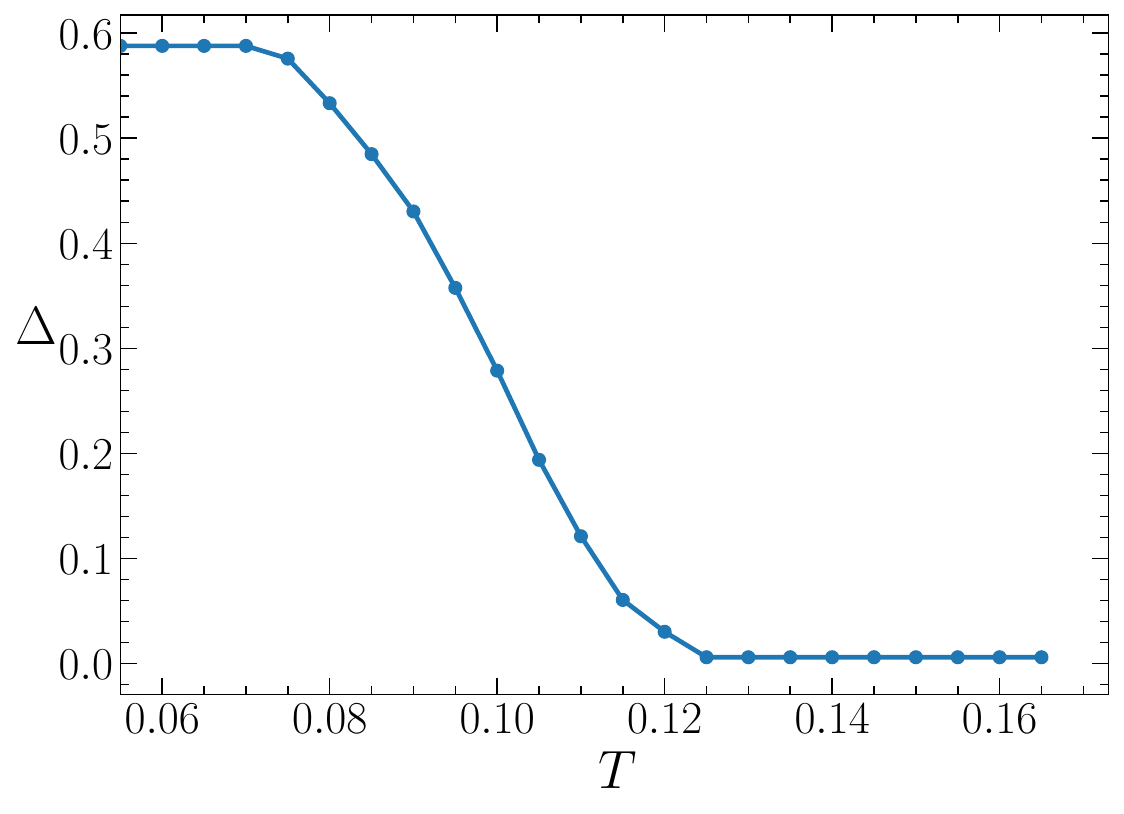} \label{gap vs T for 0D}} \hfill
	\subfloat[$T$-dependence of spectral weight]{ \includegraphics[width=0.45\linewidth,height = 0.4\linewidth]{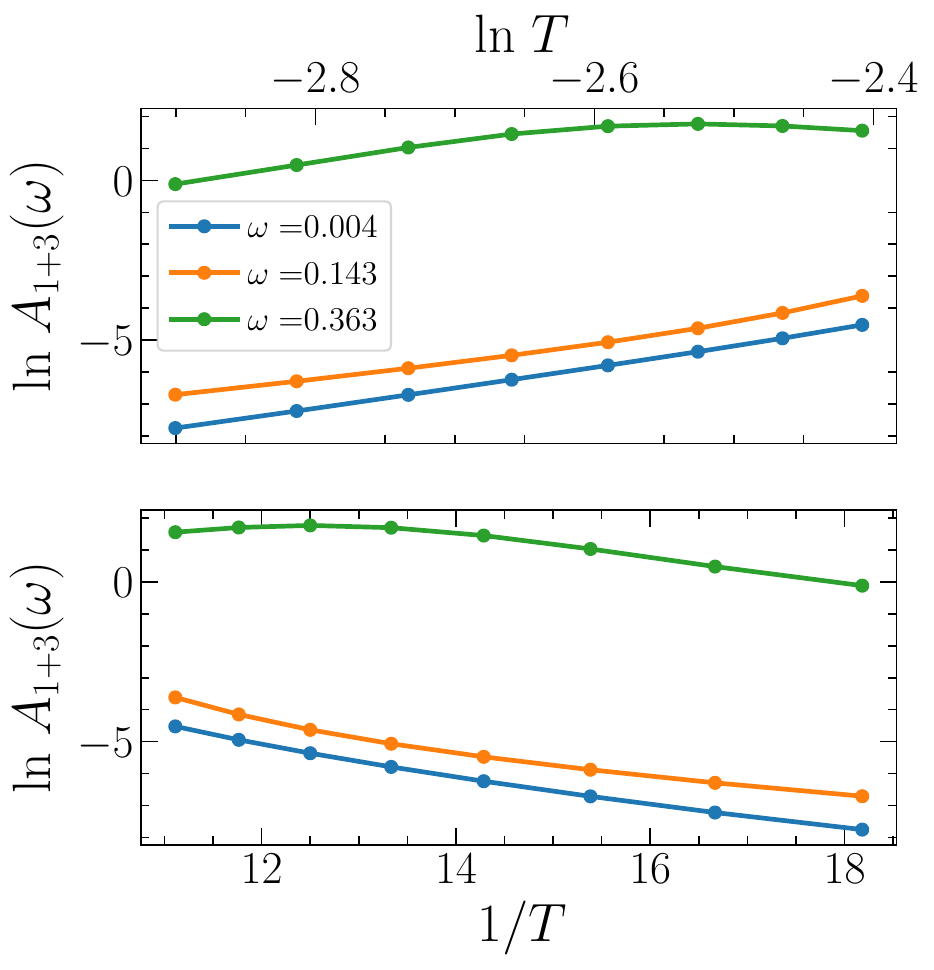} \label{Spectral weights vs T for 0D} } \hfill
 \caption{ (a) The gap, $\Delta$, as a function of temperature for 0D at $V = 1.2$, with the parameters matching Fig.~\ref{Examples of spectral functions for 0D model}. (b) The logarithm of the corresponding spectral weights $A_{1+3}(\omega)$ at three selected values of $\omega$ as a function of logarithmic temperature and inverse temperature respectively. These values of $\omega$ are chosen such that one is closest to $\omega = 0$, one lies near the gap boundaries, and one is outside the gap.  }
\end{figure}

\begin{figure}[H]
	\centering
	\subfloat[gap $\Delta(T)$]{ \includegraphics[width=0.49\linewidth,height = 0.4\linewidth]{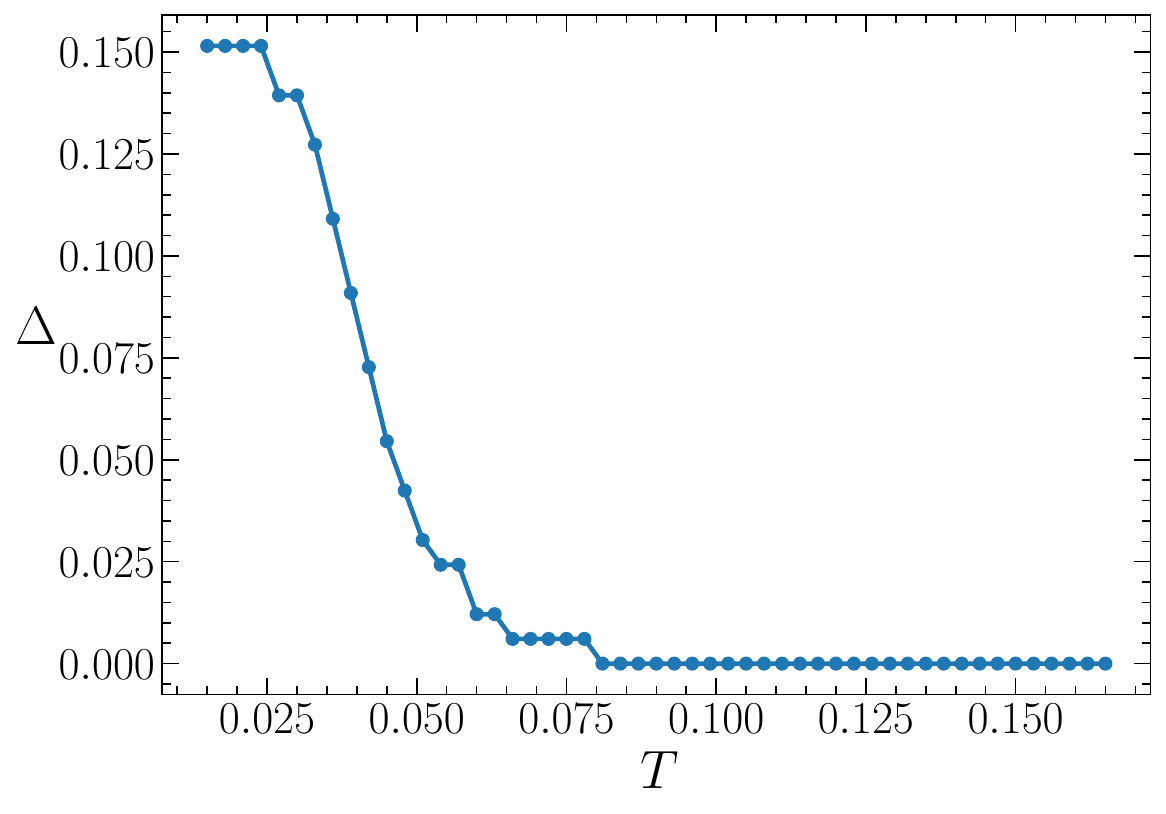}  \label{gap vs T for cubic lattice} }\hfill
 \subfloat[$T$-dependence of spectral weight]{ \includegraphics[width=0.45\linewidth,height = 0.4\linewidth]{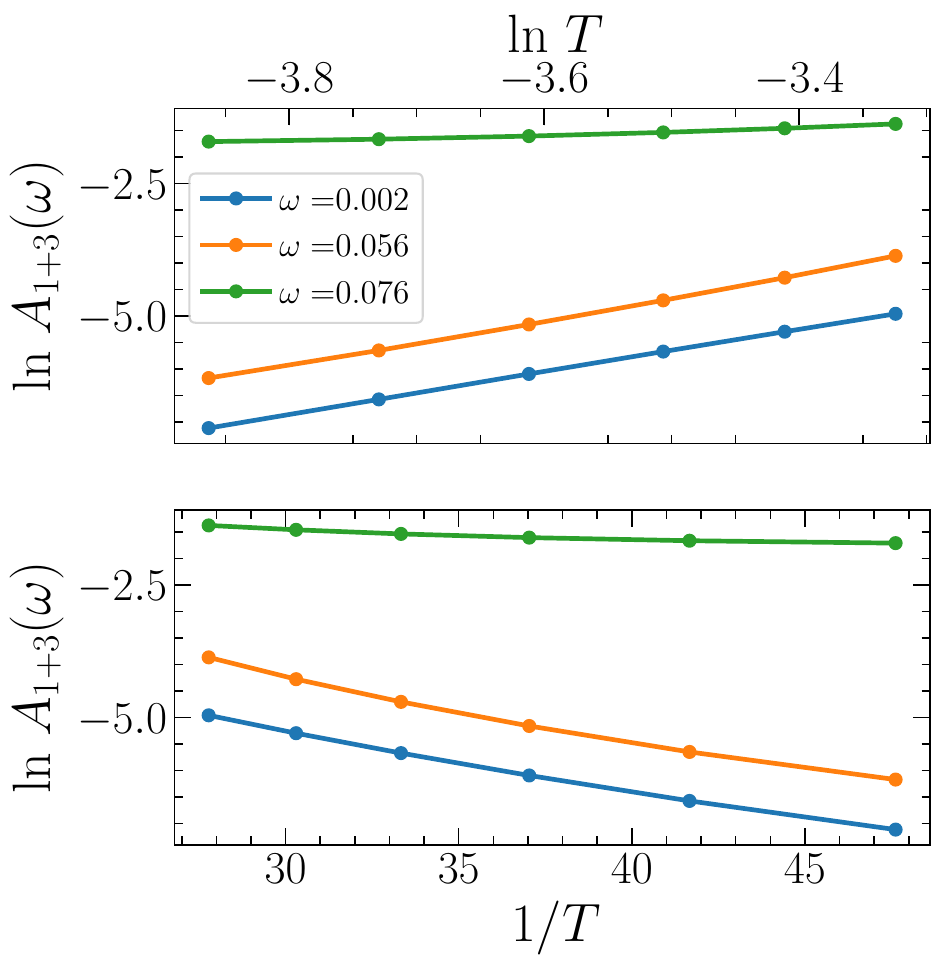} \label{Spectral weights vs T for cubic lattice} } \hfill
 \caption{ (a) The gap, $\Delta$, as a function of temperature for the cubic lattice model, with the parameters  $V = 1.2, t = 0.095, \delta t/t = 0.8$  matching Fig.~\ref{Examples of spectral functions for cubic lattice model with V = 1.2J.}. (b) The logarithm of the corresponding spectral weights $A_{1+3}(\omega)$ at three selected values of $\omega$ as a function of logarithmic temperature and inverse temperature respectively. }
\end{figure}

In Sec.~\ref{Section I C: Spectral Functions}, \ref{section IV: C Spectral Functions} for the 0D and the cubic lattice model, we argue that two of the orbitals become gapped in the $[+,0,-]$ phase at low temperature. To better quantify the gap, we define the following quantity $\Delta$ as the definition of the gap:
\begin{gather}
\Delta(T) : \quad  \int_{-\Delta/2}^{\Delta/2}d\omega  A_{1+3}(\omega,T)/2  = \delta ,
\end{gather}
where $A_{1+3}(\omega, T)$ is the sum of spectral weights from the two orbitals, $\delta_{0D} = 0.01$ and $\delta_{3D} = 0.001$.  Due to a finite value of broadening $\eta = 10^{-4}$,  we expect finite spectral weights within the gap.  The gaps $\Delta$ for both 0D and cubic lattice model as a function of temperature are shown in Fig.~\ref{gap vs T for 0D} and \ref{gap vs T for cubic lattice} respectively. For both models, $\Delta(T)$ saturates at low temperatures, indicating a formation of a gap. The temperature dependence of the spectral weights are shown in Fig.~\ref{Spectral weights vs T for 0D} and \ref{Spectral weights vs T for cubic lattice}. At low temperatures, the spectral weight within the gap is suppressed, while the spectral weight outside the gap saturates to some finite values. For both models, this suppression can be well-fitted by either a power-law or an exponential decay. Nevertheless, we argue that the power-law decay provides a better fit, and therefore we refer to the gap as a (pseudo)gap.

\begin{figure}[H]
	\centering
	\includegraphics[width=0.49\linewidth,height = 0.4\linewidth]{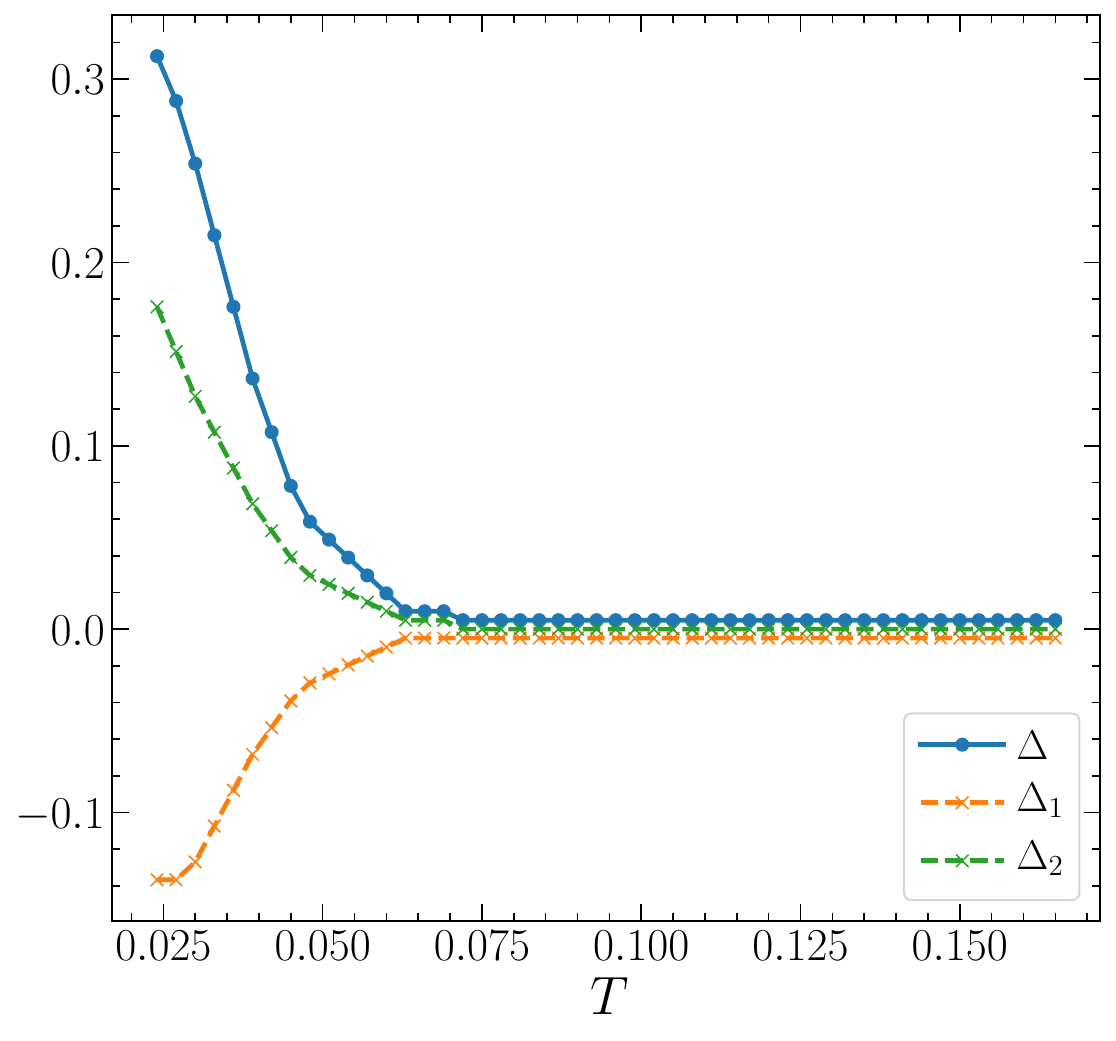} 
 \caption{ The gap $\Delta$ as a function of temperature for the triangular lattice model, with the parameters $V = 1.2, t = 0.095, \delta t/t = 0.8$ matching Fig.~\ref{Examples of spectral functions for triangular lattice model with V = 1.2J.}. }
	\label{gap vs T for triangular lattice}
\end{figure}

In Sec.~\ref{section V: C Spectral Functions} on the triangular lattice, only one orbital of the $[a,b,c]$ phase becomes gapped. We define the gap to be
\begin{gather}
\Delta(T) =  \text{max} \left \{ |\Delta_2 - \Delta_1|: \quad \int_{\Delta_1}^{\Delta_2}  d\omega A_{1}(\omega,T) = \delta \text{ and } 0\in [\Delta_1, \Delta_2] \right \},
\end{gather}
with the chosen value of $\delta = 0.001$, and $[\Delta_1,\Delta_2]$ represents the size of the gap. The magnitude of the gap $\Delta, \Delta_1, \Delta_2$ as a function of temperature is shown in Fig.~\ref{gap vs T for triangular lattice}. Since the gap has not saturated at the lowest accessible temperature, we omit the analysis of the $T$-dependence of the spectral weight.

\clearpage
\end{document}